\documentclass[a4paper,10pt]{article}
\pdfoutput=1 % if your are submitting a pdflatex (i.e. if you have
\usepackage{jheppub} % for details on the use of the package, please
\usepackage{float}
\usepackage{graphicx}
\usepackage{subcaption}
\usepackage{bbm}
\usepackage{siunitx}
\usepackage{listings}

\usepackage{graphicx,epsf}
\usepackage{amsmath,amsfonts,amssymb,amsbsy}
\usepackage{mathtools}
\usepackage{physics}

\usepackage[dvipsnames]{xcolor}

\usepackage{hyperref}
\usepackage{xcolor}

\usepackage{multirow}
\usepackage{stackrel}
\usepackage{tikz}
\usetikzlibrary{calc}

\usepackage{ifdraft}
\usepackage[obeyFinal]{easy-todo}

\usepackage{placeins}
\usepackage{slashed} % For Dirac slashed notation

\usepackage{booktabs}
\usepackage{adjustbox}

\usepackage[utf8]{inputenc}

\usepackage{graphicx}
\usepackage{pdfpages}
\usepackage{makecell}

\usepackage{cleveref}
\usepackage{pdflscape}

\usepackage{pifont}

\usepackage{enumitem}

\newcommand{\be}{\begin{equation}}
\newcommand{\ee}{\end{equation}}

\newcommand{\bea}{\begin{eqnarray}}
\newcommand{\eea}{\end{eqnarray}}

\def\eqn#1{eq.~(\ref{#1})}

\def\spa#1.#2{\left\langle#1\,#2\right\rangle}
\def\spb#1.#2{\left[#1\,#2\right]}
\def\spash#1.#2{\spa{\smash{#1}}.{\smash{#2}}}
\def\spbsh#1.#2{\spb{\smash{#1}}.{\smash{#2}}}
\def\sand#1.#2.#3{%
\left\langle\smash{#1}{\vphantom1}^{-}\right|{#2}%
\left|\smash{#3}{\vphantom1}^{-}\right\rangle}
\def\sandpp#1.#2.#3{%
\left\langle\smash{#1}{\vphantom1}^{+}\right|{#2}%
\left|\smash{#3}{\vphantom1}^{+}\right\rangle}
\def\sandpm#1.#2.#3{%
\left\langle\smash{#1}{\vphantom1}^{+}\right|{#2}%
\left|\smash{#3}{\vphantom1}^{-}\right\rangle}
\def\sandmp#1.#2.#3{%
\left\langle\smash{#1}{\vphantom1}^{-}\right|{#2}%
\left|\smash{#3}{\vphantom1}^{+}\right\rangle}

\def\trFive{{\rm tr}_5}

\def\NC{{N_c}}
\def\NF{{N_f}}

\def\degn{{\rm deg}_n}
\def\degd{{\rm deg}_d}

\def\Bdeg{\mathcal{B}^{\rm deg}}
\def\Bimp{\mathcal{B}^{\rm imp}}

\newcommand{\ii}{\,\mathrm{i}\,}
\newcommand{\gluon}{\mathrm{g}}
\newcommand{\quark}{q}
\newcommand{\Quark}{Q}

\def\trFive{{\rm tr}_5}

\newcommand{\pval}[1]{\nu_p\qty[#1]}

\DeclareUnicodeCharacter{27E8}{\langle}
\DeclareUnicodeCharacter{27E9}{\rangle}
\DeclareUnicodeCharacter{0394}{\Delta}
\DeclareUnicodeCharacter{27EA}{\langle\langle}
\DeclareUnicodeCharacter{27EB}{\rangle\rangle}

\newcolumntype{C}[1]{>{\centering\arraybackslash}m{#1}}
\newcolumntype{M}{>{\centering\arraybackslash$}l<{$}}

\title{Compact Two-Loop QCD Corrections for $Vjj$ Production in Proton Collisions}

\preprint{PSI-PR-24-29, ZU-TH 14/25}

\author[1]{Giuseppe~De~Laurentis,}
\author[2,3]{Harald~Ita,}
\author[4]{Ben~Page,}
\author[5]{Vasily~Sotnikov}

\affiliation[1]{Higgs Centre for Theoretical Physics, University of Edinburgh, Edinburgh, EH9 3FD, United Kingdom}
\affiliation[2]{Paul Scherrer Institut, 5232 Villigen PSI, Switzerland}
\affiliation[3]{Department of Astrophysics, University of Zurich, Winterthurerstrasse 190, 8057 Zurich, Switzerland}
\affiliation[4]{Department of Physics and Astronomy, Ghent University, 9000 Ghent, Belgium}
\affiliation[5]{Physik-Institut, University of Zurich, Winterthurerstrasse 190, 8057 Zurich, Switzerland}

\emailAdd{
  giuseppe.delaurentis@ed.ac.uk,
  harald.ita@psi.ch,
  ben.page@ugent.be,
  vasily.sotnikov@physik.uzh.ch
}

\abstract{
  We present compact two-loop QCD corrections in the leading-color approximation
  for the production of an electroweak vector boson, $V = \{W^{\pm},
  Z,\gamma^\star\}$, in association with two light jets ($Vjj$) at hadron
  colliders. Leptonic decays of the electroweak boson are included at the
  amplitude level.
  Working in the analytic-reconstruction approach, we develop two techniques to
  build compact partial-fraction forms for individual rational functions. One
  approach exploits their analytic structure. In the other, we iteratively
  construct subtraction terms that match the rational functions in singular
  limits.
  Moreover, we show how the singular behavior of the rational functions
  can be systematically used to find a more compact basis of the space that they
  span.
  We apply our techniques to the $Vjj$ amplitudes, yielding a representation
  that is three orders of magnitude smaller than previous results.
  We then use these compact expressions to provide an efficient and stable
  \texttt{C++} numerical implementation suitable for phenomenological applications.
}

\begin{document}
\maketitle

\section{Introduction}
\label{sec:introduction}

The production of an electroweak (EW) vector bosons $V = \{W^{\pm}, Z, \gamma^\star\}$ in association with two jets ($Vjj$)
at hadron colliders is an important scattering process, with
significant implications for both precision studies of the Standard Model (SM) and
searches for new physics. These processes serve as critical backgrounds for
top-quark production, Higgs physics, and direct searches for
beyond-the-Standard-Model phenomena. Furthermore, they act as a
laboratory for testing the robustness of perturbative quantum chromodynamics
(QCD) in multi-scale processes, where the interplay of strong and
electroweak interactions manifests in complicated kinematics. 
Theoretically, $Vjj$ production is particularly interesting due to its
anticipated rapid convergence within perturbation theory. At this jet
multiplicity, all contributing production channels are already present at leading order (LO), 
with next-to-leading-order (NLO) QCD corrections having a
relatively moderate impact across phase space and only limited dependence
on higher-order effects
\cite{Campbell:2003hd,Berger:2009ep,Anger:2017nkq,Kallweit:2015dum,Azzurri:2020lbf}.
As such, calculating the next-to-next-to-leading-order (NNLO) corrections
is a crucial step toward advancing the accuracy and evaluating the reliability of
perturbative QCD predictions.

An analytic computation of the two-loop amplitudes for $Vjj$ production in the
leading-color approximation (LCA) has been available in the literature for some
time~\cite{Abreu:2021asb}. However, their application in LHC phenomenology has
been
limited due to the complexity of the results. To date, only the simplest
partonic subprocesses, such as $q \bar{q}^\prime \to W b \bar{b}$
\cite{Hartanto:2022qhh,Buonocore:2022pqq}, and more recently $q \bar{q} \to Z b
\bar{b}$ and $g g \to Z b \bar{b}$ \cite{Mazzitelli:2024ura}, have been studied,
making use of private numerical codes.
In this work, we develop novel techniques for constructing compact analytic
representations of amplitudes, which we apply to the
case of two-loop amplitudes for $pp \rightarrow Vjj$ in LCA.
We present both results for these amplitudes, as well as an
efficient and publicly available numerical implementation, ready for
phenomenological application.

Obtaining analytic representations of two-loop scattering amplitudes is
notoriously challenging. In the standard approach, amplitudes are expressed as a
combination of master integrals with rational functions as coefficients.
However, due to the complex intermediate stages of the calculation, obtaining
these coefficients often proves an onerous task.
In recent years, we have seen great progress in our ability to perform these
computations, due to a paradigm shift toward analytic reconstruction
techniques~\cite{Peraro:2016wsq,vonManteuffel:2014ixa,Abreu:2017xsl,Badger:2018enw,Abreu:2018zmy,Laurentis:2019bjh}.
In this framework,
rational functions are computed by using numerical evaluations over finite
fields to constrain an ansatz.
Remarkably, this approach has allowed the community to push the frontier of
analytic two-loop amplitude calculations to those involving six kinematic
scales~\cite{Abreu:2021asb,Hartanto:2022qhh,Badger:2021ega,Badger:2022ncb,Badger:2024sqv,Badger:2024awe}.
In practice, the computational cost of such calculations is primarily driven by
the number of required numerical evaluations, which in turn is controlled by the
size of the ansatz.
In many applications an ansatz of a rational function in
least-common-denominator (LCD) form has been used. However, the number of terms
in an LCD ansatz grows rapidly with an increasing number of variables, and so
this approach has proven difficult to directly apply beyond massless five-point
two-loop scattering processes.
Recent progress in analytic reconstruction techniques for high-multiplicity
amplitudes has been driven by the well-known observation that rational
functions in scattering amplitudes admit compact partial-fraction representations.
While it is in principle well understood how to construct such a representation
given an analytic expression for a rational
function~\cite{leinartas1978factorization,raichev2012leinartas,Meyer:2016slj,
Pak:2011xt, Abreu:2019odu,Heller:2021qkz,Bendle:2021ueg}, it remains a challenge
to a priori construct a similarly compact partial-fraction ansatz when the
rational function is unknown. 
Additionally, computations of Gröbner bases that are required in these approaches can become 
impractical when non-linear denominator factors of higher degrees are present \cite{Badger:2024sqv,Badger:2024awe,Becchetti:2025osw}.
Several proposals have been explored to address this
issue~\cite{Abreu:2018zmy,Badger:2023mgf,Chawdhry:2023yyx,DeLaurentis:2022otd,Laurentis:2019bjh,Liu:2023cgs,DeLaurentis:2023nss,DeLaurentis:2023izi,Smirnov:2024onl}.
For instance, it has been shown that reconstruction using a \emph{univariate}
partial-fractioned form can often capture sufficient simplifications to render
the reconstruction successful
\cite{Abreu:2021asb,Badger:2023mgf,Badger:2024sqv,Badger:2024awe}.
However, it is already clear that, at the five-point one-mass frontier,
challenges arise for this approach~\cite{Badger:2024sqv}.
Moreover, even in cases where reconstruction techniques provide an analytic
expression, these results may exhibit features that make them non-trivial to
apply in phenomenological applications, such as spurious poles or very large
expression sizes which diminish numerical stability.
Nevertheless, a number of recent calculations of two-loop five-point massless
amplitudes~\cite{
DeLaurentis:2020qle, % 3-jet, LC
DeLaurentis:2022otd, % 3-photon LC.
DeLaurentis:2023nss,DeLaurentis:2023izi, % 3-jet SLC
Abreu:2023bdp} % 3-photon SLC
suggest that compact and stable representations of the rational functions in
two-loop amplitudes should exist.

In this work, we therefore explore analytic reconstruction
approaches for two-loop multi-scale amplitudes that yield ansätze with a small
number of unknowns, facilitating both analytic reconstruction beyond the current
state of the art as well as compact representations relevant for broad
phenomenological applications.
Our guiding principle is to disentangle the pole factors of an amplitude's
integral coefficients. We first consider how to do this amongst the full set of
rational functions which arise in an amplitude.
Building on a suggestion from ref.~\cite{Abreu:2018zmy}, we develop a method
(recently applied in ref.~\cite{DeLaurentis:2023izi}) to construct a new basis
of rational coefficients with numerators of lower degree by systematically
disentangling poles whose residues exhibit linear relations. This decreases the
complexity in common denominator form.
Next, we explore methods of disentangling pole structures within a single
coefficient via a multivariate partial-fraction decomposition.
In a first approach, we explore how experimental observations about the analytic
structure of the rational functions can be used to perform a partial-fraction
decomposition with relatively small ansätze, which we fit with a limited number
of finite field samples.
In a second approach, based on that of refs.~\cite{Laurentis:2019bjh,
DeLaurentis:2022otd}, we use $p\kern0.2mm$-adic numbers to numerically study
rational functions in limits where one or two denominator factors are small.
This allows us to construct compact ansätze for pole residues, which we
iteratively subtract until the coefficient is fully
determined~\cite{Laurentis:2019bjh}, resulting in compact partial-fraction
forms of the functions.
In order to exploit the well-known observation that gauge-theory amplitudes
admit compact representations when expressed in spinor-helicity variables,
throughout this paper we work in the spinor-helicity formalism.
This requires us to tackle the problem of constructing spinor-helicity ansätze that
respect the five-point one-mass kinematics, which we achieve by extending the
Gröbner basis method of ref.~\cite{DeLaurentis:2022otd} to this case.
Parts of this framework have been already applied at two loops with
five-point massless kinematics~\cite{DeLaurentis:2020qle,DeLaurentis:2023nss,DeLaurentis:2023izi,Abreu:2023bdp},
and at one-loop with a higher number of scales~\cite{Campbell:2022qpq, Campbell:2024tqg}.
Here, we apply it to a six-scale two-loop process for the first time.

Applying our method to the $Vjj$ amplitudes in LCA we obtain expressions that
are three orders of magnitude smaller than the ones from
ref.~\cite{Abreu:2021asb}. This allows us to provide a stable public C++ code
\cite{FivePointAmplitudes} which will allow broad phenomenological applications
of this process.
Moreover, given the recent availability of the master
integrals~\cite{DiVita:2014pza, Canko:2021xmn} and the three-loop
amplitudes for $Vj$ production in LCA~\cite{Gehrmann:2023jyv}, we also foresee
applicability of our results in studies of third-order corrections for this process.
While not all components of our approach are algorithmic, we expect that the
features that we describe will not be limited to $Vjj$ process and
will be useful to break through more complex processes in the future. Our method
is applicable for the functional reconstruction from suitable sets of numerical
evaluations or, as done in this work, for post processing of an available
amplitude.

The paper is organized as follows. In \cref{sec:notation}, we define our
notation and conventions. In \cref{sec:basischange}, we discuss a basis
change algorithm, which disentangles the coefficient poles in order to obtain
optimized coefficient bases in common denominator form.  In
\cref{sec:optimalfrom}, we discuss our various techniques to derive a compact
parameterization of coefficient functions in terms of partial fractions.
In \cref{sec:result}, we discusses the application of our approach to
the to planar $Vjj$ amplitudes and its impact on the
simplification of helicity amplitudes. We moreover present ancillary
files for the results, and close with demonstrating the efficiency and
numerical stability of the C++ implementation of the squared matrix elements.
Reference values for the squared matrix elements are provided in
\cref{sec:referenceEvaluations}.
Further appendices contain useful input for infra-red subtraction terms and a
table of primary decompositions used in our partial-fraction analysis.

\section{Notation and conventions}
\label{sec:notation}

\begin{figure}[ht]
  \centering

  \begin{tabular}{cc}
    \includegraphics[width=0.31\linewidth]{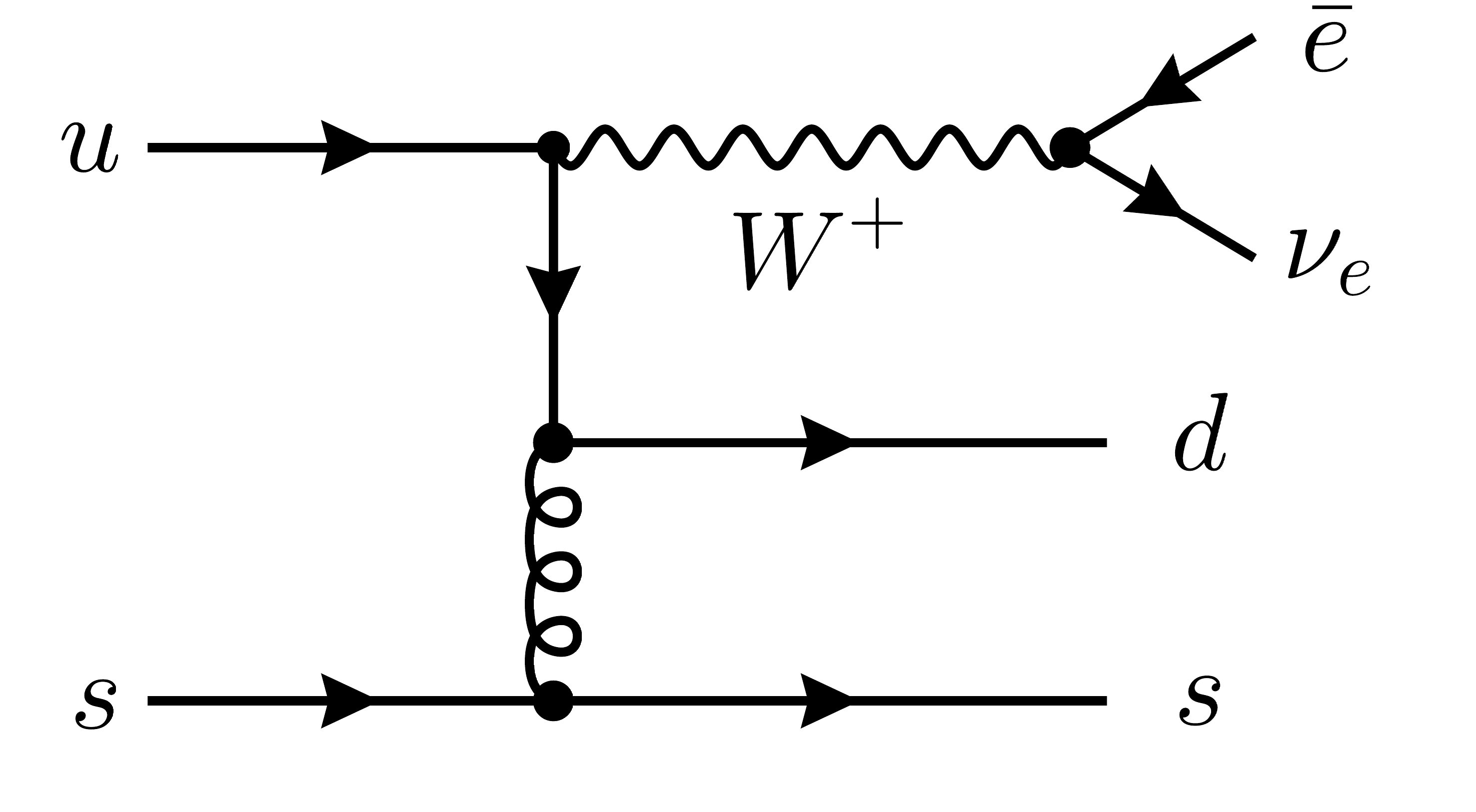} & \includegraphics[width=0.31\linewidth]{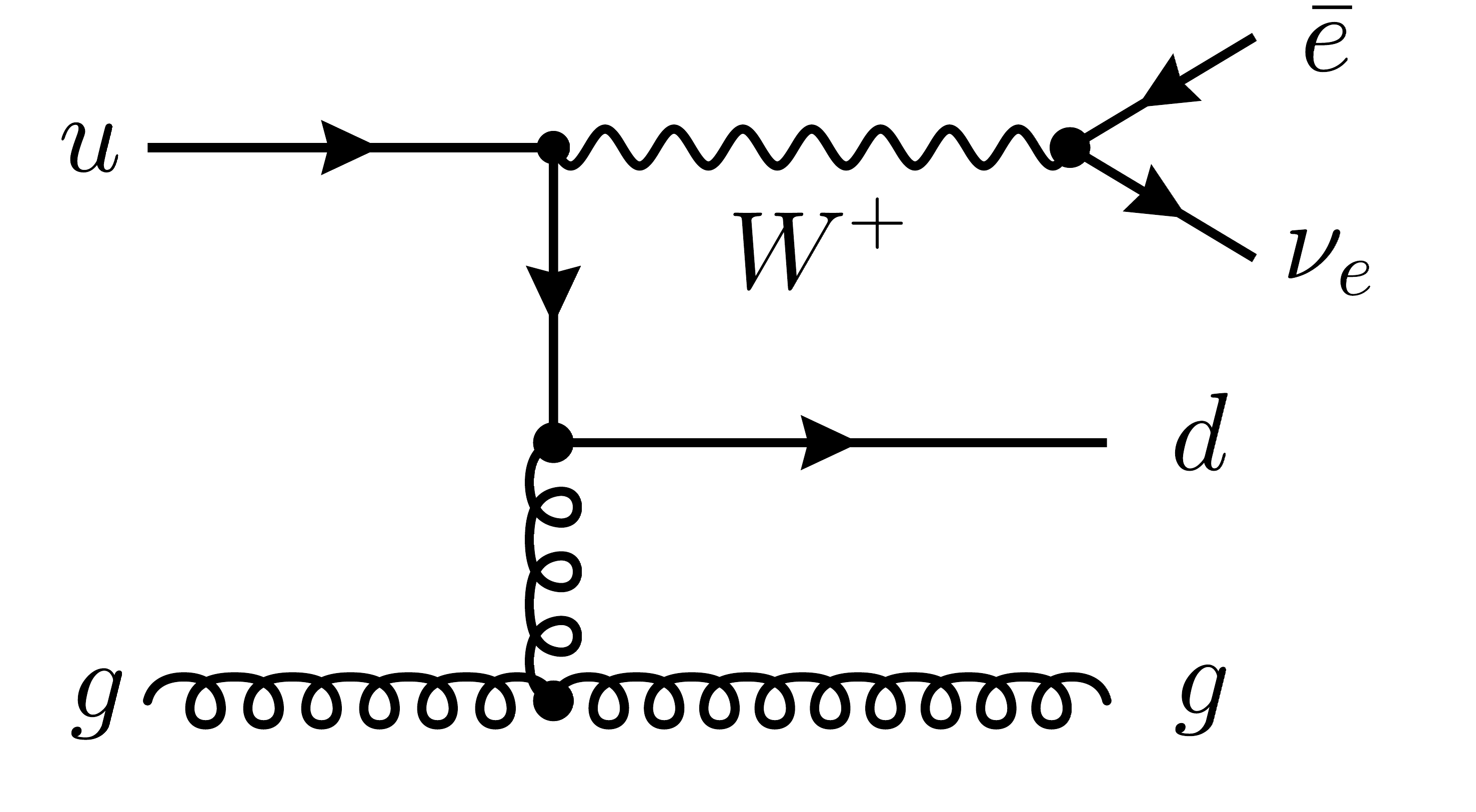} \\
  \end{tabular}

  \caption{Representative leading-order diagrams contributing to $pp \to W^+(\bar{\ell}\nu)jj$ processes.}
  \label{fig:Wjj-Feyn}
\end{figure}

\begin{figure}[ht]
  \centering

  \begin{tabular}{cc}
    \includegraphics[width=0.31\linewidth]{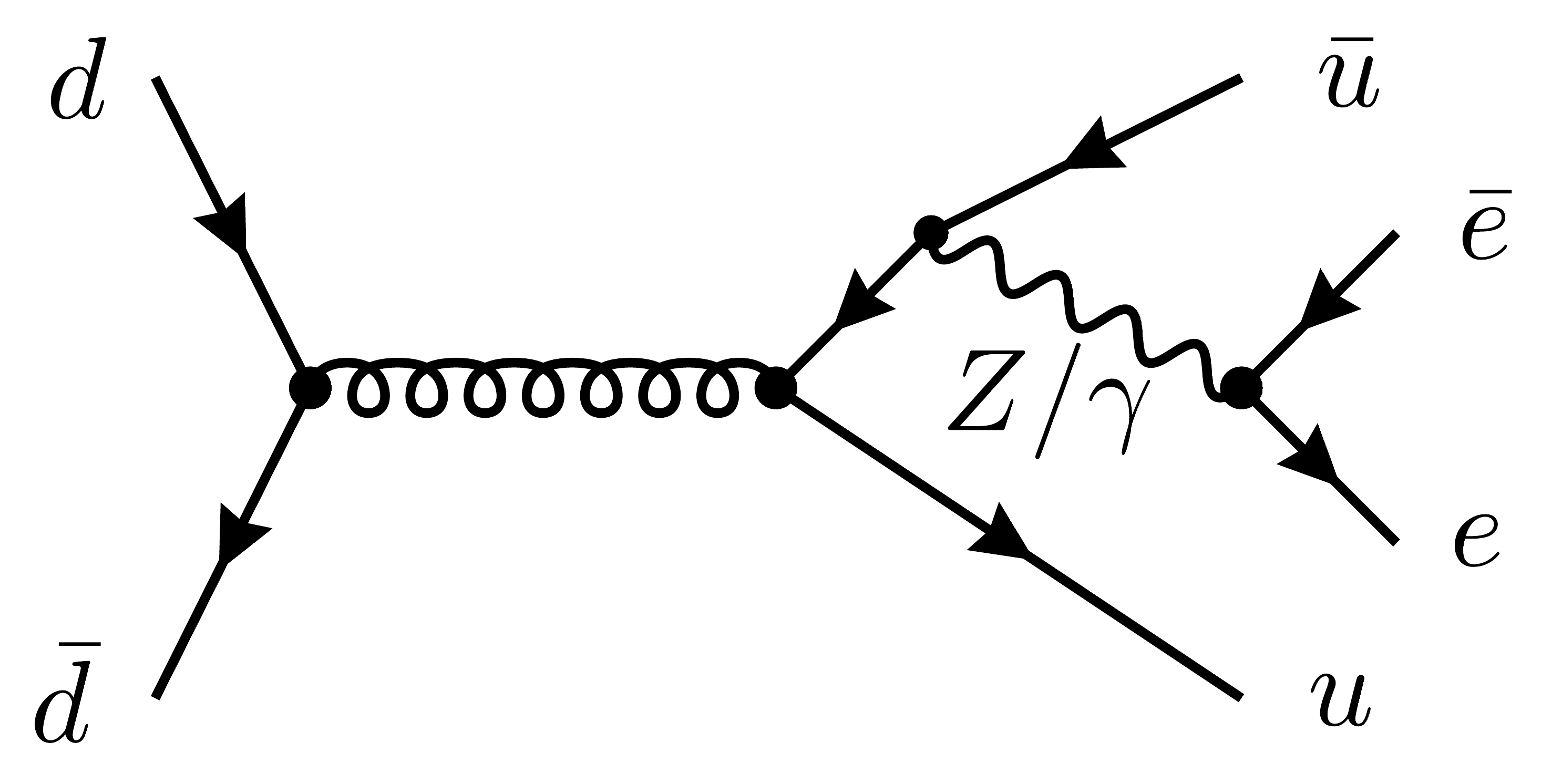} & \includegraphics[width=0.31\linewidth]{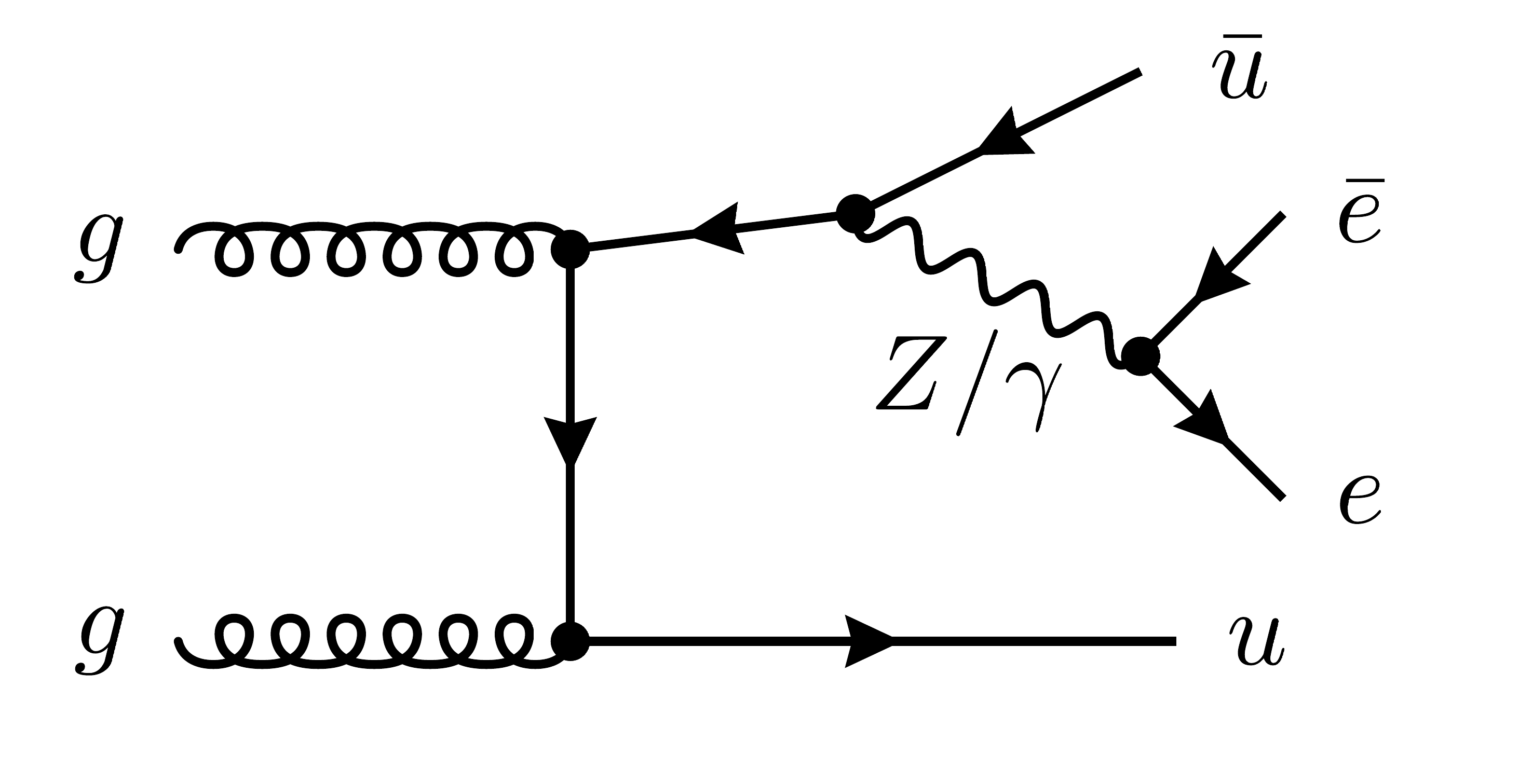} \\
    \includegraphics[width=0.31\linewidth]{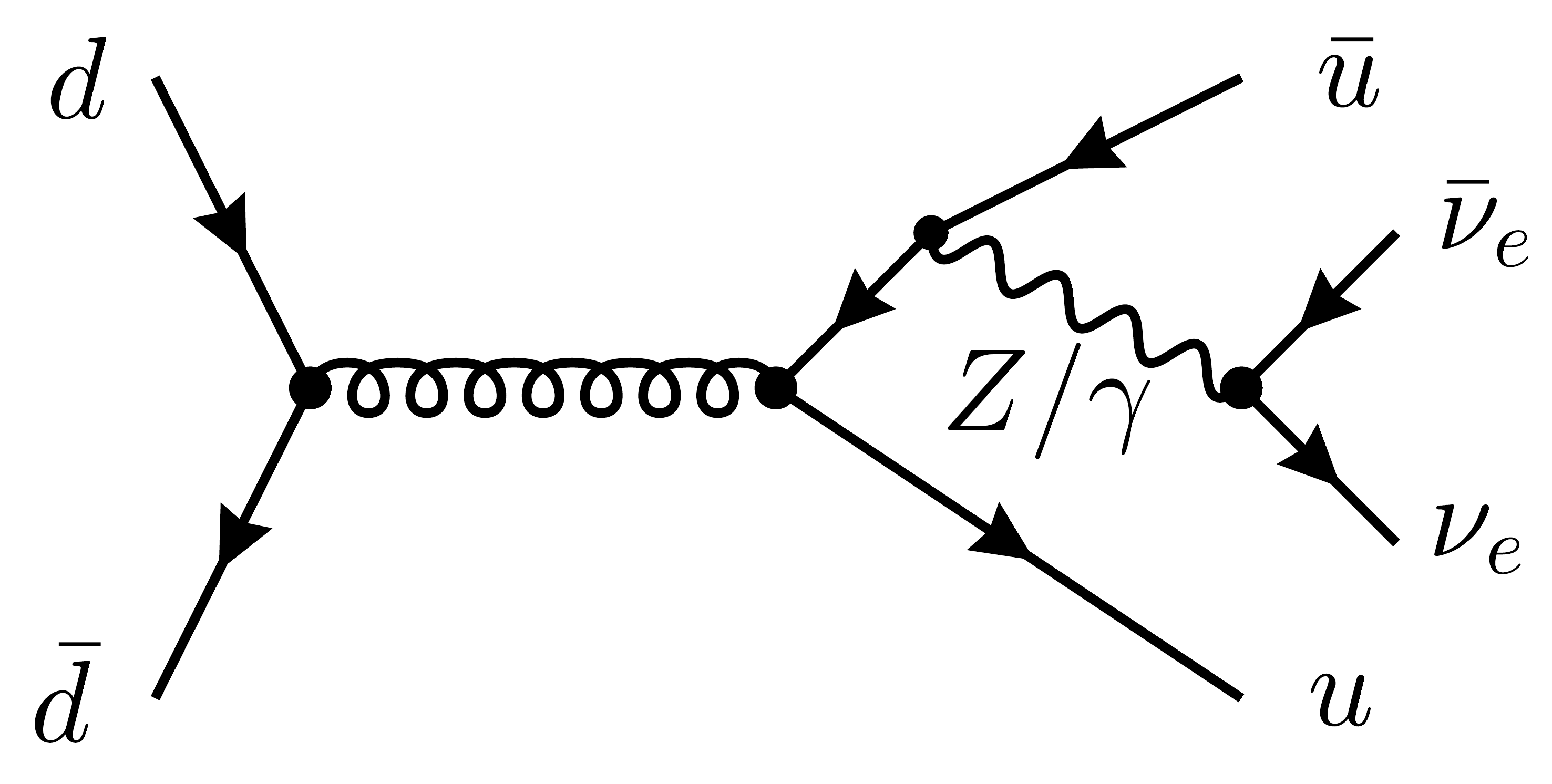} & \includegraphics[width=0.31\linewidth]{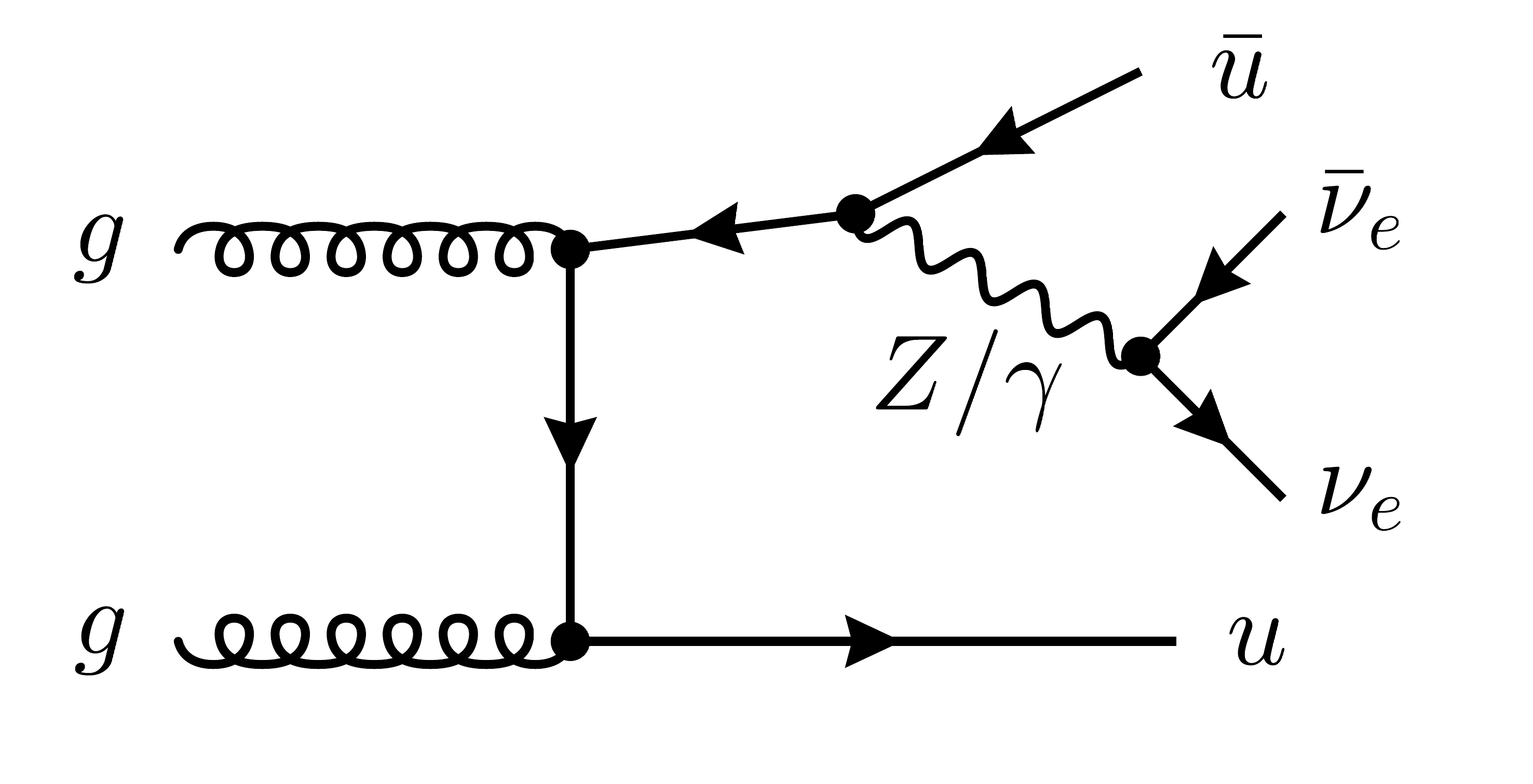} \\
  \end{tabular}

  \caption{Representative leading-order diagrams contributing to $pp \to Z(\bar{\ell}\ell)jj$ and $pp \to Z(\bar{\nu}\nu)jj$ processes.}
  \label{fig:Zjj-Feyn}
\end{figure}

In this work, we provide compact analytic results and an efficient numerical
code for amplitudes for the production of a leptonically decaying heavy
electroweak vector boson in association with two jets at hadron colliders. We
consider all partonic channels: in the SM there are 18 distinct such
channels for $pp\to W^+ (\bar{\ell} \nu)\,jj$ process and $30$ for $pp\to
Z(\bar{\ell} \ell)\,jj$ and $pp\to Z(\bar{\nu} \nu)\,jj$ processes each, where
the flavor of massless leptons is irrelevant and the mixing of $Z$ with $\gamma$
is assumed where appropriate. The full list of partonic channels can be found in
\cref{sec:referenceEvaluations} in
\cref{tab:wjj-benchmark,tab:zjj-charged-benchmark,tab:zjj-neutral-benchmark}. Up
to crossings and exchange of up and down quark charges, the $W^+$ channels are
\begin{equation} \label{eq:Wchannels}
  \begin{aligned}
    u_1\,\gluon_2  &\rightarrow \gluon_3\,d_4\;{\bar{\ell}_5}\,{\nu_{6}}\,, \\
    u_1\,s_2  &\rightarrow s_3\,d_4\;{\bar{\ell}_5}\,{\nu_{6}}\,, \\
    u_1\,u_2  &\rightarrow u_3\,d_4\;{\bar{\ell}_5}\,{\nu_{6}}\,, \\
    u_1\,d_2  &\rightarrow d_3\,d_4\;{\bar{\ell}_5}\,{\nu_{6}}\,,
  \end{aligned}
\end{equation}
with $W^-$ production channels obtained from these by $\ell \leftrightarrow \nu$ exchange and a suitable momentum relabelling,
and the $Z$ channels are 
\begin{equation} \label{eq:Zchannels}
  \begin{aligned}
    u_1\,{\bar u}_2 & \rightarrow \gluon_3\,\gluon_4        \;{\bar{\ell}_5}\,{\ell_6}\,, \\
    u_1\,{\bar u}_2 & \rightarrow {\bar d}_3\,d_4 \;{\bar{\ell}_5}\,{\ell_6}\,, \qquad\qquad \mbox{and} \qquad  \ell \leftrightarrow \nu \,. \\
    u_1\,u_2        & \rightarrow u_3\,u_4        \;{\bar{\ell}_5}\,{\ell_6}\,,
  \end{aligned}
\end{equation}
We denote  momentum assignments by subscripts, 
and assign to the initial and final-state partons labels $\{1,2\}$ and $\{3,4\}$ respectively, and to the leptons $\{5,6\}$.
The momentum labels for all or some of the particles may be suppressed for clarity.
In this paper we use symmetric all-outgoing convention for momenta except when
we describe channels with "$\to$" notation, where the two incoming particles are crossed to have positive energy. 
Representative leading-order diagrams are shown in \cref{fig:Wjj-Feyn,fig:Zjj-Feyn}.

In this work, we adopt the same approximations as those used in
ref.~\cite{Abreu:2021asb}. Specifically, we employ the leading-color
approximation, which keeps the leading terms in the large-$\NC$ limit
(with the ratio of the number of light flavors $\NF$ to $\NC$ fixed). The
contributions with the vector-boson coupling to closed quark loops (both vector
and axial-vector), as well as with massive top-quark loops are also omitted.

\subsection{Kinematics}

We consider the scattering amplitudes of four partons and a vector boson,
including its decays to leptons. We work with six massless momenta $p_i$,
which we consider in the fully outgoing convention. We therefore have
\begin{equation}
  \sum_{i=1}^6 p_i^\mu = 0, \qquad \text{and} \qquad p_i^2 = 0.
\end{equation}
To specify the kinematics in terms of Lorentz invariants we use the
Mandelstam variables $s_{ij}=(p_i+p_j)^2$ and $s_{ijk}=(p_i+p_j+p_k)^2$, as
well as the parity-odd contraction of four momenta $\trFive =
\trace(\gamma^5\slashed{p}_1\slashed{p}_2\slashed{p}_3\slashed{p}_4)$, which is
singled out by the five-point one-mass kinematics of the $Vjj$ process.
To specify the dependence on the particles' helicities we use the spinor
helicity formalism. Following the conventions of ref.~\cite{Maitre:2007jq}, we denote the two-component spinors with
$\lambda_{i}^{\alpha}$ and $\tilde\lambda_{i}^{\dot\alpha}$ and define the invariant spinor brackets as
\begin{equation}
  \langle i j \rangle = \epsilon_{\alpha\beta} \lambda^\alpha_i \lambda^\beta_j \quad \text{and} 
\quad [ij] =   \epsilon^{\dot\alpha\dot\beta}  \tilde\lambda_{i,\dot\alpha}\tilde\lambda_{j\dot\beta}\, ,
\end{equation}
where the tensors $\epsilon_{\alpha\beta}$ and $\epsilon^{\dot\alpha\dot\beta}$
are anti-symmetric and $\epsilon_{12}=-\epsilon^{12}=-1$.
The two-particle Mandelstam invariants $s_{ij}=\langle ij\rangle [ji]$ factorize in terms of spinor brackets.
We will require a number of spinor chains for which we use the notation,
\begin{equation}
  \langle i | j \pm k | m ] = \langle i j \rangle [jm] \pm \langle ik
    \rangle [km] \, .
\end{equation}
Further spinor chains are defined recursively, e.g.
\begin{equation}
\langle i | j | 5+6 | k | i ] = 
\langle i | j | 5] \langle 5 | k | i ] + \langle i | j | 6] \langle 6 | k | i ] \, .
\end{equation}
Another recurring kinematic variable is the Gram determinant associated to three
momenta, $K_1$, $K_2$, $K_3$. We define it as
\begin{equation}
  \Delta_{K_1|K_2|K_3} = (K_1 \cdot K_2)^2 - K_1^2 K_2^2 \, .
\end{equation}
Given the momentum-conservation constraint $K_1+K_2+K_3=0$,
$\Delta_{K_1, K_2, K_3}$ is invariant under permutations of the
arguments. We will make use of the case built out of pairs of massless momenta,
which we denote as
\begin{equation}
  \Delta_{ij|kl|mn} = \Delta_{K_1|K_2|K_3} = \frac{1}{4} (s_{ik}+s_{il}+s_{jk}+s_{jl})^2 - s_{ij} s_{kl}
\end{equation}
with $K_1 = p_i + p_j$, $K_2 = p_k + p_l$ and $K_3 = p_m + p_n$. Note that this
is a degree-four polynomial in spinor brackets.
Finally, we express $\trFive$ as a polynomial in spinor brackets
as
\begin{equation}
  \trFive = [12]\langle23\rangle[34]\langle41\rangle-\langle12\rangle[23]\langle34\rangle[41] \, .
\end{equation}

Helicity amplitudes transform covariantly under little-group transformations 
and redefinition of helicity states.
The little-group transformation of the $i^{th}$ leg with helicity $h_i$
reads $(\lambda_i,\tilde\lambda_i)\rightarrow (z_i
\lambda_i,\tilde\lambda_i/z_i)$. Under this transformation,
helicity amplitudes $A$ transform as $A \rightarrow z_i^{-2h_i} A$. We
refer to the exponent of the $z_i$ as the little-group weight.
Similarly, under a simultaneous rescaling of all spinors of the form
$(\lambda_i,\tilde\lambda_i)\rightarrow (z \lambda_i,z
\tilde\lambda_i)$ for all $i$, $n$-point amplitudes with dimensionless couplings
transform as $A
\rightarrow z^{2(4-n)} A$.

To relate simpler functions to a minimal set, we often consider parity conjugation on functions in spinor variables.
This transformation is implemented by %$\lambda_i^\alpha \leftrightarrow \tilde \lambda_i^{\dot\alpha}$, 
exchanging spinor brackets and labels as $\langle i j \rangle \leftrightarrow [j i]$.
Furthermore, as a shorthand notation we will use joint relabelling and parity transformations
for expressions in spinor-helicity notation, defining
\begin{align}
  \begin{split}
f(1,2,3,4,5,6)
\; &+ \; (123456\rightarrow \overline{i_1i_2i_3i_4i_5i_6})  =   \\ 
&f(1,2,3,4,5,6)
+ f(i_1,i_2,i_3,i_4,i_5,i_6)|_{\langle ij\rangle\leftrightarrow [ji]} \,.
  \end{split}
\end{align}

\subsection{Decomposition into partial amplitudes}
\label{sec:partials}

\def\manualalign{\hskip 15ex}

The SM vector-boson scattering amplitudes for processes in \cref{eq:Wchannels,eq:Zchannels} can be written in terms of
amplitudes with auxiliary photon-like off-shell vector boson $V$ exchanged
between a single quark line and the lepton line.
Recall that the amplitudes with vector and axial-vector couplings are equal up to a sign (which also holds in our regularization scheme \cite{Abreu:2021asb}),
apart from the contributions with $V$ coupling to closed quark loops which we omit in this work.
It is therefore sufficient to consider only amplitudes with vector couplings.

We consider the complete set of helicity amplitudes (with all external particles having definite helicity) for each of the processes.
We employ the 't Hooft-Veltman (HV) scheme of dimensional regularization where
the space-time dimension is $D=4-2\epsilon$
and external states are treated as four-dimensional. 
The external four-dimensional fermion states are defined as in refs.~\cite{Abreu:2021asb,Abreu:2018jgq}.

Following the notation of refs.~\cite{Bern:1996ka,Bern:1997sc,Abreu:2021asb},
the decomposition of SM amplitudes $\mathcal{M}$ with distinct-flavor quarks to $V$-amplitudes $\mathcal{A}$ is as follows.
For $W^+$ bosons we have
\begin{subequations} \label{eq:ampW}
  \begin{align}
    \label{eq:ampW-g}
    \mathcal{M}( \bar u_{1}^{+},  \gluon_{2}^{h_2}, \gluon_{3}^{h_3}, d_{4}^{-} ,\bar \ell_{5}^+ , \nu_{6}^-) &= 
    e^2 v_W^2\; {\cal P}_W(s_{56})\, \mathcal{A} \left(\bar{q}^{+}_{1},\gluon_{2}^{h_2},\gluon_{3}^{h_3},{q}^{-}_{4}; \bar{\ell}^{+}_{5},{\ell}^{-}_{6}\right)\,,\\
    \label{eq:ampW-q}
    \mathcal{M}( \bar u_{1}^{+},  c_{2}^{h}, \bar{c}_{3}^{-h}, d_{4}^{-} ,\bar \ell_{5}^+ , \nu_{6}^-) &= 
    e^2 v_W^2 \;\,{\cal P}_W(s_{56})\, \mathcal{A} \left(\bar{q}^{+}_{1},Q_{2}^{h},\bar{Q}_{3}^{-h},{q}^{-}_{4};
    \bar{\ell}^{+}_{5},{\ell}^{-}_{6}\right)\,,
  \end{align}
\end{subequations}
with other choices of helicities of $\bar{u},d, \bar{\ell}$ and $\nu$ vanishing due to the left-chiral coupling of the $W$-boson to leptons and quarks.
Partonic channels originating from non-diagonal CKM-matrix entries can be obtained by multiplying \cref{eq:ampW} with appropriate CKM matrix elements.
For charged-lepton $Z$-decays we have
\begin{subequations}\label{eq:ampZ}
  \begin{align} 
    \notag
    &\mathcal{M}( \bar u_{1}^{-h},  \gluon_{2}^{h_2}, \gluon_{3}^{h_3}, u_{4}^{h} ,\bar e^{-h'}_{5} , e^{h'}_{6})= \\ 
    & \manualalign e^2 \left(- Q_q + v_e^{h'} v_q^{h} \; {\cal P}_Z(s_{56})\right)  \mathcal{A} \left(\bar{q}^{-h}_{1},\gluon_{2}^{h_2},\gluon_{3}^{h_3},{q}^{h}_{4}; \bar{\ell}^{-h'}_{5},{\ell}^{h'}_{6}\right)\,,\\
    \notag
    & \mathcal{M}( \bar u_{1}^{-h},  d_{2}^{\tilde h}, \bar d_{3}^{-\tilde h}, u_{4}^h ,\bar e_{5}^{-h'} , e_{6}^{h'})= \\ 
    \notag
    & \manualalign e^2 \left(- Q_u + v_e^{h'} v_u^{h}  \; {\cal P}_Z(s_{56})\right)\mathcal{A} \left(\bar{q}^{-h}_{1},Q_{2}^{\tilde h},\bar{Q}_{3}^{-\tilde h},{q}^{h}_{4};
    \bar{\ell}^{-h'}_{5},{\ell}^{h'}_{6}\right) ~ + \\
    & \manualalign e^2 \left( - Q_d + v_e^{h'} v_d^{\tilde h}\;{\cal P}_Z(s_{56})\right)  \mathcal{A} \left(\bar{q}^{-\tilde h}_{3},Q_{4}^{h},\bar{Q}_{1}^{-h},{q}^{\tilde h}_{2};
    \bar{\ell}^{-h'}_{5},{\ell}^{h'}_{6}\right)\,,
  \end{align}
\end{subequations}
while for neutral decays,
\begin{subequations}\label{eq:ampZneutral}
  \begin{align}
  & \mathcal{M}^Z( \bar u_{1}^{-h}, \gluon_{2}^{h_2}, \gluon_{3}^{h_3}, u_{4}^{h} ,\bar \nu^{+}_{5} , \nu^{-}_{6}) =  e^2 v_\nu v_q^{h} \; {\cal P}_Z(s_{56}) \, \mathcal{A} \left(\bar{q}^{-h}_{1},\gluon_{2}^{h_2},\gluon_{3}^{h_3},{q}^{h}_{4};
  \bar{\ell}^{+}_{5},{\ell}^{-}_{6}\right)\,,\\
  \notag
  & \mathcal{M}^Z( \bar u_{1}^{-h}, d_{2}^{\tilde h}, \bar d_{3}^{-\tilde h}, u_{4}^h ,\bar \nu_{5}^{+} , \nu_{6}^{-})=\\
  \notag
  & \manualalign e^2  v_\nu v_u^{h}  \; {\cal P}_Z(s_{56}) \, \mathcal{A}\left(\bar{q}^{-h}_{1},Q_{2}^{\tilde h},\bar{Q}_{3}^{-\tilde h},{q}^{h}_{4};
  \bar{\ell}^{+}_{5},{\ell}^{-}_{6}\right) ~ + \\
  &  \manualalign e^2 v_\nu v_d^{\tilde h} \; {\cal P}_Z(s_{56}) \, \mathcal{A} \left(\bar{q}^{-\tilde h}_{3},Q_{4}^{h},\bar{Q}_{1}^{-h},{q}^{\tilde h}_{2};
  \bar{\ell}^{+}_{5},{\ell}^{-}_{6}\right)\,.
  \end{align}
\end{subequations}
Here on the right-hand side $q \bar{q}$ and $Q \bar{Q}$ are distinct-flavored quarks, and the $V$-boson couples only to the former and with unit charge.
The function $\mathcal{P}_{V}(s)$ replaces the massless propagator by that of the corresponding EW boson through
\begin{equation}
  \mathcal{P}_{V}(s)=\frac{s}{s-M_{V}^2+\ii\Gamma_{V} M_{V}},
\end{equation}
and the EW coupling constants are given in terms of the Weinberg angle $\theta_w$ and particle charges as
\begin{align}\label{eq:EW-couplings}
\begin{split}
%%%%%%%%%%%%%%%%%%%%%%%%%%%%%%%
%%%%%%%%%%%%%%%%%
v_e^{-}&=v_e^{L} = \frac{ -1+2 \sin^2\theta_w}{\sin{2\theta_w}} \,, \quad 
v_e^{+}=v_e^{R} = \frac{ 2 \sin^2\theta_w}{\sin{2\theta_w}} \,, \\
v_{u,d}^{-}&=v_{u,d}^{L}=\frac{\pm 1-2 Q_{u,d} \sin^2{\theta_w}}{\sin{2 \theta_w}}\,,\quad 
v_{u,d}^{+}=v_{u,d}^{R}=\frac{-2 Q_{u,d} \sin^2{\theta_w}}{\sin{2 \theta_w}}\,,\\
%%%%%%%%%%%%%%%%
v_W^2&=\frac{1}{2 \sin^{2}{\theta_w}}\,, \quad 
v_\nu= \frac{1}{\sin{2\theta_w}} \,.
\end{split}
\end{align}
The $Q_{u,d}$ are respectively the charges of the up-/down-type quarks, $Q_u=2/3$ and $Q_d=-1/3$. 
The lepton pair that couples to the photon with charge $-1$ in units of $e$ is visible 
in the expression $-Q_{u/d}$. In $v_{u,d}^{h}$ 
the upper sign corresponds to the up-type coupling $v_{u}^{h}$ and
the lower one to the down-type coupling $v_{d}^{h}$. 
$M_{W,Z}$ and $\Gamma_{W,Z}$ are the masses and the widths of the vector bosons,
and the Weinberg angle is calculated either through the on-shell prescription $\cos^2\theta_w = M_W/M_Z$,
or in the complex-mass scheme \cite{Denner:1999gp}.

The partonic channels with identical quarks are obtained as anti-symmetric sums over the corresponding distinct-flavor channels.
For example,
\begin{align} \label{eq:ampW-ident}
 \mathcal{M}( \bar u_{1}^{h_1},  u_{2}^{h_2}, \bar{u}_{3}^{h_3}, d_{4}^{-} ,\bar \ell_{5}^+ , \nu_{6}^-) = 
 \mathcal{M}( \bar u_{1}^{h_1},  c_{2}^{h_2}, \bar{c}_{3}^{h_3}, d_{4}^{-} ,\bar \ell_{5}^+ , \nu_{6}^-) - \mathcal{M}( \bar u_{3}^{h_3},  c_{2}^{h_2}, \bar{c}_{1}^{h_1}, d_{4}^{-} ,\bar \ell_{5}^+ , \nu_{6}^-),
\end{align}
where it is understood that on the right-hand side the amplitudes vanish if they do not match the helicity pattern on the left-hand side of \cref{eq:ampW-q}.

Finally, notice that for each of the amplitudes $\mathcal{A}$ on the right-hand side of \cref{eq:ampZ} we can exchange the momenta $5\leftrightarrow 6$ such that the lepton pair is always $\{\bar{\ell}_5^+, \ell_6^-\}$.
We therefore established that all SM processes in \cref{eq:Wchannels,eq:Zchannels} can be calculated through \cref{eq:ampW,eq:ampZ,eq:ampZneutral} from permutations of 
the two sets of $V$-boson amplitudes
\begin{subequations} \label{eq:partProcesses}
  \begin{align}
    \mathcal{A}_{\gluon} (1^{h_1},2^{h_2},3^{h_3},4^{-h_1}) &= \mathcal{A}\left(\bar \quark^{h_1}_{1},\gluon^{h_2}_{2},\gluon^{h_3}_{3},\quark^{-h_1}_{4};
    \bar\ell^{+}_{5},{\ell}^{-}_{6}\right)\,,\\
      \mathcal{A}_{\quark} (1^{h_1},2^{h_2},3^{-h_2},4^{-h_1}) & = \mathcal{A}\left(\bar\quark^{h_1}_{1},\Quark^{h_2}_{2},\bar{\Quark}^{-h_2}_{3},\quark^{-h_1}_{4};
    \bar{\ell}^{+}_{5},{\ell}^{-}_{6}\right)\,.
  \end{align}
\end{subequations}
In the following the helicity and/or momentum labels will sometimes be suppressed for clarity.

%%%%%%%%%%%%% FIGURE %%%%%%%%%%%%%%%%%%
\begin{figure}[ht]
\centering
  \begin{subfigure}{0.31\textwidth}
  	\centering
  	\includegraphics[scale=0.5]{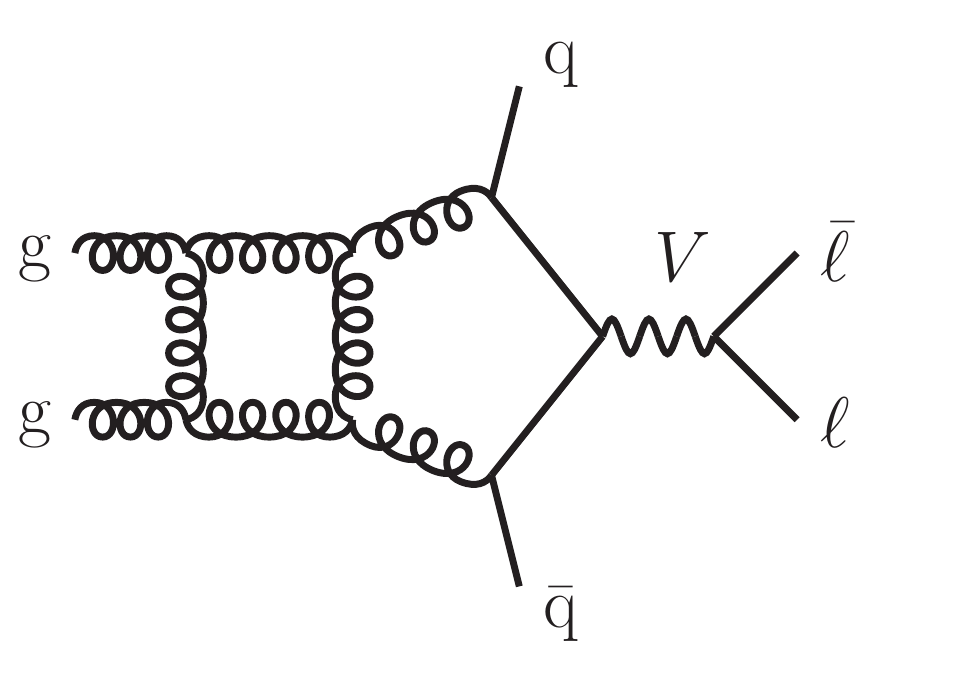}
  \end{subfigure}
  \begin{subfigure}{0.31\textwidth}
  	\centering
  	\includegraphics[scale=0.5]{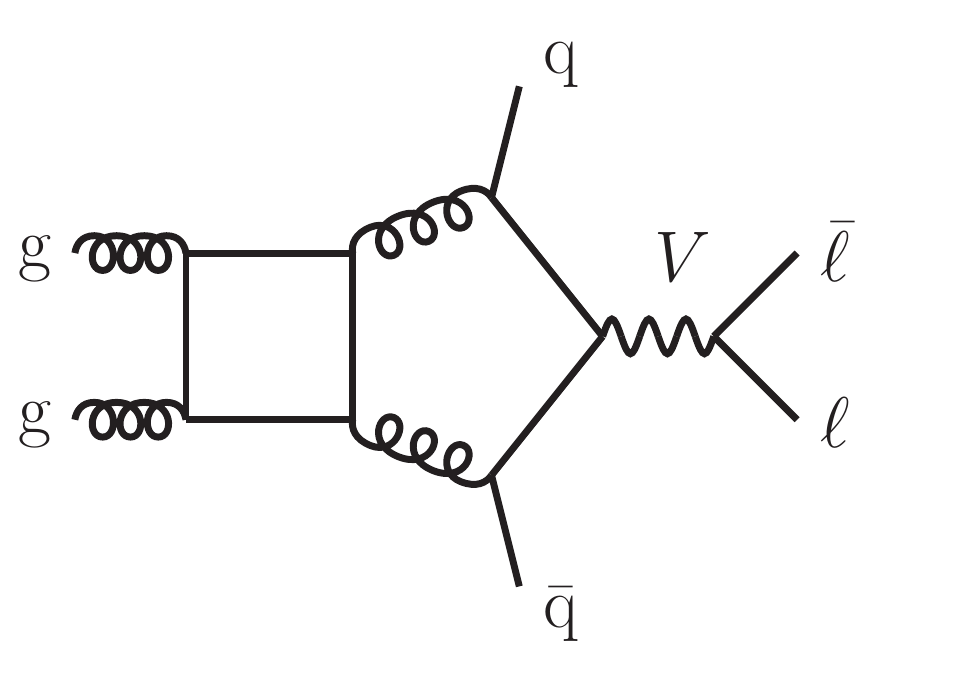}
  \end{subfigure}
   \begin{subfigure}{0.31\textwidth}
  	\centering
  	\includegraphics[scale=0.5]{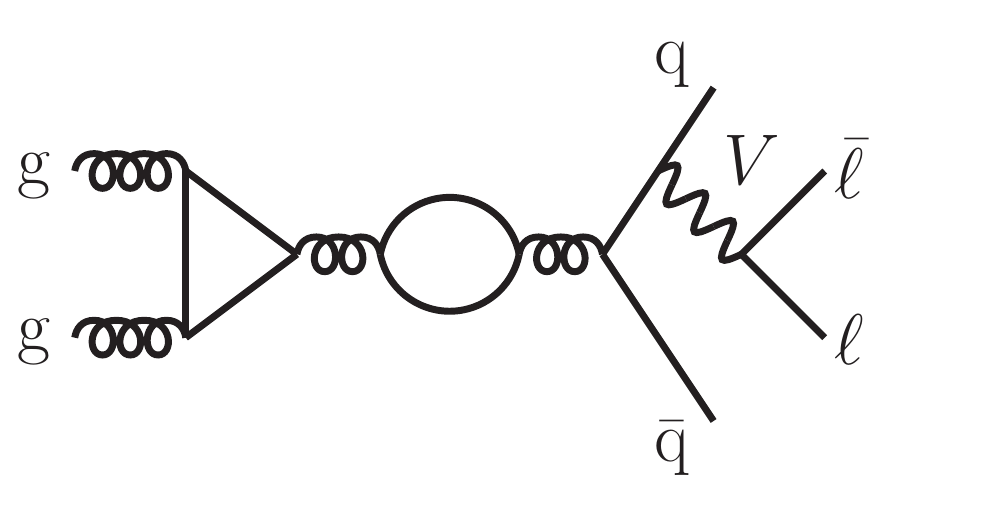}
  \end{subfigure}
\caption{Sample diagrams for the different contributions to 
\cref{eq:nfDecomp} for $A_g^{(2)[0,1,2]}$.}
\label{fig_parents2q2g}
\end{figure} 

%%%%%%%%%%%%% FIGURE %%%%%%%%%%%%%%%%%%
\begin{figure}[ht]
\centering
  \begin{subfigure}{0.31\textwidth}
  	\centering
  	\includegraphics[scale=0.5]{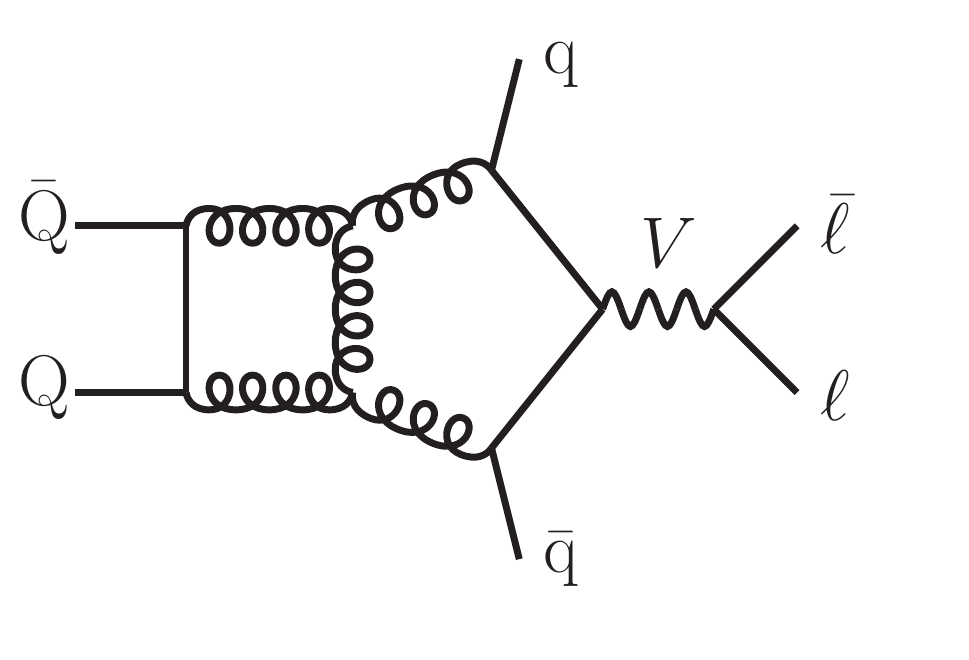}
  \end{subfigure}
  \begin{subfigure}{0.31\textwidth}
  	\centering
  	\includegraphics[scale=0.5]{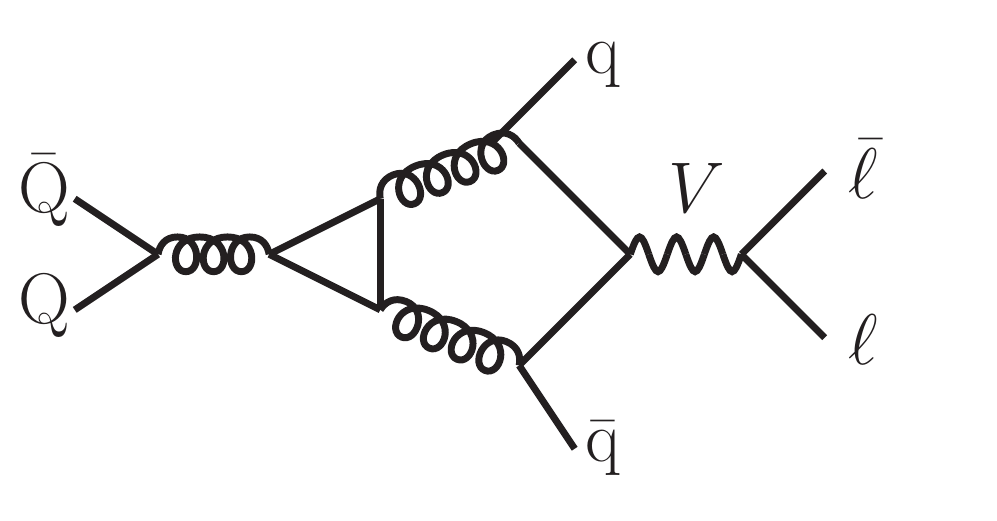}
  \end{subfigure}
   \begin{subfigure}{0.31\textwidth}
  	\centering
  	\includegraphics[scale=0.5]{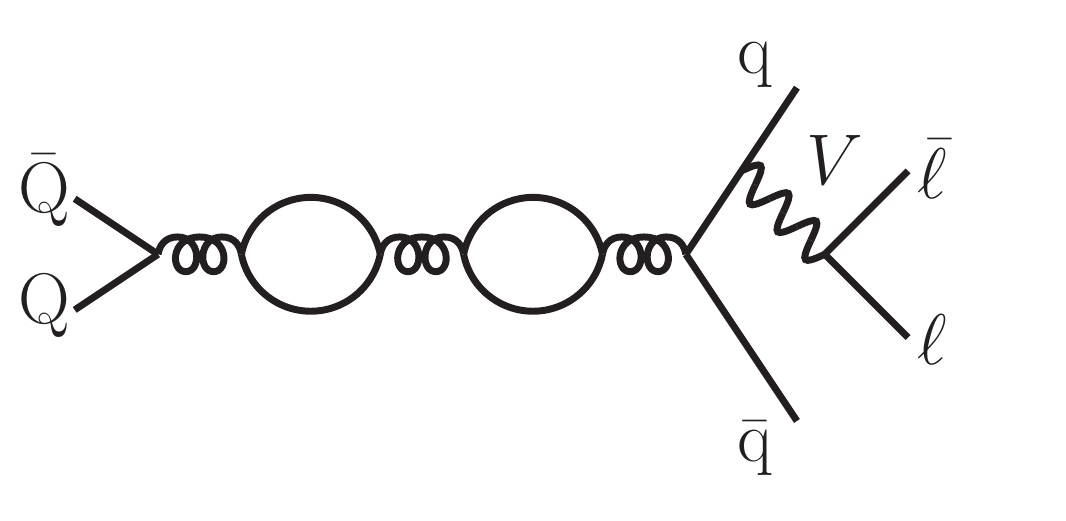}
  \end{subfigure}
\caption{Sample diagrams for the different contributions to 
\cref{eq:nfDecomp} for $A_q^{(2)[0,1,2]}$.}
\label{fig_parents4q}
\end{figure}

In the LCA, the $V$-boson amplitudes $\mathcal{A}_{\gluon}$ and $\mathcal{A}_\quark$ receive contributions only from planar diagrams,
and their color decomposition is independent of the loop order:
\begin{subequations}\label{eqn:colorAmp}
  \begin{align}
    \mathcal{A}_{\gluon} (1,2,3,4) &= \sum_{\sigma\in S_2(2,3)}  \left(T^{a_{\sigma(3)}}T^{a_{\sigma(2)}}\right)^{\,\,\bar{i}_1}_{i_4} \; A_\gluon\left(1,\sigma(2),\sigma(3),4\right)  \,,\\
    \mathcal{A}_{\quark}(1,2,3,4) &= \delta_{i_2}^{\bar{i}_1}\delta_{i_4}^{\bar{i}_3} \; A_\quark \left(1,2,3,4\right)  \,,
  \end{align}
\end{subequations}
where $a$ and $i$ are adjoint and fundamental indices of the color group $SU(\NC)$ respectively,
and $(T^{a})_i^{\;\bar j}$ are the fundamental generators satisfying  
\begin{equation}
  \begin{split}
    \left[T^a,T^b\right] & =i f_{abc} T^c \, , \\
    i f_{abc} & =\trace(T^a T^b T^c) - \trace(T^b T^a T^c) \,,
  \end{split}
  \qquad 
  {\trace}(T^a T^b) = \delta^{ab}\,.
\end{equation}
We will refer to $A_{\kappa}$ with $\kappa=\{\gluon,\quark\}$ as partial amplitudes.

We expand $A_{\kappa}$ in powers of the bare coupling $\alpha_s^0=(g_s^0)^2/(4\pi)$ through two loops as
\begin{equation} \label{eq:as-expansion}
	A_\kappa=(g_s^0)^2 \left({A}_\kappa^{(0)}+
	\frac{\alpha_s^0}{2\pi} \frac{\NC}{2} {A}_\kappa^{(1)}+
	\left(\frac{\alpha_s^0}{2\pi}\right)^2 \qty(\frac{\NC}{2})^2 {A}_\kappa^{(2)}+
	\ldots{}
	\right)\,,
\end{equation}
and further decompose ${A}_\kappa^{(l)}$ into powers of $\NF/\NC$ as 
\begin{equation}\label{eq:nfDecomp}
	{A}_\kappa^{(l)}= \sum_{j=0}^l \left(\frac{\NF}{\NC}\right)^j{A}_\kappa^{(l)[j]}\,.
\end{equation}
We show representative diagrams in \cref{fig_parents2q2g,fig_parents4q}.

Since we are interested in analytic results for the partial helicity amplitudes
$A_\kappa (1^{h_1},\ldots{},4^{h_4})$, we can apply the following
transformations on the expressions: momentum permutations, charge, and parity
conjugation. This allows us to define a minimal generating set from which all
other partial helicity amplitudes are obtained by application of those
transformations.
We find that three different helicity configurations for $A_\gluon$ and two for $A_\quark$ suffice,
and we choose the generating set to be
\begin{subequations} \label{eq:gen-hel-ampl}
  \begin{align}
    &A_\gluon (1^{+}, 2^{+}, 3^{+}, 4^{-}) \, , \label{eq:gluon-mhv} \\
    &A_\gluon (1^{+}, 2^{+}, 3^{-}, 4^{-}) \, , \label{eq:gluon-nmhv1} \\
    &A_\gluon (1^{+}, 2^{-}, 3^{+}, 4^{-}) \, , \label{eq:gluon-nmhv2} \\[2ex]
    &A_\quark (1^{+}, 2^{+}, 3^{-}, 4^{-}) \, , \label{eq:quark-nmhv1} \\
    &A_\quark (1^{+}, 2^{-}, 3^{+}, 4^{-}) \, , \label{eq:quark-nmhv2} 
  \end{align}
\end{subequations}
where we recall that we already fixed the choice of lepton helicities in \cref{eq:partProcesses}.

The corresponding tree amplitudes are
\begin{subequations} \label{eq:trees}
  \begin{align}
    {A}_\gluon^{(0)[0]}(1^{+}, 2^{+}, 3^{+}, 4^{-}) &= \frac{⟨46⟩^2}{⟨12⟩⟨23⟩⟨34⟩⟨65⟩} \, , \\
    {A}_\gluon^{(0)[0]}(1^{+}, 2^{+}, 3^{-}, 4^{-}) &= \frac{⟨13⟩⟨3|1+2|5]^2}{⟨12⟩⟨23⟩[65]⟨1|2+3|4]s_{123}} \; + \; \frac{[24]⟨6|1+5|2]^2}{[23][34]⟨65⟩⟨1|2+3|4]s_{234}} \, , \\
    {A}_\gluon^{(0)[0]}(1^{+}, 2^{-}, 3^{+}, 4^{-}) &= \frac{[13]^3⟨46⟩^2}{[12][23]⟨65⟩⟨4|2+3|1]s_{123}} \; + \; \frac{[15]^2⟨24⟩^3}{⟨23⟩⟨34⟩[65]⟨4|2+3|1]s_{234}} \, , \\[3ex]
    {A}_\quark^{(0)[0]}(1^{+}, 2^{+}, 3^{-}, 4^{-}) &= \frac{-[12]^2⟨46⟩^2}{[23]⟨56⟩⟨4|2+3|1]s_{123}} \; + \; \frac{[15]^2⟨34⟩^2}{⟨23⟩[56]⟨4|2+3|1]s_{234}} \, , \\
    {A}_\quark^{(0)[0]}(1^{+}, 2^{-}, 3^{+}, 4^{-}) &= \frac{[13]^2⟨46⟩^2}{[23]⟨56⟩⟨4|2+3|1]s_{123}} \; + \; \frac{-[15]^2⟨24⟩^2}{⟨23⟩[56]⟨4|2+3|1]s_{234}} \, .
  \end{align}
\end{subequations}
Note that the two ${A}_\quark^{(0)[0]}$ amplitudes are related by a  $2\leftrightarrow 3$ swap, which makes them dependent.
This property is violated by loop corrections.

\subsection{Finite remainders and pentagon-function decomposition}

The generating amplitudes in \cref{eq:gen-hel-ampl} contain UV and IR divergences which, in dimensional regularization, manifest as poles in $\epsilon$.
These divergences are universal and can be predicted from lower orders in the perturbation series.
This allows us to define \emph{finite remainders}
\begin{equation}\label{eqn:remainder}
  {\cal M}_{\mathcal{R}} = \lim_{\epsilon\to 0} ~ {\bf I}(\epsilon) {\cal M(\epsilon)} \,, 
\end{equation}
where the UV divergences are cancelled by expanding the bare coupling $\alpha^0_s$ in terms of the renormalized coupling 
$\alpha_s(\mu)$ in the $\overline{\text{MS}}$ scheme,
and the IR divergences are cancelled by the color-space operator $\mathbf{I}$ \cite{Catani:1998bh,Sterman:2002qn,Becher:2009cu,Gardi:2009qi}.
Following the discussion in \cref{sec:partials}, we define the finite remainders $\mathcal{R}_\kappa$ for representative channels, which
are related to $\mathcal{M}_{\mathcal{R}}$ as the amplitudes $\mathcal{M}$ are related to $\mathcal{A}_\kappa$ through \cref{eq:ampW,eq:ampZ,eq:ampZneutral,eq:ampW-ident}.
Their decomposition into partial finite remainders is the same as the one for the associated amplitudes in \cref{eqn:colorAmp},
\begin{subequations} \label{eq:partial-remainders}
  \begin{align}
    \mathcal{R}_g  = \lim_{\epsilon\to 0} ~ {\bf I}_g(\epsilon){\cal A}_g(\epsilon) &= \sum_{\sigma\in S_2(2,3)}  \left(T^{a_{\sigma(3)}}T^{a_{\sigma(2)}}\right)^{\,\,\bar{i}_1}_{i_4} \; R_\gluon\left(1,\sigma(2),\sigma(3),4\right)  \,,\\
    \mathcal{R}_q = \lim_{\epsilon\to 0} ~ {\bf I}_q(\epsilon){\cal A}_q(\epsilon) &= \delta_{i_2}^{\bar{i}_1}\delta_{i_4}^{\bar{i}_3} \; R_\quark \left(1,2,3,4\right)  \,,
  \end{align}
\end{subequations}
where the operators $\mathbf{I}_{\kappa}$ are diagonal in color space in LCA, and their explicit definitions are given in \cref{sec:subtraction}.

The kinematic components of the finite remainders, $R_{\kappa}$ can then be
perturbatively expanded in the strong coupling, as well as powers of $\NF$ as
\begin{align}
	  R_{\kappa}&= (g_s)^2 \left({R}_\kappa^{(0)}+
	\frac{\alpha_s}{2\pi} \frac{\NC}{2} {R}_\kappa^{(1)}+
	\left(\frac{\alpha_s}{2\pi}\right)^2 \qty(\frac{\NC}{2})^2 {R}_\kappa^{(2)}+
	\ldots{}
	\right)\,, \\
	{R}_\kappa^{(l)} &= \sum_{j=0}^l \left(\frac{\NF}{\NC}\right)^j{R}_\kappa^{(l)[j]}\,,
\end{align}
where the expansion is given in terms of the $\overline{\text{MS}}$ renormalized
couplings $g_s$ and $\alpha_s$.
We therefore focus our attention on the generating set of finite remainders,
\begin{equation} \label{eqn:finRemainder}
  \begin{split}
    & R_\gluon (1^{+}, 2^{+}, 3^{+}, 4^{-}) \, , \\
    & R_\gluon (1^{+}, 2^{+}, 3^{-}, 4^{-}) \, , \\
    & R_\gluon (1^{+}, 2^{-}, 3^{+}, 4^{-}) \, , 
  \end{split}
  \qquad\qquad
  \begin{split}
    & R_\quark (1^{+}, 2^{+}, 3^{-}, 4^{-}) \, , \\
    & R_\quark (1^{+}, 2^{-}, 3^{+}, 4^{-}) \, , 
  \end{split}
\end{equation}
and their components $R_\kappa^{(l)[j]}$.

\paragraph{Spaces of rational coefficients}

We represent the finite remainders as combinations of transcendental special
functions $\{G_i\}$, the one-mass pentagon
functions~\cite{Chicherin:2021dyp,Abreu:2023rco}, with coefficients that are
rational functions of spinor brackets,
\begin{align} \label{eq:pentagon-functions}
  R_\kappa^{(l)[j]} (1^+,  2^{h_2}, 3^{h_3}, 4^{-})= \sum_{i} r_{\kappa, i}^{(l)[j] h_2 h_3} \; G_i \,.
\end{align}
For each generating finite remainder from \cref{eqn:finRemainder} we define the corresponding $\mathbb{Q}$-vector spaces of rational coefficients
\begin{equation}\label{eqn:basis}
  B_\kappa(1^+,  2^{h_2}, 3^{h_3}, 4^{-}) = \text{span}_{i,l,j}\{r_{\kappa, i}^{(l)[j] h_2 h_3}\},
\end{equation}
where the subscript of the span indicates that the argument is the list of
rational functions with all possible values of $i$ and $j$ and $l$, such that $0 \le
l \le 2$, whereas the helicity and $\kappa$ are fixed by the left-hand side. The
vector spaces defined in \cref{eqn:basis} are the central objects studied in
this work.

It is advantageous to consider together the rational coefficients that transform
uniformly under the little group and define the vector-spaces sums
\begin{subequations} \label{eq:B-generating_set}
  \begin{align}
    B_\gluon^{\text{MHV}} = & \;  B_\gluon(1^{+}, 2^{+}, 3^{+}, 4^{-}) \, , \label{eq:generating_set_first} \\
    B_\gluon^{\text{NMHV}} = & \;  B_\gluon(1^{+}, 2^{+}, 3^{-}, 4^{-})\, + \, B_\gluon (1^{+}, 3^{-}, 2^{+}, 4^{-}) \, + \, (123456 \rightarrow \overline{432165}) \,, \\[2ex]
    B_\quark^{\text{NMHV}} = & \; B_\quark(1^{+}, 2^{+}, 3^{-}, 4^{-})\,+\,  B_\quark(1^{+}, 3^{-}, 2^{+}, 4^{-}) \, + \, (123456 \rightarrow \overline{432165}) \,, \label{eq:generating_set_last}
  \end{align}
\end{subequations}
where in the last two lines we use the momentum swap $2\leftrightarrow 3$ to align the little-group weights.
These three vector spaces correspond to maximal helicity violating (MHV) and next-to-MHV (NMHV) helicity configurations.
Moreover, we have further closed the two NMHV spaces under
the transformation $123456 \rightarrow \overline{432165}$, which is a symmetry
of the NMHV amplitudes.

\begin{table}[ht]
\renewcommand{\arraystretch}{1.2}
\centering
\begin{tabular}{C{30ex} C{15ex} C{15ex}}
  \toprule
  \raisebox{2mm}{\makecell{Rational-Function \\ Vector Space}} & Vector-Space Dimension & Generating-Set Size \\
  \midrule
  $B_\gluon(1^{+}, 2^{+}, 3^{+}, 4^{-})\sim B_g^{\text{MHV}}$ & 54 & 54 \\
  \midrule
  $B_\gluon(1^{+}, 2^{+}, 3^{-}, 4^{-}) $ & \textcolor{gray}{110} & \textcolor{gray}{74} \\
  $B_\gluon(1^{+}, 2^{-}, 3^{+}, 4^{-}) $ & \textcolor{gray}{120} & \textcolor{gray}{81} \\
  $B_\gluon^{\text{NMHV}}$ & 219 & 139 \\
  \midrule
  $B_\quark(1^{+}, 2^{+}, 3^{-}, 4^{-})$ & \textcolor{gray}{70} & \textcolor{gray}{45} \\
  $B_\quark(1^{+}, 2^{-}, 3^{+}, 4^{-})$ & \textcolor{gray}{98} & \textcolor{gray}{66} \\
  $B_\quark^{\text{NMHV}}$ & 142 & 86 \\
  \bottomrule
  %%%%%%%
\end{tabular}
\caption{This table compares the dimension of the vector spaces of
  rational functions before and after merging them into MHV and
  NMHV spaces, respectively, as well as the number of functions in their generating
  sets.  We provide results for the spaces in black (see section~\ref{sec:ancillaries}), while the greyed out spaces were computed at intermediate steps.}
\label{table:overlap}
\end{table}

It is well known that the coefficients $r_{\kappa, i}^{(l)[j] h_2 h_3}$ (\ref{eq:pentagon-functions}) are not linearly independent,
that is the dimensions of the $B_\kappa$ vector spaces are smaller than the
number of elements considered in the span of \cref{eqn:basis}.
We show explicit values for the dimensions in the second column of \cref{table:overlap} for each of the generating helicity remainders,
as well as for the three combined vector spaces defined in \cref{eq:B-generating_set}.
We see that the latter is beneficial, as it reduces the number of rational functions we need to consider.
A non-trivial observation about the spaces of rational functions is that they
can be organized by the action of permutations of momenta and the parity
transformation. That is, similarly to how the $A_\quark$ tree amplitudes in
\cref{eq:trees} are related by a $2\leftrightarrow 3$ swap, a subset of
coefficients may belong to such permutation orbits and we can chose a
representative from each orbit. We show in the third column of
\cref{table:overlap} that the size of the resulting generating sets is indeed
smaller than the corresponding vector-space dimension.

Finally, we note that the three vector spaces $B_g^{\text{MHV}}$,
$B_g^{\text{NMHV}}$ and $B_q^{\text{NMHV}}$ are internally partitioned
into two disjoint spaces, because rational functions with two distinct mass
dimensions appear. Most functions have dimension $-2$ since they multiply
dimensionless pentagon functions. However, some functions have
dimension $0$ instead, since pentagon functions that are odd under square-root sign changes
are always normalized by the corresponding square roots and therefore are not dimensionless.%
\footnote{
  In the conventions of \cite{Abreu:2023rco} the same would apply to the square root of Gram determinant of four momenta $\sqrt{\Delta_5} = \sqrt{\trFive^2}$,
  which would introduce an additional disjoint space of mass dimension $2$.
  In this work we use the conventions of \cite{Chicherin:2021dyp}, where the relevant pentagon functions are defined through $\trFive$ instead of $\sqrt{\Delta_5}$.
  Since $\trFive$ is polynomial in spinor brackets, we may consider the coefficients of such pentagon functions to have mass dimension $-2$ instead.
}
The sub-spaces with mass dimension 0
have dimension 11, 2, and 17 for $B_q^{\text{NMHV}}$,
$B_g^{\text{MHV}}$ and $B_g^{\text{NMHV}}$, respectively. Throughout
the rest of this work we do not treat these
sub-spaces as disjoint, although this may be beneficial in future
computations, especially if the vector spaces become bigger.

\paragraph{Spinor-helicity form of coefficient functions}

The considered helicity amplitudes are naturally 
expressed in terms of rational functions in spinor variables.
Consequently, we consider the rational functions $r_{\kappa,i}^{(l)[j] h_2 h_3}$ of
\eqn{eq:pentagon-functions} which can be expressed in the form
     \begin{equation} \label{eqn:spinorLCD}
       r_i(\lambda, \tilde{\lambda}) = \frac{\mathcal{N}_i(\lambda, \tilde{\lambda})}{\prod_{\mathcal{D}_k}\mathcal{D}_k^{q_{ik}}(\lambda, \tilde{\lambda})},
     \end{equation}
where we suppress helicity labels, $h_i$, the loop-order and $N_f$ 
labels, $(l)$ and $[j]$, as well as the process label, $\kappa$.
The numerator and denominator factors are Lorentz-invariant polynomials in
spinor variables and the $q_{ik}$ take on integer values. The denominator factors $\mathcal{D}_k$ 
are distinguished polynomials, which will be introduced in the following paragraphs. 
We require a measure of the complexity of the rational function in \cref{eqn:spinorLCD}, 
which controls the number of terms in the numerator polynomials $\mathcal{N}_i$. We use the notion 
of the polynomial degree of the function's numerator and denominator in spinor variables, 
which we denote by $\degn(r_i)$ and $\degd(r_i)$, respectively. 
Both notions are used interchangeably since they differ by a constant.

It has been observed \cite{Abreu:2018zmy} that the collection of denominator factors can be
determined from the so-called ``letters'' $W_i$ of the ``symbol alphabet''
associated to the processes' Feynman integrals.
The alphabet for the process at hand was computed in ref.~\cite{Abreu:2020jxa},
and there it is expressed in terms of Mandelstam variables. Such expressions can
naturally be written in terms of spinor variables. However, in performing such a
rewriting, a number of the symbol letters factorize in spinor variables. That
is, for some letters we find that $W_i = \mathcal{D}_i
\mathcal{D}_{i'}$ for two lower-degree spinor polynomials $\mathcal{D}_i$ and
$\mathcal{D}_{i'}$, which appear in \eqn{eqn:spinorLCD}.
Under this rewriting, we find the set of denominator factors $\mathcal{D}_i$ to
be
\begin{align}
\begin{split}
  \{\mathcal{D}_1, \ldots, \mathcal{D}_{44}\}
  = \{⟨12⟩, ⟨13⟩, ⟨14⟩, ⟨23⟩, ⟨24⟩, ⟨34⟩, ⟨56⟩, [12], [13], [14], [23], [24], [34], [56], \,\,\,\,\\
  ⟨1|2+3|1], ⟨1|5+6|1], ⟨2|1+3|2], ⟨2|3+4|2], ⟨2|5+6|2], \,\,\,\,\\
  ⟨3|1+2|3], ⟨3|2+4|3], ⟨3|5+6|3], ⟨4|2+3|4], ⟨4|5+6|4], \,\,\,\,\\
  ⟨1|3+4|2], ⟨1|2+4|3], ⟨1|2+3|4], ⟨2|3+4|1], ⟨2|1+4|3], ⟨2|1+3|4], \,\,\,\,\\
  ⟨3|2+4|1], ⟨3|1+2|4], ⟨4|2+3|1], ⟨4|1+3|2], ⟨4|1+2|3], \,\,\,\,\\
  s_{123}, s_{234}, \Delta_{12|34|56}, \Delta_{13|24|56}, \Delta_{23|14|56}, \,\,\,\,\\
  ⟨2|3|5+6|1|2]-⟨3|4|5+6|1|3], ⟨3|2|5+6|4|3]-⟨2|1|5+6|4|2], \,\,\,\, \\
  ⟨2|3|5+6|4|2]-⟨3|1|5+6|4|3], ⟨3|2|5+6|1|3]-⟨2|4|5+6|1|2] \}.
\end{split}
  \label{eq:AllDenominatorFactors}
\end{align}
This set of denominator factors is the same as those found in the one-loop
amplitude, up to the last four denominators, which are new at two loops.
We note that the set of
denominator factors is larger in spinor variables compared to the
ones expressed in Mandelstam variables. This is one reason why
rational functions in spinor space are expected to be simpler: the
factorization of denominator factors can lead to additional cancellations. In
particular cancellations are not restricted to be either parity or little-group
invariant.

\section{Improved rational bases}
\label{sec:basischange}

In a reconstruction-based approach to the calculation of rational coefficients
in a scattering amplitude, an important step is to choose which
set of rational functions one should reconstruct such that the reconstruction
procedure is as efficient as possible.
For example, it has been established for some time~\cite{Badger:2018enw,
Abreu:2018zmy} that it is more efficient to avoid the computation of the rational
functions that arise in the master integral decomposition and instead
reconstructing the rational functions $r_i$ in the finite remainder of
\cref{eq:pentagon-functions}. (For simplicity we suppress helicity and particle
labels in the following.)
This follows as, when considered in the common denominator form of
\cref{eqn:spinorLCD}, the numerator functions $\mathcal{N}_i$ are typically lower degree
polynomials than what one finds for master integral coefficients.
Beyond this, a further approach that is often used in amplitude computations to
decrease numerator degrees is to exploit that the $r_i$ are linearly
dependent~\cite{Abreu:2019odu,Abreu:2021asb}.
That is, denoting a basis of $\text{span}(r_i)$ as $\Bdeg$, one can re-express each
$r_i$ in terms of this basis as
\begin{equation}\label{eqn:Mmatrix}
  r_i = \sum_{r_j \in \Bdeg} M_{ij}  r_j,
\end{equation}
where $M_{ij}$ is a rectangular matrix of rational numbers. It is then
clearly sufficient to reconstruct the set of rational functions $r_j \in
\Bdeg$.
The benefit of this observation is that there is freedom in choosing the basis of
$\text{span}(r_i)$, which one can exploit to ease reconstruction.
To date, the standard strategy for this has been to sort the $r_i$ in terms of
the number of unknowns in their ansatz, and construct a basis prioritizing the
$r_i$ with the smallest number of unknowns~\cite{Abreu:2019odu}.
For this reason we referred to this basis as
$\Bdeg$, as it is primarily based on the polynomial degree of
the numerators of the $r_i$.

An important question is if $\Bdeg$ is the optimal choice of
basis for reconstruction, and, if not, can we systematically change basis from
$\Bdeg$ to an improved basis $\Bimp$?
In this section, we show how to do just this. That is, we construct a basis
change with a $\mathbb{Q}$-valued, full-rank matrix $U$ such that
\begin{equation}
  r_i = U_{ij} \tilde r_j \quad \mbox{with} 
\quad r_i \in \Bdeg \quad \mbox{and} \quad \tilde r_j \in \Bimp
\end{equation}
and the $\tilde r_j$ are easier to reconstruct.
While suggestions have been made some time ago~\cite{Abreu:2018zmy}, here we
present a systematic procedure based on studying the
$r_i\in\Bdeg$ on surfaces where individual denominators vanish. This algorithm,
initially developed for the
present $pp\rightarrow Vjj$ calculation, has already been used in the arguably
simpler case of $gg\rightarrow ggg$ \cite{DeLaurentis:2023nss}. We present here
the algorithm in greater detail.

To begin, let us consider a motivating example, illustrating an observation made
in ref.~\cite{Abreu:2023bdp}: a rational function in the basis can ``contain''
other basis elements. We construct a toy space of rational functions, with two
basis elements $r_{1}$ and $r_{2}$,
\begin{equation}\label{eq:example-vector-space}
  r_1 = \frac{⟨46⟩^2[12]}{⟨56⟩⟨4|1+2|3]s_{123}} \, , 
\quad \text{and} \quad r_2 = \frac{⟨46⟩^2⟨4|1+3|2][13][12]}{[23]⟨56⟩⟨4|2+3|1]⟨4|1+2|3]s_{123}} \, .
\end{equation}
The denominator degree of the two functions is $\degd(r_1)=10$ and $\degd(r_2)=16$.
We have constructed this example such that $r_2$ ``contains'' $r_1$. However, in this
representation, it is not manifest. Nevertheless, we can rewrite $r_2$ as
\begin{equation}
  r_2 = \frac{⟨46⟩^2[12]}{⟨56⟩⟨4|1+2|3]s_{123}}+\frac{⟨46⟩^2[12]^2}{[23]⟨56⟩⟨4|2+3|1]s_{123}} \, .
  \label{eq:organizedR2Example}
\end{equation}
Clearly, a simpler basis would involve replacing $r_2$. That is, if we were to 
take as our basis $\{\tilde r_1, \tilde r_2\} = \{r_1, r_2-r_1\}$, the
problem of reconstructing the $\tilde{r}_i$ is simplified. This follows as
the denominator degree of the improved basis $\{\tilde r_1, \tilde r_2\}$ is
$\degd(\tilde r_1)=10$ and $\degd(\tilde r_2)=12$ and hence the maximum degree of the
numerators has been reduced.
If we
carefully consider this example, we see that we have added a multiple of
$r_1$ to $r_2$ such that the resulting function has a lower degree denominator
polynomial. Written more generally, we say that a function $r_i$ ``contains''
other functions $r_j$ for $j \ne i$ if
\begin{equation}
  r_i = \tilde{r}_i + \sum_{j \ne i} c_j^{(i)} r_j \qquad \text{such that} \qquad \degd(\tilde{r}_i) < \degd(r_i),
  \label{eq:ContainmentRelation}
\end{equation}
where the $c_j^{(i)}$ are rational numbers.

Let us now consider how we can systematically construct the $c_j^{(i)}$.
While in our example we judiciously changed representation to find the linear
combination, in general, we want to be able to perform this procedure without
having access to the analytic expressions. Moreover, we wish to construct a
procedure that can handle coefficient functions with high degree 
(we encounter functions of numerator degree up to 228 and containing 
a large number of numerator monomials as shown in \cref{tab:ansatze-sizes}).
To this end, let us exploit the following observation: in order for the
$\tilde{r}_i$ to have lower denominator degree than the input $r_i$, it must be
the case that there is some denominator factor in $r_i$ with greater exponent
than in $\tilde{r}_i$. It must, therefore, be cancelled by the unknown linear
combination of the $r_j$. Let us denote this denominator as $\mathcal{D}$.
We are therefore motivated to consider \cref{eq:ContainmentRelation} to
leading power in $\mathcal{D}$. Let us first expand the $r_i$ themselves.
Denoting the leading power in the expansion
as $n$, we write
\begin{equation}
  r_i = \frac{\text{R}(r_i, \mathcal{D})}{\mathcal{D}^n} + \mathcal{O}\left( \frac{1}{\mathcal{D}^{n-1}} \right),
  \label{eq:CodimensionOneExpansion}
\end{equation}
where we introduce $R(r_i, \mathcal{D})$ the numerator of the leading pole in
$\mathcal{D}$ of $r_i$. Slightly abusing terminology, we will call this
the ``residue'' of $r_i$ on $\mathcal{D} = 0$. (We also use the notion ``co-dimension-one residue'' 
to emphasize the dimensionality of the $\mathcal{D} = 0$ variety.)
Inserting such expansions into \cref{eq:ContainmentRelation} we
find
\begin{equation}
  R(r_i, \mathcal{D}) =  \sum_{j \ne i} c_j^{(i)} R(r_j, \mathcal{D}) + \mathcal{O}\left( \mathcal{D} \right),
\end{equation}
where we have explicitly dropped the $\tilde{r}_i$ term, as it is sub-leading in $\mathcal{D}$.
We therefore conclude that we can construct a combination of the $r_j$, $j \ne
i$ which subtracts the leading pole in $\mathcal{D}$ by considering linear
dependencies of the $\text{R}(r_i, \mathcal{D})$ on the variety $\mathcal{D} =
0$. That is, despite the $r_i$ being linearly independent, they become linearly
dependent on $\mathcal{D} = 0$, and we can use these linear dependencies to
choose a simpler basis.

Let us now return to the above example. If we consider
eq.~\eqref{eq:organizedR2Example}, we see that the first term has a pole at
$\langle  4 | 1 + 2| 3]$, while the second term does not. Hence, expanding around
$\langle 4 | 1 + 2| 3]$ it is clear to see that $r_1$ and $r_2$ develop a linear
dependency, i.e. $\text{R}(r_1, \langle 4 | 1 + 2| 3]) = \text{R}(r_2, \langle 4
| 1 + 2| 3])$. This then tells us that the combination $r_2 - r_1$ will have a
lower degree pole in $\langle 4 | 1 + 2| 3]$, leading us systematically to the
basis choice made earlier by inspection.

\subsection{An algorithm for basis improvement}

Having motivated that it is possible to use linear dependencies of the
co-dimension one residues (\ref{eq:CodimensionOneExpansion}) to find a new basis of simpler rational functions, let
us now describe how one can systematically organize this.
We start from the initial basis of the space of rational functions, $\Bdeg$.
Our aim is to find a new collection of functions $\Bimp =
\{\tilde{r}_1, \ldots \tilde{r}_{|\Bdeg|}\}$ with lower numerator degrees
in least-common-denominator form, subject to the constraint that
the change of basis matrix form the $r_i$ to the $\tilde{r}_i$ is of
full rank.
To this end, we consider linear combinations of rational functions $r_i \in
\Bdeg$ whose pole degree in $\mathcal{D}_k$ is at most $\nu$. That
is, we consider $\alpha_i \in \mathbb{Q}$ such that
\begin{equation}
  \sum_{r_i \in \Bdeg} \alpha_i r_i = \mathcal{O}\left(\frac{1}{\mathcal{D}_k^{\nu}}\right).
\end{equation}
We denote the vector space of such 
coordinate vectors $\vec \alpha =\{\alpha_i\}_{i=1,|\Bdeg|}$ by $N_{\mathcal{D}_k}^{(\nu)}$, 
i.e. $\vec \alpha\in N_{\mathcal{D}_k}^{(\nu)}$.
Explicitly, such vector spaces represent combinations of the $r_i$ with a prescribed
maximum pole order in $\mathcal{D}_k$. A simple observation is that intersections of these vector
spaces contain elements which correspond to functions with a specified pole
order in multiple $\mathcal{D}_k$. We therefore consider the intersection of
spaces
\begin{equation}
  N^{(\vec{\nu})} = \bigcap_{\mathcal{D}_k} N_{\mathcal{D}_k}^{(\nu_k)} \,.
\end{equation}
%In this formalism, our approach is simple. 
Using this formalism, we can now construct the improved basis $\Bimp$.
Starting from a set of functions
$\{r_j\}$, we choose a function $r_i$ and look to replace it with
\begin{equation}
  \tilde{r}_i = \frac{\tilde{\mathcal{N}}_i}{\prod_k \mathcal{D}^{\tilde{q}_{ik}}_k} = r_i + \sum_{j \ne i} c_j^{(i)} r_j  
  \label{eq:ReplacementForm}
\end{equation}
such that $\sum_k \tilde{q}_{ik}$ is minimal.
That is, we look for linear combinations of the input basis $r_i$ such that the
basis change minimizes the total number of denominator factors in the least
common denominator form of the $\tilde{r}_i$.
In order to find such a replacement, we search through the collection of spaces
$N^{(\vec{\nu})}$ to find a linear
combination of elements of $\Bdeg$ with a non-zero coefficient for $r_i$.
The minimality of $\sum_k \tilde{q}_{ik}$ is then achieved by demanding that 
there doesn't exist such a space $N^{(\vec{\nu'})}$ with $\sum_k 
\nu_k' < \sum_k \nu_k$.
The full rank of the basis-change matrix is assured by the constraint that
the coefficient of $r_i$ is non-zero. We then repeat this for all of the
functions of the basis.
Naturally, the output of the procedure will depend on
the ordering of the functions, but we leave the investigation of such dependence for
future work.
This procedure involves two important steps. First is the construction of the
collection of vector spaces $N_{\mathcal{D}_k}^{(\nu_k)}$.
Next, is searching through the large collection of spaces $N^{(\vec{\nu})}$ for
our replacement function. We will now discuss both of these problems in
turn.

\paragraph{Spaces of pole-cancelling combinations}

\begin{figure}[t]
  \centering
  \includegraphics[width=0.6\textwidth, trim={0cm 3cm 25cm 3cm},clip]{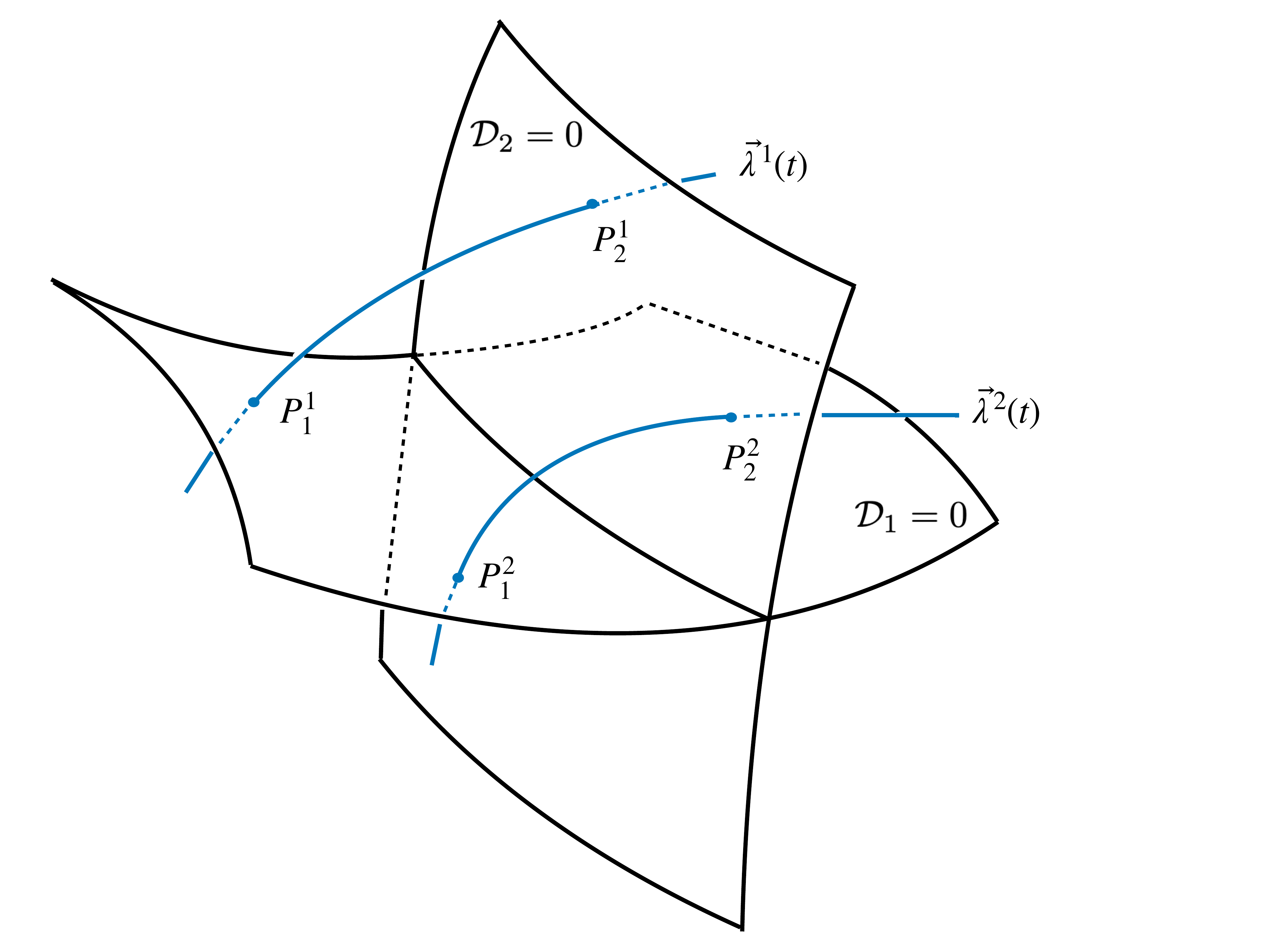}
\caption{
Displayed is the setup for the residue computation of a rational coefficient
function $r_i$ with poles at $\mathcal{D}_k=0$.
A family of two slices $\{\vec\lambda^j(t)\}_j$ intersects the 
singular surfaces at the four points $P_k^j$.
Expansion of the function $r(t)=r_i[ \vec\lambda^j(t)]$ around the
intersection points $P_k^j$ yields the residue functions $R(r_i,\mathcal{D}_k)$ 
evaluated at these points.
}
\label{fig:Slice}
\end{figure}

%\paragraph{Linear Dependencies of Residues}
Let us consider the construction of the $N_{\mathcal{D}_k}^{(\nu)}$. We denote
by $n_k$ the highest pole order of $\mathcal{D}_k$ across all of the $r_i$.
By definition, $N_{\mathcal{D}_k}^{(n_k)}$ corresponds to $\text{span}(r_i)$, so 
we begin by considering the case of $N_{\mathcal{D}_k}^{(n_k-1)}$.
A natural approach would be to evaluate on phase-space points where
$\mathcal{D}_k$ is small, and use the standard technique of constructing linear
dependencies between functions by repeatedly evaluating these functions. In
practice, such a strategy would require working in a field with a non-trivial
absolute value such as the complex, or $p$-adic numbers. For practical purposes,
we choose to limit ourselves to work in a finite-field, where no such absolute
value is available and, instead, use a semi-numerical approach. Specifically, we
will extract evaluations on co-dimension-one surfaces from regular evaluations at
points on a one-dimensional curve that intersects the surface. 
This setup is displayed in \cref{fig:Slice}, and we introduce it now.

To begin, we construct a particular ``univariate slice'' in spinor
space~\cite{Elvang:2008vz, Abreu:2021asb, Abreu:2023bdp, PageSAGEXLectures}.
Specifically, introducing arbitrarily chosen reference spinors $\eta,
\tilde{\eta}$, we build a line in spinor space, parameterized as
\begin{align}
  \vec\lambda(t) &=  \{ \lambda_1(t) , \tilde\lambda_1(t),  ... , \lambda_m(t), \tilde\lambda_m(t) \} \,, \\
  \lambda_i(t) & = \lambda_i + t x_i \eta,
  \qquad
  \tilde{\lambda}_i(t) = \tilde{\lambda}_i + t y_i \tilde{\eta} \quad \mbox{for} \quad i=1,... m\,.
  \label{eq:SpinorSlice}
\end{align}
Here we assume an m-point phase space. The spinors $\eta$, $\tilde{\eta}$, $\{\lambda_i, \tilde\lambda_i\}_{i=1,m}$ and the parameters
$\{x_i, y_i\}_{i=1,m}$ take on fixed numerical values and $t$ is a free parameter. A non-trivial
step in constructing such a slice is choosing the $\eta$, $\tilde{\eta}$, $\lambda_i$,
$\tilde{\lambda}_i$, $x_i$ and $y_i$ such that
momentum
conservation is satisfied for all values of $t$. This can efficiently be
accomplished using the methods of ref.~\cite{DeLaurentis:2022otd} and is
described in more detail in ref.~\cite{Campbell:2024tqg}. We note that, as the
spinors are linear in $t$, the coefficient functions $r_i$ are rational
functions in the parameter $t$.
As $\lambda_i(t)$ and $\tilde{\lambda}_i(t)$ spinor in \cref{eq:SpinorSlice}
each depend on the same reference spinors $\eta$ and $\tilde{\eta}$, it is easy
to see that the degree of the numerator and denominator polynomials in $t$ is
half of $\degn(r_i)$ and $\degd(r_i)$ respectively.

On a slice of the form of \cref{eq:SpinorSlice}, every denominator factor
$\mathcal{D}_k$ of the $r_i$ becomes a polynomial in $t$.
As a finite field is not algebraically closed, this polynomial will not
necessarily factor linearly on a univariate slice.
Nevertheless, it is simple to randomly generate slices until each
$\mathcal{D}_k$ has the form
\begin{equation}
  \mathcal{D}_k(t) = \overline{\mathcal{D}}_k(t) (t - t_k),
  \label{eq:LinearFactorSlice}
\end{equation}
that is, such that $\mathcal{D}_k(t)$ has a single linear factor. In this way,
we see that the denominator $\mathcal{D}_k$ vanishes for $t=t_k$, 
i.e. at the phase-space point
\begin{align}
P_k = \vec\lambda(t_k)
\end{align}
%this is true for all of the denominators of the $r_i$.
%
The importance of this is that expanding $r_i(t)$ around $t = t_k$ effectively
expands around $\mathcal{D}_k=0$. Explicitly, by comparison with
eq.~\eqref{eq:CodimensionOneExpansion}, we can write
\begin{equation}
  %r_i(t) = \frac{R(r_i, \mathcal{D}_k)[\lambda(t_k),\tilde\lambda(t_k)]}{[\overline{\mathcal{D}}_k(t_k)(t-t_k)]^{n_k}} + \mathcal{O}\left(\frac{1}{(t-t_k)^{n_k-1}}\right),
  r_i(t) = \frac{R(r_i, \mathcal{D}_k)[P_k]}{[\overline{\mathcal{D}}_k(t_k)(t-t_k)]^{n_k}} + \mathcal{O}\left(\frac{1}{(t-t_k)^{n_k-1}}\right),
  \label{eq:ResidueFromExpansion}
\end{equation}
where $n_k$ is the highest power of $\mathcal{D}_k$ across the full set of $r_i$. 
In other words, by considering $r_i$ on the univariate slice, and expanding
around $t=t_k$, we are able to evaluate the $R(r_i, \mathcal{D}_k)$ at the point
$P_k$ on the surface $\mathcal{D}_k=0$.

To practically apply this evaluation procedure, for a given univariate slice, we
reconstruct $r_i(t)$ from numerical evaluations along the slice
(\ref{eq:SpinorSlice}). The expansion around $t=t_k$ is then easily achieved in
a modern computer algebra system. An important feature of this procedure is that
it allows us to numerically evaluate all of the co-dimension one residues
$R(r_i, \mathcal{D}_k)$ of the $r_i$ while using only a single univariate
slice. We can see this geometrically in \cref{fig:Slice}, where each univariate slice
intersects multiple surfaces.

From this evaluation strategy, it is now easy to build the set of
$\mathbb{Q}$-linear relations experienced by the leading-pole residues. In
analogy to the standard method of determining $\mathbb{Q}$-linear relations
between the functions $r_i$, we consider sampling the co-dimension one residues
associated to $\mathcal{D}_k$ over a set of phase-space points satisfying
$\mathcal{D}_k = 0$. To obtain these evaluations we use a collection of slices 
$\{ \vec \lambda^j(t)\}_j$ which we label by $j$.
Phase-space points satisfying $\mathcal{D}_k=0$ are then
\begin{align}
P_k^j= \vec\lambda^j(t_k^j) \,.
\end{align}
To determine linear 
dependence of the residues we consider the equation
\begin{equation}
  %\sum_{r_i \in \Bdeg} \alpha_i \mathrm{R}(r_i, \mathcal{D}_k)[\lambda^{j}(t_k^j), \tilde{\lambda}^{j}(t_k^j)] = 0, \qquad \alpha_i \in \mathbb{Q}
  \sum_{r_i \in \Bdeg} \alpha_i \, \mathrm{R}(r_i, \mathcal{D}_k)[P_k^j] = 0, \qquad \alpha_i \in \mathbb{Q} , \quad \forall j\,.
  \label{eq:LinearDependencyFromEvaluations}
\end{equation}
By evaluating \cref{eq:LinearDependencyFromEvaluations} on as many points as the
dimension of $\mathrm{span}(r_i)$ we end up with a set of linear equations that
define the space in which the $\vec{\alpha}$ live.
This space of linear relations is exactly $N_{\mathcal{D}_k}^{(n_k - 1)}$, and a
basis can easily be calculated using standard finite-field and linear algebra
methods.
In our experience, it is practically important to
lift the $\alpha_i$ that solve \cref{eq:LinearDependencyFromEvaluations} from finite-field to $\mathbb{Q}$ at this stage of the
algorithm.
We remark that, in this  approach, we practically
require at most as many slices as the dimension of
$\text{span}_i[\mathrm{R}(r_i, \mathcal{D}_k)]$. Interestingly, we often find that
this is much smaller than the dimension of $\mathrm{span}(r_i)$.

Let us now consider the construction of the spaces $N_{\mathcal{D}_k}^{(\nu - 1)}$
for $\nu \le n_k - 1$, that is, combinations of the $r_i$ which have even lower
pole order than we have just constructed.
If we proceed iteratively, we can assume that we have access to a basis of 
$N_{\mathcal{D}_k}^{(\nu)}$. Such a basis allows us to construct explicit
rational functions whose pole order at $\mathcal{D}_k = 0$ is less than or equal
to $\nu$.
Specifically, for any element $\vec{\alpha}$ of $N_{\mathcal{D}_k}^{(\nu)}$, one
constructs such a function $r(\vec{\alpha})$ as
\begin{equation}
  \vec \alpha \in N_{\mathcal{D}_k}^{(\nu)} \quad \longrightarrow \quad r(\vec \alpha) = \sum_{r_i \in \Bdeg} \alpha_i r_i = \mathcal{O}\left(\frac{1}{\mathcal{D}_k^{\nu}}\right)\,.
  \label{eq:CombinationConstruction}
\end{equation}
Given this construction, it is not difficult to see that that we can further
apply our linear
relation approach to build a basis of $N_{\mathcal{D}_k}^{(\nu - 1)}$. Specifically, given a basis $\{ \vec \beta_j \}_j$ of
$N_{\mathcal{D}_k}^{(\nu)}$, by using \cref{eq:CombinationConstruction} we
construct a set of functions $\{ r(\vec \beta_j) \}_j$ whose leading pole at $\mathcal{D}_k$ is
order at most $\nu$. We can then use the sampling logic from before to find combinations of
the $\{r(\vec \beta_j)\}_j$ which cancel their leading pole at $\mathcal{D}_k = 0$.
A basis of this space of combinations, together with the basis $\{\vec \beta_j\}_j$ of
$N_{\mathcal{D}_k}^{(\nu)}$ can then be combined to yield a basis of the space
of combinations with pole order in $\mathcal{D}_k$ at most $\nu - 1$, i.e.
$N_{\mathcal{D}_k}^{(\nu-1)}$.
This procedure can clearly be iterated all the way down to a space of
combinations of the $r_i$ which do not have any pole at $\mathcal{D}_k=0$.

To close, we comment that the constraint of requiring linear factors on the
slices of \cref{eq:LinearFactorSlice} while convenient, can become problematic
if a large number of non-linear-in-$t$ $\mathcal{D}_k$ exist.
Alternatively, one could accept any slice, even with non-linear
factors, by generalising \cref{eq:ResidueFromExpansion} to a $p(t)$-adic
expansion (which has recently been applied in the context of intersection
theory~\cite{Fontana:2023amt}).

\paragraph{Searching the graph of pole cancellations}

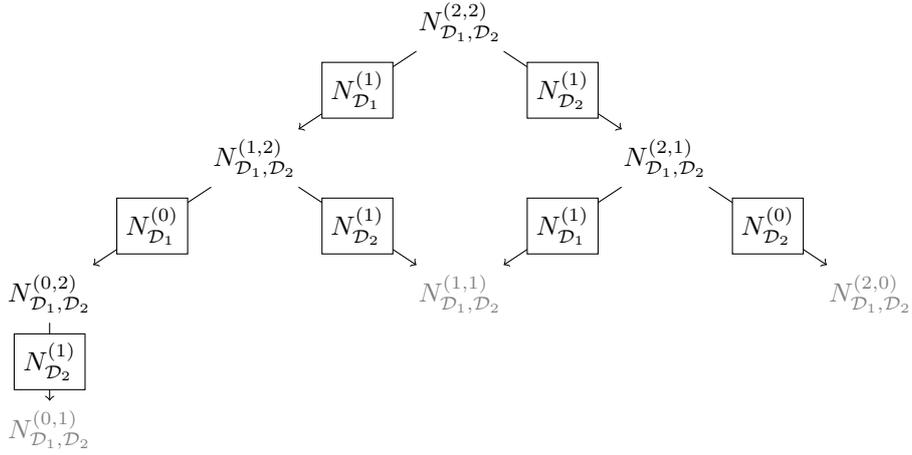
\begin{figure}
  \hspace{-3mm}
  \centering
\begin{tikzpicture}
[
  scale=0.9,
	level 1/.style = {draw, ->, black, sibling distance = 6cm},
  level distance = 20mm
	%level 2/.style = {black, sibling distance = 2.5cm},
	%level 3/.style = {black, level distance = 1.5cm},
	%edge from parent /.style = {draw, ->} path = {(\tikzparentnode\tikzparentanchor) -- (\tikzchildnode\tikzchildanchor)}
]

\node {$N^{(2,2)}_{\mathcal{D}_1, \mathcal{D}_2}$}
      child {
           node {$N^{(1,2)}_{\mathcal{D}_1, \mathcal{D}_2}$}
           child {
             %node {$\tilde{r}_i \in N^{(0,2)}_{\mathcal{D}_1, \mathcal{D}_2}$}
             node {$N^{(0,2)}_{\mathcal{D}_1, \mathcal{D}_2}$}
               child{
                 node [opacity=0.5]{$N^{(0,1)}_{\mathcal{D}_1, \mathcal{D}_2}$}
                 edge from parent node [midway, fill=white, draw] {$ N_{\mathcal{D}_2}^{(1)}$}
               }
               edge from parent node [midway, fill=white,draw] {$ N_{\mathcal{D}_1}^{(0)}$} }
            child {node(shared)[opacity=0.5] {$N^{(1,1)}_{\mathcal{D}_1, \mathcal{D}_2}$}
                  edge from parent node [midway, fill=white,draw] {$N_{\mathcal{D}_2}^{(1)}$} }
	edge from parent node [midway, fill=white, draw] {$N_{\mathcal{D}_1}^{(1)}$}
        }
      child {
            node(2ndparent) {$N^{(2,1)}_{\mathcal{D}_1, \mathcal{D}_2}$}
	    child {node {$\phantom{N^{(1,1)}_{\mathcal{D}_1, \mathcal{D}_2}}$}
                  edge from parent node [midway, fill=white,draw] { $N_{\mathcal{D}_1}^{(1)}$} }  % shared child
	    child {node[opacity=0.5] {$N^{(2,0)}_{\mathcal{D}_1, \mathcal{D}_2}$}
                  edge from parent node [midway, fill=white,draw] {$N_{\mathcal{D}_2}^{(0)}$} }
	edge from parent node [midway, fill=white, draw] {$N_{\mathcal{D}_2}^{(1)}$}
      };
%      \draw (shared) -- (2ndparent) [out=0, in=180, dashed];
\end{tikzpicture}
\caption{\label{FigSearchTree}Toy example of a simple search graph for
  finding a replacement for the function $r_i$.
  Any edge represents the intersection of the space entering the edge with the
  space along the edge. Faded vector spaces are empty since they have no overlap with the function
  $r_i$. Descendants of faded vector spaces are not constructed.
  This example graph has both a local minimum, $N^{(2,1)}_{\mathcal{D}_1,
  \mathcal{D}_2}$, and a global minimum, $N_{\mathcal{D}_1
  \mathcal{D}_2}^{(0,2)}$, from where we pick the linear combination of $r_i$ 
  to obtain the replacement function $\tilde r_i$ for $r_i$.
}
\vspace{-2mm}
\end{figure}

Having constructed the basic spaces $N_{\mathcal{D}_k}^{(\nu)}$, it remains to
consider their intersections in order to find a replacement for each $r_i$ as
in \cref{eq:ReplacementForm}. A priori, this task is somewhat daunting as,
defining $n_k$ as the largest pole order of denominator $\mathcal{D}_k$ in the
set $\Bdeg$, there are $\prod_{\mathcal{D}_k}(n_k + 1)$ such spaces, a
count that grows exponentially with the number of denominators. The calculation
of an exponentially large number of vector-space intersections, even of low
dimension, seems prohibitive. However, in practice, we find that a suitable
organization of the construction of these vector spaces makes this tractable.

Our organization is exemplified by fig.~\ref{FigSearchTree}. First, we
organize the spaces into ``rows'' of total denominator count $n$. That is, each
element $N^{(\vec{\nu})}_{\vec{\mathcal{D}}}$ in a row satisfies $\sum_i \nu_i =
n$. The starting row contains a single node, which allows all possible poles
in the original set of functions, and therefore corresponds to the space
$\text{span}(r_i)$.
Next, we observe that there is a simple operation to generate row $n-1$ from
row $n$. Specifically, each element can be constructed through
\begin{equation}
  N_{\vec{\mathcal{D}}}^{(\nu_1, \ldots \nu_k - 1, \ldots)} = N_{\vec{\mathcal{D}}}^{(\vec{\nu})} \cap N_{\mathcal{D}_k}^{(\nu_k - 1)}.
  \label{eq:RowStep}
\end{equation}
In fig.~\ref{FigSearchTree}, this operation is denoted by a directed edge
between an element of each row. We therefore see that our organization as the
structure of a directed acyclic graph.
The vector-space intersection in \cref{eq:RowStep} is a very cheap linear
algebra operation due to the low dimensionality of the spaces involved.
Importantly, this construction avoids a large amount of redundancy
in constructing the $N_{\vec{\mathcal{D}}}^{(\vec{\nu})}$.

An important property of the graph is that nodes on the graph which are
``descended'' from other nodes, i.e.~related by walking a path backwards
against the arrows, necessarily correspond to vector spaces whose dimension
cannot be larger. In other words, we have that
\begin{equation}
  \nu_k' \le \nu_k  \qquad \Rightarrow \qquad N_{\vec{\mathcal{D}}}^{(\vec{\nu'})} \subseteq N_{\vec{\mathcal{D}}}^{(\vec{\nu})} \,.
  \label{eq:DescendentContainment}
\end{equation}
This point becomes important when combined with a practical observation: at a
relatively early stage in the construction, the vector spaces become
zero-dimensional. (In the example of fig.~\ref{FigSearchTree}, such spaces are
grayed out.)
From a physical perspective, this is to be expected, as not all poles can be
eliminated from the rational functions.
This leads to an essential practical improvement in our approach: by
\cref{eq:DescendentContainment}, when constructing the graph of spaces row
by row, we need only to construct those which are not descended from zero-dimensional
vector spaces. This results in a ``pruning'' procedure for our graph, which no
longer contains all such spaces. In the example of fig.~\ref{FigSearchTree}, we
see that the final row only has a single element as a result of this pruning
procedure.
In practice, we observe that this structure has a large effect, pruning graphs that
naively have $\mathcal{O}(10^9)$ nodes down to $\mathcal{O}(10^4)$.

Having constructed our graph, it is now possible to find a ``simple'' function
according to the criteria of \cref{eq:ReplacementForm}.
Let us denote the function we wish to replace as $r_1$. To construct a
replacement, we start from the lowest non-empty row of the graph and ask for a
function which is linearly dependent on $r_1$. We take this function as our
replacement, and naturally construct the associated basis-change matrix $U_1$ that
performs the change of basis.
To continue, we apply the same procedure to the set of functions $\{\tilde{r}_1, r_2, \ldots,
r_{|\Bdeg|}\}$, looking for a replacement for $r_2$. We stress that there
is no need to recompute bases of any space $N^{(\vec{\nu})}_{\vec{\mathcal{D}}}$
or $N^{(\nu)}_{\mathcal{D}_k}$ that was computed in the $r_1$ case, as they are
just related by application of the change of basis matrix $U_1$. Naturally, we
continue this procedure until we have constructed the full set of replacement
functions $\tilde{r}_i$. By composition of the collection of partial basis
change matrices, we therefore also compute the full basis change matrix.

While the pruning strategy is sufficient to overcome the exponential
growth of the graph, we comment on three more strategies that we employed in order
to decrease the size of the graph further.
Firstly, similar to the fact that spaces which are descended from zero-dimensional
spaces are also zero-dimensional, descendants of spaces which do not overlap with
the target function will also share this property. Hence we augment the pruning
procedure with this observation and omit such nodes.
Secondly, we choose to impose that
$\tilde{r}_i$ should not have poles that are not in $r_i$. To impose this, we
start constructing the graph from taking $N^{(\vec{q_{i}})}$
as the initial node, where $\vec{q_{i}}$ is the denominator exponent vector for
$r_i$. This naturally walks through a subset of the full graph. While this comes
at the cost of excluding possible replacements that introduce new poles, in
order to lower the total denominator degree, we do not find this cost important
in practice.
Thirdly, we perform a two-stage construction of the graph.
Note that each edge represents lowering the power of some $\mathcal{D}_k$ by one.
In this optimization, we only construct the nodes in the graph corresponding to
completely removing all powers of each $\mathcal{D}_k$.
We then take the output node of this procedure and run a second iteration with
this node as the starting point, therefore trying to lower the powers of each
$\mathcal{D}_k$ in a more fine-grained manner.
From this discussion, it is clear that the basis with which we end up is not
uniquely defined by the input space of functions. In practice, we find that this
non-uniqueness is unimportant to the impact of the basis change procedure and we
leave understanding of this point to future work.

Lastly, we observe that the final node in the graph defining the
replacement $\tilde r_i$ for $r_i$ may not correspond to a
$1$-dimensional space. In other words, there may exist multiple
distinct combinations of the $r_i$'s, that result in functions with
the same poles to the same orders. This introduces yet another
non-uniqueness, as any $\mathbb{Q}$-linear combination of these
functions is equally valid. Moreover, this may lead to large rational
numbers that are cumbersome to lift back from $\mathbb{F}_p$ to
$\mathbb{Q}$ when reconstructing $\tilde r_i$. In practice, we use
this freedom to choose a linear combination ($U_{ij}$) with reasonably
small rational entries. We observe that, at least in some cases, this
degeneracy could be lifted by supplementing the co-dimension-one pole
data with co-dimension-two data, and/or data on the zeros. While this
may also yield a better basis change, it does not guarantee the final
spaces to be $1$-dimensional, thus for the moment we do not pursue
this line of investigation further.

\section{Construction of compact analytic amplitudes}
\label{sec:optimalfrom}

Having discussed our strategy for choosing a basis of the space of rational
functions under consideration, it remains to provide analytic expressions for
the basis elements. While, in principle, this can be done by analytically
combining the expressions given in ref.~\cite{Abreu:2021asb}, this procedure is
computationally intractable. Instead, we consider an analytic reconstruction
procedure for the simplification of our target functions, directly
reconstructing them in a simplified form.
Our aim is to apply the analytic reconstruction approach of
ref.~\cite{Laurentis:2019bjh, DeLaurentis:2022otd}, making
use of studies of the rational functions on co-dimension 1 and 2 surfaces.
However, implementation of the expressions of ref.~\cite{Abreu:2021asb} in a way
that allows for efficient evaluations proved non-trivial. Instead, we
take a two-pass approach. We make an initial study, conjecturing ans\"atze based
upon observations about their analytic structure, allowing us to build
representations of the functions which are compact enough to allow efficient
p-adic evaluations.
Thereafter, we reconstruct compact forms of the rational functions by evaluating
them in singular limits with $p$-adic numbers and iteratively constructing
ansätze that match their behavior in these limits.
We discuss the analytic-structure-inspired procedure in section~\ref{sec:FpRec}
and the singular-structure-based partial-fraction reconstruction procedure in
section~\ref{sec:QpRec}. First, in section~\ref{sec:reducedansatz}, we discuss a
common feature of both procedures: constructing ans\"atze for individual terms
in the partial-fraction decomposition.

\subsection{Spinor ans\"atze for a single massless decay current}
\label{sec:reducedansatz}

An important aspect of all of the analytic reconstruction methods that follow is
the ability to build an ansatz for the numerator of a rational function of
spinor variables.
Specifically, in spinor space the numerator is a polynomial of
spinor brackets and the brackets are related by the Schouten identities and momentum
conservation. Importantly, these relations are non-linear and building a set of functions
which are linearly independent given the relations is therefore a non-trivial
problem.

In ref.~\cite{DeLaurentis:2022otd}, this problem was dealt with systematically
for massless $n$-point spinor space by identifying an underlying quotient ring
and using Gr\"obner basis techniques.
While the $pp\rightarrow V(\bar{l}l) jj$ amplitudes under discussion are naturally defined
within a six-point phase-space, the direct use of the techniques of
ref.~\cite{DeLaurentis:2022otd} dramatically overparametrizes the space of
numerator polynomials. Due to the structure of the underlying Feynman diagrams,
the dependence of the rational functions on the momenta of the leptons takes
the form of a decay current. That is, all rational functions are
expressable in the form
\begin{equation}
  r_i = r_{i, \mu}(\lambda_1, \tilde{\lambda}_1, \ldots, \lambda_4, \tilde{\lambda}_4) \langle 6|\bar\sigma^\mu|5].
  \label{eq:DecayStructure}
\end{equation}
Indeed, making use of this structure was important for the calculation
of the amplitudes of ref.~\cite{Abreu:2021asb}. In this section, we
discuss how to exploit this structure in spinor space, while handling the
non-linear relations between the variables. Our approach is analogous to that of
ref.~\cite{DeLaurentis:2022otd}, and makes use of quotient rings and Gr\"obner
bases.

To begin, let us note that, given eq.~\eqref{eq:DecayStructure}, the numerator
of $r_i$ is not a general polynomial in spinor variables. Specifically, it is a
polynomial in spinor variables which is independent of $\tilde{\lambda}_6$ and
$\lambda_5$. More precisely, we see that the numerator of $r_i$ is
a member of the polynomial ring
\begin{equation}
\overline{S}_6 = \mathbb{F}[\lambda_1^{\alpha}, \tilde{\lambda}_1^{\dot{\alpha}}, \ldots, \lambda_4^{\alpha}, \tilde{\lambda}_4^{\dot{\alpha}}, \tilde{\lambda}_5^{\dot{\alpha}}, \lambda_6^{\alpha}] \, .
\label{eq:ComponentRing}
\end{equation}
The ring $\overline{S}_6$ is a polynomial ring in 6-point spinor variables,
excluding $\lambda_5^{\alpha}$ and $\tilde{\lambda}_6^{\dot{\alpha}}$ and is
defined over some coefficient field $\mathbb{F}$. Here the notation
$\lambda_i^{\alpha}$ and $\tilde\lambda_j^{\dot\alpha}$ is used to emphasize
that spinor components are used as independent variables and the bar notation
emphasizes the absence of $\lambda_5^{\alpha}$ and $\tilde{\lambda}_6^{\dot{\alpha}}$.

Physically, polynomials of spinors which differ by momentum
conservation are considered equivalent, and hence we wish to introduce this
equivalence into our polynomial algebra. However, naively momentum conservation
depends on $\lambda_5^{\alpha}$ and $\tilde{\lambda}_6^{\dot{\alpha}}$, so this
requires some care. It can be shown\footnote{Specifically, recognizing
  $\overline{S}_6$ as a subring of the complete set of spinor polynomials
$S_6 = \mathbb{F}[\lambda_1^{\alpha}, \tilde{\lambda}_1^{\dot{\alpha}}, \ldots,
\lambda_6^{\alpha}, \tilde{\lambda}_6^{\dot{\alpha}}]$, one finds
$\overline{J}_{\Lambda_6}$ by eliminating the excess variables from
$J_{\Lambda_6} = \langle \sum_{i} \lambda_i^{\alpha}
\tilde{\lambda}_i^{\dot{\alpha}} \rangle_{S_6}$. That is, one computes
$\overline{J}_{\Lambda_6} = J_{\Lambda_6} \cap \overline{S}_6$.
} that we can implement momentum conservation by working modulo the ideal of relations
\begin{equation}\label{eq:mom_cons_ideal_restricted}
  \overline{J}_{\Lambda_6} = \Bigg\langle \sum_{i=1}^4 \langle 6i\rangle[i5] \Bigg\rangle_{\overline{S}_6} \, ,
\end{equation}
where we use the subscript notation to indicate that $\overline{J}_{\Lambda_6}$
is an ideal of the polynomial ring $\overline{S}_6$.
This leads us to define the ring of physical spinor polynomials as
\begin{equation}
\overline{R}_6 = \overline{S}_6 \, \big/ \, \overline{J}_{\Lambda_6} \, .
\label{eq:ComponentQRing}
\end{equation}

Next, we impose that we wish to work with Lorentz invariant polynomials. To this
end, we choose to work not with an arbitrary element of $\overline{R}_6$, but
within the Lorentz-invariant sub-ring of spinor products, which we here denote
by $\overline{\mathcal{R}}_6$. Following
the logic of ref.~\cite{DeLaurentis:2022otd}, it can be shown that
$\overline{\mathcal{R}}_6$ is isomorphic to a polynomial quotient ring.
Specifically,
\begin{equation}
   \overline{\mathcal{R}}_6 \simeq \overline{\mathcal{S}}_6/(\overline{\mathcal{J}}_{\Lambda_6} + \mathcal{K}^{\langle \rangle}_{6} + \mathcal{K}^{[]}_{6}),
\end{equation}
where $\overline{\mathcal{S}}_6$ is the polynomial ring in spinor brackets that
do not involve $\lambda_{5}$ and $\tilde{\lambda}_6$,
\begin{equation}
  \overline{\mathcal{S}}_6 = \mathbb{F} \big[ \langle ij \rangle \,\, : \,\, 1 \le i < j \le 6, \,\, i \ne j \ne 5, \quad  [ij] \,\, : \,\, 1 \le i < j \le 5 \big]\, ,
  \label{eq:InvariantRing}
\end{equation}
$\overline{\mathcal{J}}_{\Lambda_6}$ is the ideal of eq.~\eqref{eq:mom_cons_ideal_restricted} now taken in the ring $\overline{\mathcal{S}}_6$
\begin{equation}
  \overline{\mathcal{J}}_{\Lambda_6} = \Bigg\langle \sum_{i=1}^4 \langle 6i\rangle[i5] \Bigg\rangle_{\overline{\mathcal{S}}_6} \, ,
\end{equation}
and $\mathcal{K}^{\langle \rangle}_6, \mathcal{K}^{[]}_6$ are the ideals of Schouten
identities between the appearing angle and square brackets respectively,
\begin{align}
  \mathcal{K}^{\langle \rangle}_6 &= \bigg\langle
    \langle ij \rangle \langle kl \rangle - \langle ik \rangle \langle jl \rangle + \langle il \rangle \langle jk \rangle \quad : \quad 1 \le i < j < k < l \le 6, i \ne j \ne k \ne l \ne 5
    \bigg\rangle_{\overline{\mathcal{S}}_6},
  \\
  \mathcal{K}^{[]}_6 &= \bigg\langle
    [ij] [kl] \,\,-\,\, [ik] [jl] \,\,+\,\, [il] [jk] \,\quad : \quad 1 \le i < j < k < l \le 5
    \bigg\rangle_{\overline{\mathcal{S}}_6}.
\end{align}
The ring of functions $\overline{\mathcal{R}}_6$ is where we shall consider the
numerators of rational functions with the decay structure of
eq.~\eqref{eq:DecayStructure} to live. We note that $\overline{\mathcal{R}}_6$
is strictly larger than the set of functions with the decay structure. For
example, $\overline{\mathcal{R}}_6$ can parameterize functions with multiple
powers of $\lambda_6$ in the numerator, not just one. However, we find that this
causes no issues in practice.

Having identified an appropriate ring of functions, it remains to construct a set of linearly
independent spinor brackets. We use the Gr\"obner basis approach of
ref.~\cite{DeLaurentis:2022otd}, which we briefly review. We refer to the
aforementioned reference for more details.
We consider generating a linearly independent set of spinor bracket monomials for a
function $\mathcal{N}$, with mass dimension $d$, and $k^{\text{th}}$ little group
weight $\phi_k$. We define a general monomial in spinor brackets as
\begin{equation}
  m_{\alpha, \beta} = \left( \prod_{j=1, j \ne 5}^6 \prod_{i=1, i \ne 5}^{j-1} \langle ij \rangle^{\alpha_{ij}} \right)\left( \prod_{l=1}^5 \prod_{k=1}^{l-1} [kl]^{\beta_{kl}} \right),
\end{equation}
where the involved collection of $\alpha_{ij}$ and $\beta_{kl}$ are non-negative
integers. Given the mass dimension and little group constraints, the exponents
of $m_{\alpha, \beta}$ are restricted to satisfy both
\begin{equation}
  \sum_{j=1, j \ne 5}^6 \sum_{i=1, i \ne 5}^{j-1} \alpha_{ij} + \sum_{l=1}^5 \sum_{k=1}^
  {l-1} \beta_{kl} = d,
\end{equation}
and
\begin{equation}
\sum_{j=1, j \ne 5}^6 \sum_{i=1, i \ne 5}^{j-1} \alpha_{ij} (\delta_{i w} + \delta_{jw}) - \sum_{l=1}^5 \sum_{k=1}^{l-1} \beta_{kl} (\delta_{k w} + \delta_{lw}) = \phi_w.
\end{equation}

In order to generate a linearly independent set of spinor brackets, we further
require that all monomials are irreducible modulo the lead terms (LTs) of the Gröbner
basis of $\overline{\mathcal{J}}_{\Lambda_6} + \mathcal{K}^{\langle \rangle}_{6} + \mathcal{K}^{[]}_{6}$. That is, we
require that
\begin{equation}
  \label{eq:IrreducibilityConstraint}
\text{LT}(g) \,\, \nmid \,\,  m_{\alpha, \beta} \quad \text{for all} \quad g \in \mathcal{G}(\overline{\mathcal{J}}_{\Lambda_6} + \mathcal{K}^{\langle \rangle}_{6} + \mathcal{K}^{[]}_{6}).
  \end{equation}
These conditions give a finite collection of solutions which can easily be
enumerated. In practice, we make use of the \texttt{CP-SAT} solver from
\texttt{OR-Tools} \cite{ortools} to enumerate the solutions and
Singular~\cite{DGPS} to compute the Gröbner bases.
We stress that this procedure is highly efficient and easily runs on a modern
laptop computer. 
In table~\ref{tab:ansatze-sizes}, we demonstrate the importance of our
procedure, by comparing the size of the ansätze that it generates with the
ansätze generated with the procedure of ref.~\cite{DeLaurentis:2022otd} for $n
= 5$ and $n=6$.
Finally, we comment that analogous procedures have been applied in cases
involving a single massive scalar boson~\cite{Budge:2020oyl} or multiple massive
scalars~\cite{Campbell:2024tqg}.

\begin{figure}[]
  \centering
  	\includegraphics[scale=0.7]{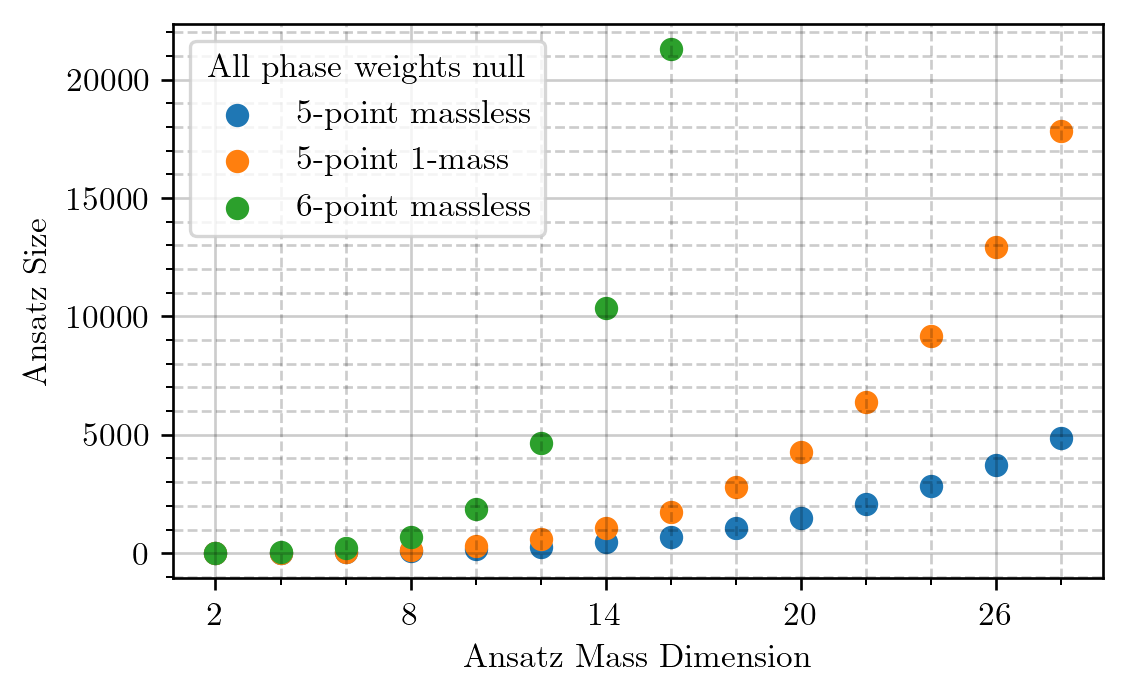}
  	\caption{Comparison of the size of ans\"atze for different
          kinematic configurations, as a function of mass dimension
          (polynomial degree) when all phase weights are zero.}
\label{fig_np}
\end{figure}

\subsection{Partial-fraction ansatz from analytic structure}
\label{sec:FpRec}

After performing the basis change on each vector space of rational functions, we
need to reconstruct the newly identified basis elements.
In practice, while complexity is significantly reduced after the basis change, the
rational coefficients in LCD form are still remarkably complex, making a direct
reconstruction procedure highly demanding.
In order to remedy this, we wish to make use of the iterative partial-fraction
approach that we present in the next subsection. However, this requires the
ability to efficiently evaluate the rational functions on phase-space points
where denominators are small.
In order to obtain such
efficient numerical evaluations, we therefore undertake an analytic
reconstruction procedure, which leads to analytic results that can be easily
evaluated in limits.
To this end, we write an ansatz for the rational functions in a 
partial-fraction form as
\begin{equation}
  \tilde r_i  = \sum_{k=1}^{k_{\text{max}}} \frac{\tilde{\mathcal{N}}_{ik}}{\prod_{j} \mathcal{D}_j^{q_{ijk}}} \quad \text{with} \quad q_{ijk} \leq q_{ij} \, \forall \, k \, ,
  \label{eq:SchematicPartialFractions}
\end{equation}
where the $\tilde{\mathcal{N}}_{ik}$ are unknown polynomials in spinor brackets.
The non-trivial task is to determine a set of exponents $q_{ijk}$ such that the
number of free parameters in the $\tilde{\mathcal{N}}_{ik}$ is small, while the
decomposition in eq.~\eqref{eq:SchematicPartialFractions} is still valid.
Our approach to this is heuristic: we construct an ansatz based upon intuition
about the analytic structure of the rational functions.
We note that a similar approach was already successfully used in
ref.~\cite{Abreu:2023bdp} to reconstruct the full-color two-loop amplitudes for
$pp \rightarrow \gamma \gamma \gamma$.
Once we have an ansatz of the form eq.~\eqref{eq:SchematicPartialFractions}, we
then sample the function randomly over the phase-space in order to constrain the
numerator polynomials.
Importantly, while the numerical evaluations are obtained from the known
analytic results of~\cite{Abreu:2021asb}, they could equally as well have been obtained from
a numerical code for the reduction of an amplitude such as \texttt{Caravel} \cite{Abreu:2020xvt}.
Beyond the goal of providing evaluations in singular limits for
section~\ref{sec:QpRec}, our approach therefore also provides a valuable
exploration for the construction of ans\"atze for new amplitude calculations.

\paragraph{From analytic structure of the rational functions to an ansatz}

\label{sec:Properties}
In order to construct partial-fraction ans\"atze, we make use of a collection of
conjectural properties that are motivated by the structure of the known one-loop
amplitudes and already reconstructed simple two-loop functions.
That is, we identify several recurring features of the rational coefficients of
the finite remainders, and by extension of the finite remainders themselves,
that can be manifested by suitable partial-fraction decompositions.
These features can be easily verified once the analytic expressions are
available, and, if predicted, can be used to significantly simplify the
reconstruction procedure. We expect that most of these features hold true for
any process with five-point one-mass kinematics, meaning these observations will
be beneficial for future computations.

In practice, our observations about the analytic structure of the rational functions in
an amplitude come in two classes. Firstly, for a rational function with
denominators $\mathcal{D}_i^{q_i}$ and $\mathcal{D}_j^{q_j}$ we say that these
denominators can be ``cleanly separated'' if it is possible to write the
rational function in a partial-fraction form where $\mathcal{D}_i$ and
$\mathcal{D}_j$ do not appear in the same fraction. That is,
\begin{equation}
  \frac{\mathcal{N}_{ij}}{\mathcal{D}_i^{q_i} \mathcal{D}_j^{q_j} \mathcal{D}_{\text{rest}}} = \frac{\mathcal{N}_i}{\mathcal{D}_i^{q_i} \mathcal{D}_{\text{rest}}} + \frac{\mathcal{N}_j}{\mathcal{D}_j^{q_j} \mathcal{D}_{\text{rest}}}.
  \label{eq:CleanSeparation}
\end{equation}
Secondly, we say that they can be ``separated to order $\kappa$'' if it is possible
to rewrite the fraction as a sum of terms where the total power of
$\mathcal{D}_i$ and $\mathcal{D}_j$  in each term is $\kappa$. That is, there exists
a $\kappa$ such that $\kappa \ge q_i$, $\kappa \ge q_j$ and that we can rewrite the rational function as
\begin{equation}
  \frac{\mathcal{N}_{ij}}{\mathcal{D}_i^{q_i} \mathcal{D}_j^{q_j} \mathcal{D}_{\text{rest}}} = \sum_{\kappa - q_j \le m \le q_i} \frac{\mathcal{N}^{(m)}_{ij}}{\mathcal{D}_i^{m} \mathcal{D}_j^{\kappa-m} \mathcal{D}_{\text{rest}}}.
  \label{eq:SeparationToOrder}
\end{equation}
Naturally, this separation is only interesting if $\kappa < q_i + q_j$.  
In terms of these concepts, we observe the following
separability properties of the rational functions (presented in no particular
order):
\begin{enumerate}[label=c\arabic*]
\item Distinct denominator pairs $\{s_{ijk}, s_{lmn}\}$ can be cleanly separated,
      mirroring the analogous statement on the structure of Feynman diagrams at
      tree level. \label{mandelstamconstraint}
\item Distinct denominator pairs $\{\Delta_{ab|cd|ef}, \Delta_{ij|kl|mn}\}$ can be
      cleanly separated. \label{cons:gramseparation}
\item Pairs of denominator that are of the form
      $(\langle2|3|5+6|1|2]-\langle3|4|5+6|1|3])$ or any permutation can be cleanly
      separated.
\item 
  Denominator pairs $\{ \Delta_{ij|kl|mn}, \langle ab \rangle\}$, $\{
      \Delta_{ij|kl|mn}, [ab]\}$, $\{ \Delta_{ij|kl|mn}, \langle a|b+c|d]\}$, and \\
      $\{\Delta_{ij|kl|mn}, [a|b+c|d \rangle\}$ can all be cleanly separated unless $\{b, c\}$ is
      $\{i,j\}, \{k,l\}$, or $\{m, n\}$.
      \label{cons:other-poles-with-delta}
    \item Denominator pairs $\{\langle  i| j + k | l], [i| j + k | l \rangle \}$
          can be cleanly separated.
\item Denominator pairs $\{ \langle  ij \rangle, [ij]\}$ can be
      cleanly separated for $i,j$ cyclic in $\{1,2,3,4,5,6\}$.%
\footnote{We note that this constraint, when applied with $i=5$ and $j=6$,
          implies that numerators are of degree 2 in $|6\rangle$ or $[5|$,
          depending on whether the pole is $\langle 56\rangle$ or $[56]$,
          respectively.
          Non-trivially, this is still consistent with the constraint that the
          decay current takes the form $\langle 6| \bar\sigma^\mu|5]$.
      }
\label{cons:two-particle-brackets}

\item Denominator pairs of the form $\{\Delta_{ij|kl|mn}, \langle a|b+c|d]\}$
and $\{\Delta_{ij|kl|mn}, [a|b+c|d \rangle\}$ where $\{b, c\}$ is one of
$\{i, j\}, \{k, l\}$ or $\{m, n\}$ can be separated up to order 5.
  \label{cons:other-pole-with-delta-power-cons}

\end{enumerate}
We verified that each of the seven properties holds for the rational functions
under consideration when taken individually. However, it is important to note
that trying to combine multiple such properties (e.g., two clean separations) is
often a non-trivial extension of eq.~\eqref{eq:CleanSeparation}. In the language
of ref.~\cite{DeLaurentis:2022otd, DeLaurentis:2022knk}, this can be understood
as a statement that the set of denominators do not form a regular sequence and
hence the ideal intersection is non-trivial.
At this time, each of these properties are mostly empirical observation.
Nevertheless, it would be interesting to understand if they can be predicted a
priori.
For instance, one might expect that these features arise from the
structure of the hierarchy of unitarity cuts in the reduction.
As an example, constraint \ref{cons:gramseparation} is familiar from one loop,
where these Gram determinants are associated with three-mass triangle diagrams,
which implies that, at one loop, distinct such Grams cannot be in the same denominator.

Given knowledge of the properties
(\ref{mandelstamconstraint}-\ref{cons:other-pole-with-delta-power-cons}), our goal is to
write an ansatz for the rational functions in our amplitude. To this end, we make use of the properties by using
them to conjecture a partial-fraction decomposition of each rational function
by interpreting eq.~\eqref{eq:CleanSeparation} and
eq.~\eqref{eq:SeparationToOrder} as rules to apply to a partial-fraction
decomposition.
That is, to apply the constraint that $\mathcal{D}_1$ and
$\mathcal{D}_2$ are cleanly separable to a partial-fraction representation of
the rational function, we look for all such terms that contain both
denominators and develop a new partial-fraction decomposition by separating
them as
\begin{equation}
  \frac{\mathcal{N}_{12}}{\mathcal{D}_1^{q_1} \mathcal{D}_2^{q_2} \mathcal{D}_{\text{rest}}}
  \rightarrow
  \frac{\mathcal{N}_{1}}{\mathcal{D}_1^{q_1} \mathcal{D}_{\text{rest}}}
  +
  \frac{\mathcal{N}_{12}}{\mathcal{D}_2^{q_2} \mathcal{D}_{\text{rest}}} \, ,
  \label{eq:CleanSeparationRule}
\end{equation}
where the term on the left-hand-side is some term in the partial-fraction
decomposition. Naturally, eq.~\eqref{eq:CleanSeparation} is the reinterpretation
of eq.~\eqref{eq:CleanSeparation} as a rule. For constraints that a pair of
denominators is separable up to order $\kappa$, we make the analogous
replacement, looking to eq.~\eqref{eq:SeparationToOrder}.

Our procedure is to iteratively apply a sequence of the constraints
(\ref{mandelstamconstraint}-\ref{cons:other-pole-with-delta-power-cons}), starting from the
least common denominator representation, in order to generate a partial
fractions representation as eq.~\eqref{eq:SchematicPartialFractions}.
In practice, the particular sequence of constraint applications is determined
ad-hoc. As each strictly only holds for the full rational coefficient, there are sequences
of constraints which do not provide valid partial-fraction decompositions for
the amplitude. All such decompositions are therefore a priori conjectural and we
validate them by checking to see if the ansatz fits.
An important observation in this procedure is that the polynomials
$\tilde{\mathcal{N}}_{ik}$ in a partial-fraction decomposition of the form of
eq.~\eqref{eq:SchematicPartialFractions} develop partial linear dependencies.
These can be understood as syzygies between the different terms in the partial
fraction decomposition. As a simple example of such a redundancy, if we consider
\cref{eq:CleanSeparation}, it is clear that the two terms in the partial
fractions decomposition can both parameterize a rational function with
denominator $\mathcal{D}_{\text{rest}}$.
In practice, we remove such linear dependencies by explicit linear algebra.
We find that the most efficient way to remove the redundancies is via a
bi-partite strategy. That is, we first remove redundancies between pairs, then
between quartets, octets and so on.
This avoids overshooting the final ansatz size by large margins and to avoid
running the Gaussian elimination several times on systems of size close to the
final ansatz size, which happens with a sequential algorithm.

In practice, we find that our procedure, making use of only the listed
properties, is generally sufficient to reduce the ans\"atze to have
approximately $50\,000$ unknowns. Indeed, once we were able to reach
this number we stopped a systematic search for further properties.
This number is somewhat arbitrary, and motivated by the practical limitation of
the amount of memory available on the GPUs used for solving the linear systems.
Only for the most complicated functions, we supplement these with some amount of
guess-and-check work, always in the form of a statement that two poles are
separable to some non-trivial order $\kappa$. In practice, our initial partial
fractions ans\"atze consist of a modest number of terms, with $k_{\text{max}}$ at
around 10 to 20 for the most complicated cases.

\paragraph{Example partial-fraction ansatz from analytic structure}
For concreteness, let us now consider the application of the partial-fraction
from analytic-structure approach to an explicit example coefficient,
specifically one that
arose in the reconstruction of the NMHV gluon amplitude.
In least common denominator form, the expression reads
\begin{equation}
  f^{\text{ex}} = \frac{\mathcal{N}^{\text{ex}}}{⟨14⟩^2[14]^2 s_{56} ⟨1|2+4|3]^2⟨2|1+4|3]^4⟨2|1+3|4]^2Δ_{14|23|56}^4} \, .
  % \phi  = $\{1, 4, -4, -1, -1, 1\}$
\end{equation}
Here, the numerator $\mathcal{N}$ has mass dimension 36, and can be shown to
have $104\,128$ terms, using the procedure of section~\ref{sec:reducedansatz}.
Importantly, $f^{\text{ex}}$ is invariant under the combined action of parity
(angle-bracket exchange) and permuting the external legs as $123456\rightarrow
432165$. We will also see how this symmetry decreases the number of
free parameters in our ansatz.

To build a partial-fraction decomposition, we first apply
\ref{cons:other-poles-with-delta} from the above list. This results in the
tentative representation 
\begin{equation}
  f^{\text{ex}} = \frac{\mathcal{N}^{\text{ex}}_1}{⟨14⟩^2[14]^2s_{56}⟨2|1\!+\!4|3]^4Δ_{14|23|56}^4 \,} + \frac{\mathcal{N}^{\text{ex}}_2}{⟨14⟩^2[14]^2s_{56}⟨2|1+4|3]^4⟨1|2\!+\!4|3]^2⟨2|1\!+\!3|4]^2}\,.
  \label{eq:exampleFirstSplitting}
\end{equation}
The ans\"atze for $\mathcal{N}_1$ and $\mathcal{N}_2$ have $39\,221$ and $4\,368$
terms respectively. There are linear relations between parts of the ans\"atze in
the two terms of eq.~\eqref{eq:exampleFirstSplitting}, corresponding to the fact
that they both contain an ansatz for a rational function with denominator
$⟨14⟩^2[14]^2⟨56⟩[56]⟨2|1+4|3]^4$. We remove this $661$ term overlap, resulting
in an ansatz with $42\,928$ free parameters.

Next, we apply \ref{cons:two-particle-brackets}, with $i=1$, $j=4$.
That is, we split the $⟨14⟩$ from the $[14]$ pole. Note that this splitting is
not invariant under the symmetry of $f^{\text{ex}}$ and hence the symmetry now
exchanges terms. We therefore write 
\begin{equation}
\begin{aligned}
  f^{\text{ex}} =
  \frac{\mathcal{N}^{\text{ex}}_{3}}{⟨14⟩^2 s_{56} ⟨2|1+4|3]^4Δ_{14|23|56}^4}
  + \frac{\mathcal{N}^{\text{ex}}_{4}}{⟨14⟩^2 s_{56} ⟨1|2+4|3]^2⟨2|1+4|3]^4⟨2|1+3|4]^2} \, \,&
  \\
  + \,\, (123456\rightarrow \overline{432165})&\,,
  \label{eq:exampleSecondSplitting}
\end{aligned}
\end{equation}
where the terms hidden by the symmetry are fixed by those that we display. The
ans\"atze for $\mathcal{N}^{\text{ex}}_{3}$ and $\mathcal{N}^{\text{ex}}_{4}$ have $24\,882$ and $1\,921$
terms respectively, and overlap by $166$ terms. Hence, the ansatz of
eq.~\eqref{eq:exampleSecondSplitting} has $26\,737$ terms.

Next, we apply \ref{cons:other-pole-with-delta-power-cons}, and taking the
total power of $⟨2|1+4|3]$ and $Δ_{14|23|56}$ to be 5. This gives the tentative
decomposition
\begin{equation}
  \begin{aligned}
  f^{\text{ex}} = 
  \sum_{k=0}^3 \frac{\mathcal{N}^{\text{ex}}_{5,k}}{⟨14⟩^2 s_{56} ⟨2|1+4|3]^{1+k} Δ_{14|23|56}^{4-k}}
    + \frac{\mathcal{N}^{\text{ex}}_6}{⟨14⟩^2 s_{56}⟨1|2+4|3]^2⟨2|1+4|3]^4⟨2|1+3|4]^2}\, \,&
    \\
  + \,\, (123456\rightarrow \overline{432165}) &\, .
  \label{eq:exampleFinalForm}
  \end{aligned}
\end{equation}
The ans\"atze for the $\mathcal{N}^{\text{ex}}_{5,k}|_{k=0,\ldots,3}$ and
$\mathcal{N}^{\text{ex}}_6$ have size $7\,857$, $5\,064$, $2\,688$, $1\,071$,
$1\,921$, respectively. The overlap is more complex now, but can be easily
removed by linear algebra.
In summary, the tentative ansatz in eq.~\eqref{eq:exampleFinalForm} now has size
$13\,532$, almost a factor of 10 smaller than the original one. We then check
the validity of eq.~\eqref{eq:exampleFinalForm} by sampling $f^{\text{ex}}$
the corresponding number of times and fitting the ansatz.

\subsection{Reconstruction by iterated subtraction of singular behavior}
\label{sec:QpRec}

While the finite-field reconstruction approach of the previous section
successfully produces expressions that are much more compact than those of
ref.~\cite{Abreu:2021asb}, in order to further improve the numerical performance
of our results we undertake a further simplification procedure following an
approach based on understanding the singular behavior of the rational
coefficients with $p$-adic numbers and algebraic
geometry~\cite{Laurentis:2019bjh, DeLaurentis:2022otd}.
This allows us to deliver the highly compact results found in our ancillary
files, that we will describe in section~\ref{sec:ancillaries}.
Let us consider a rational function $f$, whose common denominator form is given
by
\begin{equation}
  f = \frac{\mathcal{N}}{\prod_{j}\mathcal{D}_j^{\alpha_j}}.
\end{equation}
Given an $f$, the approach that we take is to find another function which matches the
leading behavior of $f$ in the limit where some denominator factor becomes small.
Specifically, consider picking some $\mathcal{D}_0$.
Our aim is to find a numerator $\mathcal{N}^{(0)}$, and denominator exponents $\alpha_{0,j}$ such that
\begin{equation}
        f = \frac{\mathcal{N}^{(0)}}{\mathcal{D}_0^{\alpha_0} \prod_{j \ne 0} {\mathcal{D}_j}^{\alpha_{0,j}}} + \mathcal{O}(\mathcal{D}_0^{\alpha_0-1}).
    \label{eq:CodimensionOneStructure}
\end{equation}
Here $\mathcal{N}^{(0)}$ is a polynomial in spinor variables, modulo terms which
are proportional to $\mathcal{D}_0$, a point we shall return to later.
Once we have found $\mathcal{N}^{(0)}$ and $\alpha_{0,j}$ we consider the subtracted function
\begin{equation}\label{eqn:subtractionTerm}
    f' = f - \frac{\mathcal{N}^{(0)}}{\mathcal{D}_0^{\alpha_0} \prod_{j \ne 0} {\mathcal{D}_j}^{\alpha_{0,j}}}.
\end{equation}
Given that $f$ and the subtraction term match in the limit of small
$\mathcal{D}_0$ then $f'$ must have a pole in $\mathcal{D}_0$ of order at most
$\alpha_0 - 1$. We therefore see that $f'$ is
an easier function to reconstruct. We then iterate this procedure until we have
made a sufficient number of subtractions such that the result is a very simple
function that can easily be reconstructed with a low number of finite-field
samples.
Naturally, there is an order dependence to the procedure, but we leave
understanding of this to future work.
Given our iterative approach, the procedure is reduced to the determination of
the subtraction terms, which will be the remainder of our discussion.

In order to determine both the denominator exponents $\alpha_{0,j}$ and the
numerator $\mathcal{N}^{(0)}$, we make use of evaluations of $f$ on phase-space
points where $\mathcal{D}_0$ and $\mathcal{D}_j$ are small. That is, for some
small quantity $\epsilon$, we consider a phase-space point
$(\lambda^{(\epsilon)}, \tilde{\lambda}^{(\epsilon)})$, such that
\begin{align}
  \begin{split}
  \mathcal{D}_0(\lambda^{(\epsilon)}, \tilde{\lambda}^{(\epsilon)}) &= \mathcal{O}(\epsilon^{n_0}), \\
  \mathcal{D}_j(\lambda^{(\epsilon)}, \tilde{\lambda}^{(\epsilon)}) &= \mathcal{O}(\epsilon^{n_j}),
  \end{split}
  \label{eq:D0DjSmallPoint}
\end{align}
which in turn yields
\begin{align}
  \begin{split}
  f(\lambda^{(\epsilon)}, \tilde{\lambda}^{(\epsilon)}) &= \mathcal{O}(\epsilon^{n_f}) .
  \end{split}
  \label{eq:fSmallPoint}
\end{align}
In its original formulation~\cite{Laurentis:2019bjh}, this approach
used high-precision complex numbers, where $\epsilon$ was often $\sim
10^{-30}$.  In this context, eq.~\eqref{eq:D0DjSmallPoint} is only
approximate, and a value for $n_f$ can only be reliably obtained in
the limit of very small $\epsilon$.
In order to gain better control of this issue, in
ref.~\cite{DeLaurentis:2022otd} it was proposed to make use of the $p$-adic
numbers. In this formulation, the size of quantities is interpreted with the
$p$-adic absolute value and the prime $p$ is a naturally small quantity. For any
$p$-adic number $x$, we can write
\begin{equation}
  x = p^{n} \left( x_0 + \mathcal{O}(p) \right),
  \label{eq:PadicSchematic}
\end{equation}
where $x_0$ is a non-zero integer value that is coprime to $p$ and $n = \pval{x}$
is an integer known as the $p$-adic ``valuation'' of $x$.
The notation $\mathcal{O}(p)$ in \cref{eq:PadicSchematic} indicates that we omit
terms which are proportional to $p$, and are therefore ``small'' in the $p$-adic
absolute value. In this sense, $p$ can be used to play the role of $\epsilon$ in
\cref{eq:D0DjSmallPoint,eq:fSmallPoint}.
The valuation $\pval{x}$ makes this notion of size more precise:
colloquially, $\pval{x} = n$ lets us say that $x$ ``vanishes to order $n$''.
The $p$-adic numbers can roughly be regarded as power series in $p$, and
$\pval{x}$ then denotes the leading-order exponent.
The subset of $p$-adic numbers which satisfy $\pval{x} \ge 0$ (those with no
pole in $p$) are known as the ``$p$-adic integers''. In practice, all
phase-space points that we handle are $p$-adic integers, allowing us to conclude
that evaluations of polynomials on our phase-space points will always have
non-negative valuation.
Finally, we note that we work with a large prime number $p$, such that
\cref{eq:fSmallPoint} can be interpreted as a statement about the structure of
$f$, and not the valuation of rational numbers in its expression.

An important observation made in ref.~\cite{DeLaurentis:2022otd} is that
evaluations on phase-space points where a pair of denominators are small, as considered in
\cref{eq:D0DjSmallPoint}, are close to the variety $\mathcal{D}_i =
\mathcal{D}_j = 0$. 
Moreover, evaluation of $f$ on such a phase-space point
gives information about how the function $f$ behaves on this algebraic variety.
Intuitively, if one considers $\epsilon$ to be a
free-parameter, then for $\epsilon \rightarrow 0$, the points sit on the
variety. Hence, for small $\epsilon$, they can be considered close to the
variety and the power of $\epsilon$ describes the behavior of $f$ on that
variety.
An important subtlety in understanding this behavior is that the algebraic
variety may be reducible. That is,
\begin{equation}
  V(\mathcal{D}_i, \mathcal{D}_j) = \bigcup_k U_{ijk},
  \label{eq:Codim2IrreducibleDecomposition}
\end{equation}
where the $U_{ijk}$ are irreducible varieties and $V(\mathcal{D}_i,
\mathcal{D}_j)$ is the variety $\mathcal{D}_i = \mathcal{D}_j = 0$. We will
often write $U$ to denote an irreducible algebraic variety.
To systematically study the behavior of the function in such doubly-singular
limits therefore requires that we understand the decomposition onto irreducible
varieties in eq.~\eqref{eq:Codim2IrreducibleDecomposition}. For the application
of our approach to the $pp\rightarrow Vjj$ amplitudes, in
appendix~\ref{sec:PrimDec} we, therefore, report on the primary decomposition of the ideals
$\langle \mathcal{D}_i, \mathcal{D}_j \rangle$ associated to all denominators
found at one-loop.

Given this $p$-adic technology, our procedure has two major steps. First, we
determine the exponents $\alpha_{0,j}$, and secondly we construct a number
of constraints on $\mathcal{N}^{(0)}$ that lead to a compact representation.
In practice, we find this approach to be very powerful as the subtraction terms
in \eqref{eq:CodimensionOneStructure} turn out to be very simple. Specifically,
the denominator exponents $\alpha_{0,j}$ are much lower than the $\alpha_j$
in the common denominator form, and the numerator $\mathcal{N}^{(0)}$ often has
a highly compact form. Let us now consider the two steps in turn.

\paragraph{Constraining the denominator}
Consider evaluating the function $f$ on a $p$-adic phase-space point
$x_{\epsilon, U}$ which is close to an irreducible variety $U$, such that $\mathcal{D}_0$
and some $\mathcal{D}_j$ are small, i.e. $U$ is some $U_{0 j k}$ of \cref{eq:Codim2IrreducibleDecomposition}.
(We note that, for simplicity of notation, from now on we use $x$ instead of a pair of spinors
$\lambda, \tilde{\lambda}$.)
Assuming that on $x_{\epsilon, U}$ the leading pole in $\mathcal{D}_0$ dominates\footnote{While this
  assumption may not always hold, it can be shown that our final result is
  always more conservative than the true behavior.},
we find from \cref{eq:CodimensionOneStructure} that 
\begin{equation}
  \pval{f(x_{\epsilon, U})} = \pval{\mathcal{N}^{(0)}(x_{\epsilon, U})}~-~\alpha_0\,\pval{\mathcal{D}_0(x_{\epsilon, U})} ~-~ \sum_{j \ne 0} \alpha_{0,j}\,\pval{\mathcal{D}_j(x_{\epsilon, U})} \,,
\end{equation}
where we used the fact that the valuation is additive on products. Note that
usually very few terms in the sum over $j$ contribute, as the summand is only
non-zero if the associated denominator is small on the phase-space point
$x_{\epsilon, U}$.

Given that we have no explicit knowledge of the behavior of the numerator
$x_{\epsilon, U}$, we are not able to make direct use of this equality.
However, since $\pval{\mathcal{N}^{(0)}(x_{\epsilon, U})}$ must be
non-negative, we conclude that we have the bound
\begin{equation}
   \sum_{j \ne 0} \alpha_{0,j} \, \pval{\mathcal{D}_j(x_{\epsilon, U})} \ge -\pval{f(x_{\epsilon, U})} - \alpha_0 \, \pval{\mathcal{D}_0(x_{\epsilon, U})},
  \label{eq:ExponentCombinationLowerBound}
\end{equation}
We interpret this constraint as being independent of the particular choice of
$x_{\epsilon, U}$, depending only on $U$ and the way in which $x_{\epsilon, U}$
approaches $U$. 
Our strategy for constructing $\alpha_{0,j}$ which satisfies the full set of
constraints is as follows. First, we notice that integers
\begin{equation}
  \overline{\alpha}_{0,j}(x_{\epsilon, U}) = 
  \begin{cases}
  \left\lceil \frac{1}{\pval{\mathcal{D}_j(x_{\epsilon, U})}}\left( -\pval{f(x_{\epsilon, U})} - \alpha_0 \, \pval{\mathcal{D}_0(x_{\epsilon, U})}\right) \right\rceil  \quad &\text{if}\quad \pval{\mathcal{D}_j(x_{\epsilon, U})} \ne 0\\
  0  & \text{if}\quad \pval{\mathcal{D}_j(x_{\epsilon, U})} = 0
  \end{cases}
\end{equation}
are a solution to \cref{eq:ExponentCombinationLowerBound}.
However, we further need to impose that $\alpha_{0,j}$ is non-negative and not
greater than $\alpha_j$, such that the subtraction term does not introduce higher
degree poles. To this end, we take 
\begin{equation}
  \alpha_{0,j} = \max_{x_{\epsilon, U}}\left( \min\left(  \max\left(  \overline{\alpha}_{0,j}(x_{\epsilon, U}) , 0 \right), \alpha_j\right)\right),
  \label{eq:ExponentChoice}
\end{equation}
where the maximum over $x_{\epsilon, U}$ is taken over all phase-space
points that approach an irreducible variety such that $\mathcal{D}_0$ and some
$\mathcal{D}_j$ vanish.
In practice, we only pick one point per such variety $U$.
In most cases, it is sufficient to only consider $x_{\epsilon, U}$ that approach
$U$ in a symmetric fashion, but it is not hard to find examples where one needs
to use asymmetric approaches to the variety to correctly determine denominator
exponents (as we will see later in an example).
Note that eq.~\eqref{eq:ExponentChoice} satisfies
eq.~\eqref{eq:ExponentCombinationLowerBound} in a manifest way, except in the
case where the behavior of the function forces $\alpha_{0,j} = \alpha_j$. While
no longer manifest, the choice in eq.~\eqref{eq:ExponentChoice} still satisfies
eq.~\eqref{eq:ExponentCombinationLowerBound} as
\begin{equation}
  \pval{f(x)} \ge -\alpha_0 \, \pval{\mathcal{D}_0(x)} - \sum_{j \ne 0} \alpha_{0,j}\,\pval{ \mathcal{D}_j(x) },
\end{equation}
for any phase-space point $x$.
We note that it is easy to construct examples where the choice in
eq.~\eqref{eq:ExponentChoice} is not optimal, though we find that these all
involve multiple denominator factors vanishing on $U$. Moreover, we expect that,
for symmetric approaches to varieties where only a single denominator vanishes,
the choice in eq.~\eqref{eq:ExponentChoice} is optimal. It would be interesting
to rigorously understand the optimality and improvements to the choice in
eq.~\eqref{eq:ExponentChoice}, but we leave it to future work.
Nevertheless, an important feature of the choice in
eq.~\eqref{eq:ExponentChoice} is that it is often zero, indicating that the
denominator $\mathcal{D}_j$ is not necessary in the subtraction term. This
greatly decreases the mass dimension of the numerator therefore leaving far
fewer unknowns that must be determined.

\paragraph{Constraining the numerator}
Having determined the denominator of the subtraction term, we now turn to
determining the numerator. Our approach will be to write an ansatz for the
numerator $\mathcal{N}^{(0)}$ and fix the unknowns from $p$-adic evaluations
near the surface $\mathcal{D}_0 = 0$. A simple ansatz for $\mathcal{N}^{(0)}$
would be to take it as a generic polynomial in spinor brackets.
However, an alternative approach, that often leads to far fewer unknowns in the
ansatz, can be made by making use of the ideas from ref.~\cite{DeLaurentis:2022otd}.
Specifically, we make use of the fact that
the behavior of a function near a variety can be used to demonstrate that the
numerator belongs to a polynomial ideal.

The powerful observation that is exploited in refs.~\cite{Laurentis:2019bjh,
DeLaurentis:2022otd} is that, for a phase space point $x_{\epsilon, U}$ near a
variety $U$, where a set of denominator factors vanish, $f$ can diverge less
strongly than one would expect from the common denominator form. That is, 
\begin{equation}
  \pval{f(x_{\epsilon, U})} \ge - \sum_{j} \alpha_j \, \pval{\mathcal{D}_j(x_{\epsilon, U})}.
  \label{eq:DivergenceConstraint}
\end{equation}
Naturally, if the point $x_{\epsilon, U}$ is chosen arbitrarily, then
this behavior can be understood as a feature of $f$ when considered near the
whole variety $U$.
In this work, where we consider the construction of subtraction terms, we wish
to consider imposing this behavior on our subtraction terms.
While not strictly a necessary requirement, it is natural to impose that the
subtraction term does not have worse behavior than the original function in
limits where denominators vanish.

In ref.~\cite{DeLaurentis:2022otd}, it was observed that unexpectedly weak
divergence behavior of the function $f$ can be interpreted as a non-trivial
constraint on its numerator.
That is, the divergence behavior of the function $f$ in
eq.~\eqref{eq:DivergenceConstraint} translates into a vanishing behavior of the
numerator $\mathcal{N}^{(0)}$ as
\begin{equation}
   \pval{\mathcal{N}^{(0)}(x_{\epsilon, U})} \ge 0.
\label{eq:ValuationConstraint}
\end{equation}
This vanishing behavior can then be interpreted as a statement that
$\mathcal{N}^{(0)}$ belongs to a particular polynomial ideal. Naturally, the
particular class of ideal depends on how the point $x_{\epsilon, U}$
in eq.~\eqref{eq:ValuationConstraint} approaches the surface $U$. Intuitively,
for a surface $U$ defined by the vanishing of multiple polynomials, for a point
on a ``symmetric'' approach, these polynomials vanish at the same rate, while,
for a point on an ``asymmetric'' approach, some may vanish faster than others.
In ref.~\cite{DeLaurentis:2022otd}, a procedure was introduced to build $p$-adic
phase-space points that approach $U$ ``symmetrically''.
In this case, defining $\kappa_U = \pval{\mathcal{N}^{(0)}(x_{\epsilon, U})}$, one can show that
\begin{equation}
\mathcal{N}^{(0)} \in I(U)^{\langle \kappa_U \rangle},
\label{eq:SymmetricMembership}
\end{equation}
where $I(U)$ is set of polynomials in $\overline{R}_6$ that vanish everywhere
on $U$ and the superscript $\langle \kappa_U \rangle$ denotes that we take
it to the $\kappa_U^{\text{th}}$ ``symbolic power''. The symbolic power is a
refined notion of an ideal power which correctly captures vanishing to the
prescribed order (see ref.~\cite{DeLaurentis:2022otd} for further details and
computational techniques).
The procedure of ref.~\cite{DeLaurentis:2022otd} for the generation of
numerical points close to varieties has been generalised to arbitrary
quotient rings and extended to ``asymmetric'' approaches, where the
valuations of the ideal generators can be controlled separately
subject to the restriction imposed by the number of orthogonal
directions to the variety \cite{DeLaurentis:2023qhd, Syngular}.
In ref.~\cite{Campbell:2022qpq} a technique was introduced to
interpret the result of such evaluations at points $x_{\epsilon, U}$
that approach the variety asymmetrically through same framework as the
symmetric approach considered in \cite{DeLaurentis:2022otd}. In this
case, an auxiliary variable\footnote{Naturally, this approach can be
extended to multiple variables in order to gain further control over
the approach.} is introduced and we work in the ring
\begin{equation}
   \overline{R}_6^* = \overline{R}_6[r_{\mathcal{D}}]/\langle r_{\mathcal{D}}^2 - \mathcal{D} \rangle.
\end{equation}
Here, $r_{\mathcal{D}}$ is an algebraic square root of $\mathcal{D}$, and
$\overline{R}_6^*$ is an extension ring of $\overline{R}_6$\footnote{An
  important technical property of $\overline{R}_6^*$ is that it is
  ``Cohen-Macaulay'', which follows as it is the quotient of another
  Cohen-Macaulay ring $\overline{R}_6[r_{\mathcal{D}}]$ by a regular sequence.
  This importance of this property is that it often allows for simpler
  computations of symbolic powers.}.
The variety $U$ is also extended to a variety $U^*$, where we introduce a new
direction associated to $r_{\mathcal{D}}$ and constrain the associated coordinate to satisfy
$r_{\mathcal{D}}^2 = \mathcal{D}$. Naturally, on this variety we have that
$\pval{\mathcal{D}(x_{\epsilon, U^*})} = 2 \pval{r_{\mathcal{D}}(x_{\epsilon, U^*})}$, and hence this technique allows us to
control the valuation of a chosen polynomial.
Given this setup, if we generate symmetric approach points in the extended
context, one can show that
\begin{equation}
  \mathcal{N}^{(0)} \in I^*(U^*)^{\langle \kappa_{U^*} \rangle} \cap \overline{R}_6,
  \label{eq:AsymmetricMembership}
\end{equation}
where $I^*(U^*)$ is the ideal of polynomials in the extension ring
$\overline{R}_6^*$ that vanish on $U^*$ and we explicitly implement
that $\mathcal{N}^{(0)}$ does not depend the auxiliary variable $r_{\mathcal{D}}$
by intersecting with $\overline{R}_6$. Importantly, this intersection
with a subring can be understood as elimination of the auxiliary variable
$r_{\mathcal{D}}$ and can be implemented with Gr\"obner basis techniques.

The ideal membership constraints of eqs.~\eqref{eq:SymmetricMembership} and
\eqref{eq:AsymmetricMembership} are very powerful. While the construction of
a generating set of the symbolic power ideals can be systematically handled
using the techniques of ref.~\cite{DeLaurentis:2022otd}, combining all of the
constraints into a single ansatz for polynomials of high
mass dimension can be technically demanding. In practice, we follow an ad-hoc
procedure to implement the ideal membership constraints. Specifically, we
organize them in such a way that they suggest possible numerator factors for our
subtraction terms.
This approach is heuristic, and it would be interesting to understand the
algebraic geometry in more detail to  develop it more rigorously.
To construct possible numerator factors, we consider the collection of varieties
$U$ such that $\mathcal{D}_0$ and some other $\mathcal{D}_j$ vanish on $U$. That
is, we consider the set of varieties $U_{0jk}$ from \cref{eq:Codim2IrreducibleDecomposition}.
Importantly, we do not restrict to the case that $\mathcal{D}_j$ is a
denominator of $f$, as this can lead to many more constraints.
We then sample on either phase-space points which approach the varieties $U$,
heuristically choosing either symmetric or asymmetric approaches, leading to a
number of constraints of the form of eq.~\eqref{eq:SymmetricMembership} and
eq.~\eqref{eq:AsymmetricMembership}.
The involved symbolic powers are computed in an ad-hoc manner, using the
various approaches introduced in ref~\cite{DeLaurentis:2022otd}.
In order to make use of these ideal membership constraints, we take
intersections of small sets of the ideals, chosen in a heuristic manner. In
practice, we often find we find ideals that, when considered modulo
$\mathcal{D}_0$, are generated by very few polynomials, perhaps only a single
one.
Denoting such a polynomial as $g_j$, we see that $g_j$ is a natural candidate
for a numerator factor. We note that $g_j$ is sometimes, but not always, a
denominator factor. In this way, we are able to write
\begin{equation}
  \mathcal{N}^{(0)} = \overline{\mathcal{N}}^{(0)} \prod_{j} g_j,
  \label{eq:ConjecturedNumeratorFactors}
\end{equation}
for some polynomial in spinor brackets
$\overline{\mathcal{N}}^{(0)}$, whose mass dimension is lower than that of
$\mathcal{N}^{(0)}$.

It remains only to determine the unknown piece of the numerator factor,
$\overline{\mathcal{N}}^{(0)}$. To this end, we write an ansatz for
$\overline{\mathcal{N}}^{(0)}$ given by a linear combination of monomials in
spinor brackets that are linearly independent on the surface $\mathcal{D}_0 =
0$.
In order to generate this set of monomials, we extend the procedure of
section~\ref{sec:reducedansatz}, adjoining $\mathcal{D}_0$ to the set of
generators of the ideal in eq.~\eqref{eq:IrreducibilityConstraint}.
Finally, we sample $f$ on phase-space points that are $p$-adically close to
$\mathcal{D}_0 = 0$, in order to fit the unknowns in the ansatz for
$\overline{\mathcal{N}}^{(0)}$.
Success of this procedure then validates the
form of $\mathcal{N}^{(0)}$ in eq.~\eqref{eq:ConjecturedNumeratorFactors}.

\subsubsection*{Example subtraction term construction}

Let us now consider an explicit example, $f^{\text{ex}}$, of a coefficient
encountered in the computation of the NMHV basis for the gluon amplitude. We
note that this is the same function as considered in section~\ref{sec:FpRec}.
For clarity of discussion, we again recall its common denominator form,
\begin{equation}
f^{\text{ex}} = 
  \frac{\mathcal{N}^{\text{ex}}}{\langle 14 \rangle^2[14]^2 \langle 56 \rangle [56] \langle 1|2+4|3]^2 \langle 2|1+4|3]^4 \langle 2|1+3|4]^2 \Delta_{14|23|56}^4} \, ,
\end{equation}
where $\mathcal{N}^{\text{ex}}$ is polynomial in spinor brackets of
mass-dimension 36 and has $104\,128$ terms.
We begin by choosing a $\mathcal{D}_0$ to study. We take $\mathcal{D}_0 =
\Delta_{14|23|56}$ due to its high power in the denominator and its high mass
dimension.
We consider applying the denominator determination procedure outlined earlier in this
section. First, we sample on symmetric approaches to $\Delta_{14|23|45} =
\mathcal{D}_j = 0$ for any $\mathcal{D}_j$. On all such approaches, we find
that $\pval{f^{\text{ex}}} \ge -4$. If we were to take only these constraints as
input to \cref{eq:ExponentChoice}, we would conclude that, in the
leading pole around $\Delta_{14|23|45} = 0$, there are no denominator factors
other than $\Delta_{14|23|45}$.
However, it turns out we must take into account further constraints from
an asymmetric approach in the case that $\mathcal{D}_j$ is $\langle 2|1+4|3]$.
Specifically, if we consider an asymmetric approach\footnote{Such an asymmetric
approach exists as $\Delta_{14|23|45} = 4(s_{124} - s_{123})^2 - 4 \langle
2|1+4|3] \langle 3|1+4|2]$. In the language of algebraic geometry, $\langle
\Delta_{14|23|45}, \langle 2|1+4|3] \rangle_{\overline{R}_6}$ is not a radical
ideal} where $\pval{s_{124} - s_{123}} = 1$, but $\pval{\Delta_{14|23|45}} =
\pval{\langle 2|1+4|3]} = 2$, we find that that $\pval{f^{\text{ex}}}
= -9$.
Referring again to \cref{eq:ExponentChoice}, we see that this constraint
mandates a denominator factor of $\langle 2|1+4|3]$.
Altogether, we find that
\begin{equation}
f^{\text{ex}} = \frac{\mathcal{N}^{(0), \text{ex}}}{\Delta_{14|23|56}^4 \langle2|1+4|3]} + \mathcal{O}(\mathcal{D}_0^{-3}).
\label{eq:exampleLeadingPoleStructure}
\end{equation}
It is clear that the denominator of the expression in
eq.~\eqref{eq:exampleLeadingPoleStructure} has much lower mass dimension than
that of the full $f^{\text{ex}}$. In this form, an ansatz for $\mathcal{N}^{(0),
\text{ex}}$ contains 3360 terms.

We next consider constraining this numerator with ideal membership constraints.
In principle, we consider all co-dimension 2 varieties such that
$\Delta_{14|23|56} = \mathcal{D}_j = 0$, for some $\mathcal{D}_j$. For clarity,
we highlight those that give us a sufficient set of constraints.
We first consider a symmetric approach to the variety $\Delta_{14|23|56} = 0$,
$\langle 14 \rangle = 0$. A non-trivial feature of this variety is that the
ideal of polynomials that vanish on the variety is not $\big\langle \Delta_{14|23|56},
\langle 14 \rangle \big\rangle_{\overline{R}_6}$ as this ideal is not radical.
The associated radical ideal is
\begin{equation}
    \sqrt{\big\langle \Delta_{14|23|56}, \langle 14 \rangle \big\rangle_{\overline{R}_6}} = \big\langle s_{56} - s_{23}, \langle 14 \rangle \big\rangle_{\overline{R}_6},
    \label{eq:FirstExampleIdeal}
\end{equation}
where the square root operation on an ideal takes its radical. By numerically
evaluating $f^{\text{ex}}$ on a point on a symmetric approach to the associated
variety we conclude that its numerator vanishes and hence we impose that $\mathcal{N}^{(0)}$ belongs to $\sqrt{\big\langle \Delta_{14|23|56},
\langle 14 \rangle \big\rangle_{\overline{R}_6}}$. Performing a similar study of
the cases where one of $[14] = 0$, $\langle 23 \rangle = 0$ or $[23] = 0$, we
conclude that
\begin{align}
  \begin{split}
  \mathcal{N}^{(0), \text{ex}} \,\, \in \quad &\bigcap_{\{i,j\} \in \{\{1,4\}, \{2,3\}\}} \sqrt{\big\langle \Delta_{14|23|56}, \langle ij \rangle \big\rangle_{\overline{R}_6}} \!\cap\!
  \sqrt{\big\langle \Delta_{14|23|56}, [ij] \big\rangle_{\overline{R}_6}} \\
  & \,\,\, = \big\langle s_{56}-s_{23}-s_{14}, \Delta_{14|23|56} \big\rangle_{\overline{R}_6}.
  \end{split}
\end{align}
When considered modulo $\Delta_{14|23|56}$, we see that this ideal is generated
by a single polynomial, which provides a natural candidate numerator factor.

Next, we consider asymmetric approaches. First, we consider 
$\Delta_{14|23|56} = 0$, $\langle 23 \rangle = 0$ and perform an asymmetric
approach such that $\pval{\langle 23 \rangle} = 2$.
Numerical evaluation of $f^{\text{ex}}$ shows that its numerator vanishes
quadratically\footnote{In practice, the numerator of $f^{\text{ex}}$ also
vanishes to third order, but that complicates the argumentation.}.
Performing the analogous analysis for the case of $[23] = 0$ and working in the
ring extensions where we introduce $r_{\langle 23 \rangle}$ and
$r_{[23]}$ we find the constraints
\begin{align}
  \begin{split}
\mathcal{N}^{(0), \text{ex}} \,\, \in \,\, &\big\langle s_{14}-s_{56}, r_{\langle 23 \rangle} \big\rangle_{\overline{R}_6^*}^{\langle 2 \rangle} \cap
 \big\langle s_{14}-s_{56}, r_{[23]} \big\rangle_{\overline{R}_6^*}^{\langle 2 \rangle}  \cap \overline{R}_6
  \\
  &
  %= \big\langle \Delta_{14|23|56}, ⟨23⟩ \big\rangle_{\overline{R}_6} \cap \big\langle \Delta_{14|23|56}, [23] \big\rangle_{\overline{R}_6}
  = \big\langle s_{23}, \Delta_{14|23|56} \big\rangle_{\overline{R}_6} ,
  \end{split}
\end{align}
where the symbolic powers are equal to standard ideal powers as the ideals are
generated by regular sequences. Similar argumentation also leads to the ideal
membership statement that
\begin{equation}
  \mathcal{N}^{(0), \text{ex}} \in \big\langle s_{56}, \Delta_{14|23|56} \big\rangle_{\overline{R}_6}.
\end{equation}
Second, we consider an asymmetric approach to
the surface $\Delta_{14|23|56} = 0$, $\langle 6|1+4|5] = 0$ such that $\pval{\langle
6|1+4|5]} = 2$. Similar to before, finding the ideal associated to the variety
is a non-trivial operation as
\begin{gather}
  \sqrt{\big\langle Δ_{14|23|56}, ⟨6|1+4|5] \big\rangle_{\overline{R}_6}} = \big\langle s_{145}-s_{146}, ⟨6|1+4|5] \big\rangle_{\overline{R}_6}.
\end{gather}
Numerical sampling shows that the numerator of $f^{\text{ex}}$
vanishes quadratically on such a point. Working in the appropriate ring
extension, introducing $r_{⟨6|1+4|5]}$, we find the constraint
\begin{gather}
  \mathcal{N}^{(0), \text{ex}}
  \,\, \in \,\, 
  \big\langle s_{145}-s_{146}, r_{⟨6|1+4|5]} \big\rangle_{\overline{R}_6^*}^{\langle 2 \rangle}
  \cap \overline{R}_6
  = \big\langle ⟨6|1+4|5], \Delta_{14|23|56} \big\rangle_{\overline{R}_6} \, ,
\end{gather}
where we use that the symbolic power is equivalent to the normal power as the
two generators form a regular sequence.
Finally, we consider sampling near the variety $\Delta_{14|23|56} = 0$,
$\langle 4|2+3|1] = 0$, on a point such that $\pval{\langle 4|2+3|1]} = 2$. Once
again, finding the ideal associated to this variety is non-trivial as
\begin{gather}
  \sqrt{\big\langle Δ_{14|23|56}, ⟨4|2+3|1] \big\rangle_{\overline{R}_6}} = \big\langle s_{123}-s_{234}, ⟨4|2+3|1] \big\rangle_{\overline{R}_6}.
\end{gather}
Numerical sampling of $f^{\text{ex}}$ shows that the numerator vanishes to third
order on this point, which gives the ideal membership constraint
\begin{align}
  \begin{split}
  \mathcal{N}^{(0), \text{ex}} \,\, \in \,\, &\big\langle s_{123}-s_{234}, r_{\langle 4|2+3|1]} \big\rangle_{\overline{R}_6^*}^{\langle 3 \rangle}
  \cap \overline{R}_6 \, , \\
  &= \big\langle  (s_{123}-s_{234}) ⟨4|2+3|1], ⟨4|2+3|1]^2, \Delta_{14|23|56} (s_{123}-s_{234}) \big\rangle_{\overline{R}_6} \, ,
  \end{split}
\end{align}
where again the symbolic powers are equal to the standard ones as the ideals are
generated by regular sequences. Analogous logic leads to the membership
\begin{equation}
    \mathcal{N}^{(0), \text{ex}} \,\, \in \,\,
    = \big\langle  (s_{134}-s_{124}) ⟨3|1+4|2], ⟨3|1+4|2]^2, \Delta_{14|23|56} (s_{134}-s_{124}) \big\rangle_{\overline{R}_6} \, .
    \label{eq:ThreeGeneratorMembership}
\end{equation}

In order to make use of these ideal membership constraints to construct an
ansatz, we interpret each constraint as providing numerator factors of
$\mathcal{N}^{(0), \text{ex}}$.
While non-rigorous, this interpretation avoids the practical problem of
computing generating sets of the intersection of the ideals and is easily
validated by checking if the ansatz can be fit.
For most of the ideals, this factor is
the generator which is not $\Delta_{14|23|56}$ (as this is naturally sub-leading
in our chosen limit). For the membership of
eq.~\eqref{eq:ThreeGeneratorMembership}, we take $(s_{123}-s_{234}) ⟨4|2+3|1]$
as the other two generators are either proportional to $\Delta_{14|23|56}$ or
vanish to higher order than required from the numerical sampling. Altogether, we
reach an ansatz for the leading pole of $f^{\text{ex}}$ around
$\Delta_{14|23|56} = 0$ given by
\begin{equation}
f^{\text{ex}} = c_0 \frac{(s_{56} \!-\! s_{23} \!-\! s_{14})(s_{134}\!-\!s_{124})⟨6|1\!+\!4|5]⟨4|2\!+\!3|1]⟨3|1\!+\!4|2](s_{123}\!-\!s_{234})s_{23}s_{56}}{Δ_{14|23|56}^4⟨2|1\!+\!4|3]} \!+\! \mathcal{O}(\mathcal{D}_0^{-3}),
\end{equation}
where $c_0$ is a single unknown rational number. By evaluating $f^{\text{ex}}$
on a single phase-space point where $\mathcal{D}_0$ is $p$-adically small, we
can determine that $c_0 = -\frac{105}{256}$.

As we move to subleading orders in the various poles the number of available
constraints from ideal membership decrease, requiring us to increasingly rely on
ansätze with fewer or no numerator factors. However, note that the
complexity of an ansatz falls quickly as the degree of the numerator falls,
therefore it is much easier to fit a generic ansatz for subleading poles than it
is for the leading ones.
In practice, we compute the full analytic form of $f^{\text{ex}}$ by iteration
of this procedure, resulting in a representation with 62 fractions, up to
symmetry.

It is interesting to briefly compare the result obtained here with that from the approach of
section~\ref{sec:FpRec}. Using the subtraction term approach, the final result
has $346$ non-zero terms, in comparison to the $3\,504$ terms found in the
analytic-structure-based approach. The largest numerator of a rational number in
the subtraction term approach is $4\,517$, while in the analytic-structure approach it
is $6\,904\,270\,536\,241\,831\,187$. These differences culminate in the
statement that the subtraction approach generates an expression file size of
$18$kB, while for the analytic-structure-based approach it was $266$kB. Nevertheless, both
representations of the rational function are dramatically more compact than the
LCD form or the results of \cite{Abreu:2021asb}. Naturally, these differences arise from 
the large amount of freedom in expressing the rational functions. This includes
the freedom to apply momentum conservation and Schouten identities, but also
redundancies in the partial-fraction decomposition, since numerator polynomials
proportional to a denominator factor can be moved to a different fraction in the
decomposition.

\section{Implementation and results}
\label{sec:result}

\subsection{Implementation and impact of simplification procedures}

Analytic expressions for the NNLO leading-color $Vjj$ scattering amplitudes were
first computed in ref.~\cite{Abreu:2021asb}. However, the complexity of the analytic
expressions made their applications in phenomenological studies of $Vjj$
processes at NNLO difficult.
In this section, we apply the various techniques described in this paper to
express the $Vjj$ amplitudes in a compact form, thereby facilitating a stable
and efficient evaluation of the hard functions, which we demonstrate in
section~\ref{sec:performance}.
Our procedure consists of three main steps, which we perform on the basis of
rational coefficient functions \eqn{eq:pentagon-functions}:
\begin{enumerate}[label=s\arabic*]
\item
We construct and apply a constant linear basis change using the approach of
\cref{sec:basischange}.
\label{step1} 
\item
   We perform a first reconstruction of the improved basis of rational
   functions, using the analytic structure approach to partial-fraction ans\"atze
   of \cref{sec:FpRec}. All unknowns are determined using finite-field evaluations.
\label{step2} 
\item We perform a final reconstruction of this basis using the iterated
      subtraction approach described in \cref{sec:QpRec}. Unknowns are determined
      using $p$-adic evaluations.
\label{step3} 
\end{enumerate}
As inputs to steps~\ref{step1}--\ref{step2} we take the expressions of
ref.~\cite{Abreu:2021asb}. To enable more efficient evaluations of the rational
coefficients over finite fields we compile the expressions in \texttt{C++}.
The file size of the \texttt{C++} binaries was around 20 GB and they
allowed for evaluation speeds of around 1 second for all rational
functions in a given partial amplitude (in $\mathbb{F}_p$). The
compiled functions, representing form factors, were then combined into
helicity amplitudes allowing for evaluation in six-point
spinor-helicity phase space.
Equipped with efficient evaluation routines we apply the steps
\ref{step1}--\ref{step3} to obtain compact analytic expressions.

Given this broad overview of our approach, let us now discuss a number of
salient technical details arising in performing the steps.
The basis change \ref{step1} yields coefficient functions in
an optimized common-denominator form.
We apply this optimized basis construction procedure to each helicity partial
remainder, including all contributing partials in the loop, $N_c$ and $N_f$
expansion. That is, we apply it for each vector space described in
\cref{tab:ansatze-sizes}.
It would be interesting to investigate if improved results can be obtained by
applying it to the spaces $B_g^{\text{MHV}}$, $B_g^{\text{NMHV}}$ and
$B_q^{\text{NMHV}}$, but we leave that to future work.
The linear solving over finite fields in both steps \ref{step2} and
\ref{step3} was performed on GPU workstations with RTX 2080 Ti cards, each
with 11GB of VRAM. This limited the linear system sizes to a side length of
$52\,440$ and, in step \ref{step2}, systems with sizes approaching this limit
had to be solved. The results of step~\ref{step2} involved some rational numbers
whose numerator or denominator exceeded 100 digits.
On the order of 20 finite fields of size approximately $2^{31}-1$ were used to
obtain such rational numbers.
The results of step \ref{step2} were then used to provide efficient and stable
$p$-adic evaluations in kinematic limits as input to step \ref{step3}.
In practice, when constructing subtraction terms in step \ref{step3}, the
largest linear systems that we encountered involved approximately 1000 unknowns.
This allowed us to easily perform the reconstruction work on a laptop, making
use of efficient GPU-based finite-field row reduction.

Let us now quantify the effects of various stages of our simplification
procedures.
First, we consider the effect of the basis change. To this end, for each basis
function of the vector space, we consider the number of unknowns in the LCD
ansatz before and after the basis change. In table \ref{tab:ansatze-sizes} we
report the size of the largest ansatz before and after the basis change, for
each helicity partial remainder and we see that the improvement in ansatz size
varies between a factor of 3 and
90. Notably, more important simplifications are
observed for the more complex vector spaces, while the smallest is for the only
MHV contribution. We observe that
the effect of the basis change on the parameterization size is comparable to the
one of the univariate partial-fraction decomposition used in
ref.~\cite{Abreu:2021asb}, despite the functions still being in LCD form.

\begin{table*}[t!]
\renewcommand{\arraystretch}{2}
\centering
\resizebox{\textwidth}{!}{%

\begin{tabular}{c @{\hskip 5mm} c @{\hskip 5mm} c @{\hskip 5mm} c @{\hskip 5mm} c }
\toprule
Rational-Function 
& \multicolumn{2}{c}{Reference \cite{Abreu:2021asb} (Mandelstam Variables)} 
& \multicolumn{2}{c}{Max Spinor-Helicity Ansatz Size (LCD)} 
\\[-2mm]
Vector Space 
& Common Denominator  & Partial Fractioning 
& Before Basis Change & After Basis Change \\
\midrule
%\(\mathcal{R}_g ++\) 
$B_\gluon(1^{+}, 2^{+}, 3^{+}, 4^{-}) $ 
& $2\,700$ k & $230$ k 
& $930$ k & $290$ k
\\[-2mm]
%\(\mathcal{R}_g +-\) 
$B_\gluon(1^{+}, 2^{+}, 3^{-}, 4^{-}) $ 
& $18\,000$ k & $1\,000$ k 
& $13\,500$ k & $680$ k
\\[-2mm]
%\(\mathcal{R}_g -+\) 
$B_\gluon(1^{+}, 2^{-}, 3^{+}, 4^{-}) $
& $50\,000$ k & $1\,100$ k 
& $24\,800$ k & $620$ k
\\[-2mm]
%\(\mathcal{R}_q +-\) 
$B_\quark(1^{+}, 2^{+}, 3^{-}, 4^{-})$
& $4\,300$ k & $450$ k 
& $4\,700$ k & $120$ k
\\[-2mm]
%\(\mathcal{R}_q -+\) 
$B_\quark(1^{+}, 2^{-}, 3^{+}, 4^{-})$ 
& $10\,000$ k & $510$ k 
& $7\,100$ k & $87$ k
\\
\bottomrule
\end{tabular}
}
\caption{
\label{tab:ansatze-sizes}
For each partial amplitude, this table compares the effect of the
basis change to that of partial-fraction decomposition in
ref.~\cite{Abreu:2021asb}. For each amplitude we quote the maximum
ansatz sizes, i.e.~the number of numerical samples required for the
reconstruction. 
Ref.~\cite{Abreu:2021asb} gives the ansatz sizes for three form factors 
associated to a single helicity configuration, which we sum in this table.
}
\end{table*}

Next, we compare the effect of the steps \ref{step2} and \ref{step3} on the
representations of the rational functions.
In table \ref{tab:nonzero-terms}, we display the number of
contributing monomials in the original form of the remainders in
ref.~\cite{Abreu:2021asb}, with the number of non-zero terms after undertaking
the analytic-structure-inspired partial-fraction decomposition, and finally the number of
non-zero terms after the simplification by iterated subtraction procedure.
We observe a drop in unknown parameters in the coefficient parameterization by
four orders of magnitude from a naive LCD form in ref~\cite{Abreu:2021asb}, and
two orders of magnitude compared to the univariate partial-fractioning in the
same paper.
Beyond the effect on expression size, we observe that step \ref{step3} has a
dramatic effect on the size of the rational numbers in the bases of rational
functions. Specifically, in the basis of $B_{g}^{\text{NMHV}}$ the largest
integer has 12 digits, while the bases of $B_{g}^{\text{MHV}}$ and
$B_{q}^{\text{NMHV}}$ contained integers with at most 3 digits.
We note that the matrices of rational numbers in \cref{eqn:Mmatrix} can still be considerably
complicated, and we leave an analysis of their source to future work.

\begin{table*}[t!]
\renewcommand{\arraystretch}{2}
\centering
%\resizebox{\textwidth}{!}{%
\begin{tabular}{c @{\hskip 5mm} c @{\hskip 5mm} c @{\hskip 5mm} c }
\toprule
Rational-Function & \multicolumn{3}{c}{Max Non-Zero Terms} \\[-2mm]
Vector Space 
& Reference \cite{Abreu:2021asb}
& Ansatz from Analytic Structure & Iterated Subtractions 
\\
\midrule
%$B_g^{\text{MHV}} $
$B_\gluon(1^{+}, 2^{+}, 3^{+}, 4^{-}) $ 
%\(\mathcal{R}_g ++\) 
& 15 k
& 7100 & 762 
\\[-2mm]
$B_\gluon(1^{+}, 2^{+}, 3^{-}, 4^{-}) $ 
%\(\mathcal{R}_g +-\) 
& 110 k
& 8700 & \multirow{2}{*}{$\Bigg\}$1810} 
\\[-2mm]
$B_\gluon(1^{+}, 2^{-}, 3^{+}, 4^{-}) $
%\(\mathcal{R}_g -+\) 
& 130 k
& 8200 & 
\\[-2mm]
$B_\quark(1^{+}, 2^{+}, 3^{-}, 4^{-})$
%\(\mathcal{R}_q +-\) 
& 56 k
& 6800 & \multirow{2}{*}{\raisebox{2.5ex}{$\Bigg\}$134}} 
\\[-2mm]
$B_\quark(1^{+}, 2^{-}, 3^{+}, 4^{-})$ 
%\(\mathcal{R}_q -+\) 
& 57 k
& 4900 & 
\\
\bottomrule
\end{tabular}
%}
\caption{
\label{tab:nonzero-terms}
The maximum number of non-zero terms in a reconstructed rational function for each partial amplitude.
The second column shows the results of ref.~\cite{Abreu:2021asb}, which are given in terms of Mandelstam invariants.
The results of this work are displayed in the last two columns: first after the
application of the partial fractions from analytic structure approach of \cref{sec:FpRec}
and second, and after the iterated subtraction approach described in \cref{sec:QpRec}.
}
\end{table*}

\subsection{Ancillary files}\label{sec:ancillaries}

We provide analytic results for the finite remainders up to two loops
as ancillary files accompanying this article.
The partial remainders are given in the form of
\cref{eq:pentagon-functions}. The rational functions are provided in terms
of the improved bases constructed with the techniques of \cref{sec:basischange} as
\begin{equation}
  r_i = \sum_{\tilde{r}_j \in \mathcal{B}} \tilde{M}_{ij} \tilde{r}_j,
\end{equation}
for a $\mathcal{B}$ a basis of the appropriate space, $B_g^{\text{MHV}}$,
$B_g^{\text{NMHV}}$ and $B_q^{\text{NMHV}}$ that are defined in
eqs.~\eqref{eq:generating_set_first}--\eqref{eq:generating_set_last}.
The results for the spaces $B_g^{\text{MHV}}$, $B_g^{\text{NMHV}}$ and
$B_q^{\text{NMHV}}$, are respectively organised into three directories:
\begin{center}
1.~\texttt{gluons\_mhv} $\qquad$ 2.~\texttt{gluons\_nmhv} $\qquad$
3.~\texttt{quarks\_nmhv}
\end{center}
For
convenience, we provide files in two formats, for use in
\texttt{Python}, which directly reads a set of associated \texttt{LaTeX} files, and
\texttt{Mathematica}.
Within each folder, the basis of the respective vector space, written in the
spinor-helicity formalism, is provided in the following files:
\begin{enumerate}
\item[a.] \texttt{basis\_rational} (for \texttt{Mathematica})
\item[b.] \texttt{basis.txt} and \texttt{coeff\_\#.tex}
  (for \texttt{Python})
\end{enumerate}
The \texttt{Mathematica} files use the notation of \texttt{S@M}
\cite{Maitre:2007jq}, while the \texttt{coeff\_\#.tex} files are read
into \texttt{Python} using the package \texttt{antares}
\cite{Antares}. For pre-compiled human-readable results see the
documention of \texttt{antares-results} \cite{AntaresResults}. The
\texttt{Mathematica} expressions are generated from \texttt{Python}
via the \texttt{antares} function
\href{https://github.com/search?q=repo\%3AGDeLaurentis\%2Fantares+toSaM&type=code}{\texttt{toSaM}}.
The number of explicit coefficients is that of the generating set of
table \ref{table:overlap}. The bases for the NMHV coefficients with
helicities $(2^-, 3^+)$ are obtained from the provided ones via a
$2\leftrightarrow 3$ swap. The set of pentagon functions $h_i$ in
\cref{eq:pentagon-functions} is given in
\begin{enumerate}
\item[c.] \texttt{basis\_transcendental} (for \texttt{Mathematica} and \texttt{Python}).
\end{enumerate}
The \texttt{tex}/\texttt{Python} results use the notation introduced
in ref.~\cite[Section 4.3]{Laurentis:2019bjh} to express coefficients
involving transformations. We recall that the convention is that
transformations are applied to all preceding rational functions until
another block of transformations is encountered. Transformations are
composed of a permutation, a possible conjugation, and a sign (minus
for anti-symmetry). In most cases where one such transformation is
used, the entire coefficient is mapped onto itself under it (i.e.~it
is a symmetry). However, in some cases, this is only true for a subset
of the terms. For convenience, we have explicitly expanded these
transformations in the \texttt{Mathematica} files. Transformations are
also used to construct the full basis from the generating set of
rational functions. Similarly here they refer to the last
spinor-helicity coefficient preceding them.
For each partial amplitude in
eq.~\eqref{eq:gluon-mhv}--\eqref{eq:quark-nmhv2}, we provide the
matrix $\tilde{M}_{ij}$ in files labeled as
\begin{center}
  \texttt{(quarks|gluons)(pp|pm|mp)\_\#L\_Nf\#} $\, .$
\end{center}
These sparse matrices are stored in coordinate list format, where each
line contains three entries: row index, column index, and rational
value.  For convenience, we also provide the trees in this format in
the files ending in \texttt{\_tree} (instead of \texttt{\_\#L\_Nf\#}).

To demonstrate their usage, the expressions for the remainders can be
loaded and evaluated at a test kinematic point using the files
\texttt{assembly.m} or \texttt{test\_remainders.py} (to be used with
\texttt{pytest}). The files \texttt{momenta.m} and \texttt{momenta.py}
contain the four-momenta for the reference kinematic point of
ref.~\cite{Abreu:2021asb}, \texttt{target\_values.py} contains
reference results of the same paper to match against, and
\texttt{functions.m} contains auxiliary functions used in
\texttt{assembly.m}. The evaluation of the finite remainders requires
the \texttt{PentagonFunctions++} library to be installed
\cite{Chicherin:2021dyp, PentagonFunctions}. Spinor-string evaluations in
\texttt{Python} are done with the package \texttt{lips}
\cite{DeLaurentis:2023qhd}, which also allows for an easy change of
the number field used in the evaluation of the rational functions,
e.g.~to finite fields. 

Lastly, we provide an ancillary file,
\texttt{test\_primary\_decompositions.py}, for the primary
decompositions given in Appendix \ref{sec:PrimDec}. This file relies
on the primality test of ref.~\cite{DeLaurentis:2022otd}, implemented
in \cite{Syngular}. These decompositions may serve, beyond the present
application, as input for partial-fraction decomposition of more
general six-point amplitudes.

\subsection{Program for numerical evaluation of hard functions}\label{sec:cppimpl}\label{sec:performance}

Beyond analytic results for amplitudes for $pp \to V j j$, we also present an
efficient and stable numerical code for their evaluation which is suitable for
phenomenological applications.
The two-loop NNLO QCD corrections for $pp \to V j j$ cross sections are derived from squared helicity- and color-summed finite remainders, expressed as:
\begin{equation}\label{eq:hard-function}
  \mathcal{H} = \frac{1}{\mathcal{B}} \sum_{\text{color}, \text{helicity}} \mathcal{M}_{\mathcal{R}}^\star \mathcal{M}_{\mathcal{R}} \,, \qquad \mathcal{B} = \sum_{\text{color}, \text{helicity}} \abs{\mathcal{A}^{(0)}}^2 \,,
\end{equation}
where $\mathcal{H}$ denotes the hard function which we expand to $\mathcal{O}(\alpha_s^2)$.
To compute the sum, we first use \cref{eq:ampW,eq:ampZ,eq:ampZneutral,eq:partProcesses} to express $\mathcal{H}$ in terms of helicity finite remainders of the representative channels.
We then decompose each of them into the sum of generating finite remainders in \eqref{eqn:finRemainder}, evaluated with appropriately permuted momenta.
In cases when parity conjugation is required (which flips helicities of all particles) to map into one of the generating finite remainders, it is applied to the corresponding expressions by exchanging the spinor brackets $\langle ij \rangle \leftrightarrow [ji]$.
The rational coefficients from the sets $B_\gluon^{\text{MHV}}, B_\gluon^{\text{NMHV}}, B_\quark^{\text{NMHV}}$ in \cref{eq:pentagon-functions} are evaluated at the permuted phase-space point, 
while the permutations of pentagon functions $G_i$ are reduced to a single standard permutation using the relations provided in \cite{Chicherin:2021dyp}.

We comment on a subtlety regarding the parity-conjugated helicity finite remainders.
It may be convenient to implement their numerical evaluation by evaluating the corresponding helicity remainder from a generating set on a parity-conjugated phase-space-point $\{\bar{p}_i\}$
However, this introduces a convention-dependent little-group phase $\phi_{ij}$ relative to the non-conjugated spinor brackets, i.e.~$\langle i j \rangle = e^{\ii \phi_{ij}} [\bar{j}\,\bar{i}]$, which must be compensated before conjugated and non-conjugated helicity remainders can be coherently summed (as in the case of $Z(\bar{\ell}\ell)jj$).
In practice, this can be achieved through renormalizing by the factor ${\qty(A^{(0)}(\{p_i\}))^{\star}}/{A^{(0)}(\{\bar{p}_i\})}$.

We implement generating helicity partial remainders and the hard functions for all partonic channels 
for $W^+(\bar{e}\nu_e)jj$, $Z(\bar{e}e)jj$, and $Z(\bar{\nu}_e \nu_e)jj$ production in proton collisions with leptonic vector boson decays in a public C++ code \cite{FivePointAmplitudes}. 
Reference values for the complete set of channels are presented in \cref{sec:referenceEvaluations}. 
We checked that the one-loop hard functions in all channels agree with the results from \texttt{BlackHat} \cite{Berger:2008sj}. 
Additionally, we verified that the two-loop hard functions reproduce the correct renormalization scale evolution, as predicted from lower perturbative orders \cite{Catani:1998bh,Becher:2009cu}.

To control the numerical precision of evaluations, we employ a dynamical precision-rescue system. 
This system uses the stability of the result under input rounding errors as an indicator for reliable evaluation. 
First, for each phase-space point $\{p_i\}$, we perform two evaluations in double precision.
For the second evaluation, we apply a random Lorentz boost to the momenta, $p_i^\prime = \Lambda p_i$, chosen such that all denominators encountered in the rational coefficients differ as floating-point numbers due to rounding errors. 
Additionally, the second evaluation is performed with a renormalization scale $\mu^\prime = \frac{3}{2} \mu$. 
We then compare the second evaluation to the evolution of the first one from scale $\mu$ to $\mu^\prime$, reproduced by the factor 
\begin{equation}
  U(\mu^\prime,\mu) = 1 + \frac{\alpha_s}{2 \pi}~U^{(1)}\qty(\log\frac{\mu^\prime}{\mu}) ~+~ \ldots{}\,, \qquad  U(\mu,\mu) = 1\,,
\end{equation}
derived in \cite[Section 4.3]{Abreu:2021oya}. 
This process provides a relative error estimate:
\begin{equation}
  \Delta_{\mathrm{d}}\mathcal{H} = \abs{1 - \frac{U(\mu^\prime,\mu)~\mathcal{H}(\{p_i\},\mu)}{\mathcal{H}(\{p_i^\prime\}, \mu^\prime)}}\,.
  \label{eq:HardFunctionErrorEstimate}
\end{equation}
We then verify whether $\Delta_{\mathrm{d}}\mathcal{H} < \delta$, where the technical parameter $\delta$ is set to $\delta = 10^{-2}$. 
If this condition is not satisfied, we switch to evaluation in quadruple precision.

\begin{figure}[ht]
  \centering
  \includegraphics[width=0.8\textwidth]{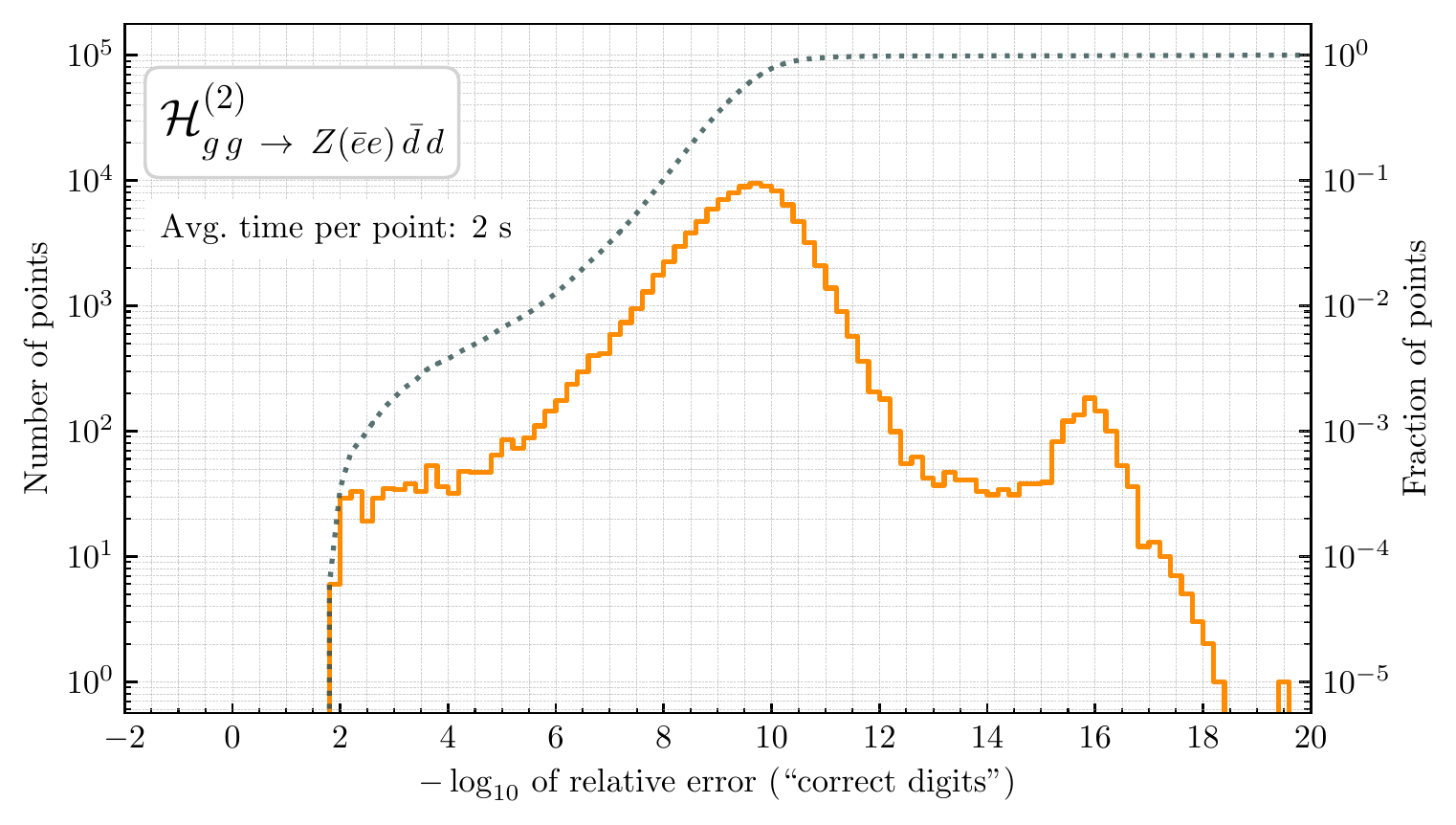}
  \caption{
      A distribution of the base 10 logarithm of the relative error (correct
digits) of the two-loop hard function for the $gg \to Z(\bar ee) \bar{d} d$ partonic channel.  
The dashed curve represents the respective cumulative distribution.
Here $\NC=3$ and $\NF=5$, and the renormalization scale is set dynamically to
invariant mass of $\bar{e}e\bar{d}d$ system.
The phase space definition is taken from ref.~\cite{Mazzitelli:2024ura}.
  }
  \label{fig:h2-stability}
\end{figure}

To illustrate the numerical performance of our implementation we consider the $gg \to Z(\bar ee) \bar{d} d$ partonic channel, 
which is among the ones with the largest number of contributing permutations of generating finite remainders.
In \cref{fig:h2-stability} we show the distribution of relative errors of the evaluations over 100k phase-space points from
a sample used in the recent computation of NNLO QCD corrections to $Z\bar{b}b$ production \cite{Mazzitelli:2024ura}.
The relative error is calculated by comparing to an evaluation on the same phase-space point in quadruple precision.
We observe that the majority of evaluations have 6 to 12 correct digits which demonstrates excellent numerical stability of our results.
Nevertheless, around $1.5\%$ of the points have less than two correct digits in double precision. 
They are successfully detected by the second evaluation and rescued through higher precision evaluation,
which explains the second small peak around 16 digits in \cref{fig:h2-stability}.
We achieve an average evaluation time per phase-space point of around 2 seconds,
making use of a single thread on a heterogeneous cluster with 2.4GHz average
clock speed. This timing includes the second evaluation and higher-precision
rescue when required. This convinces us that our results are suitable for
calculation of two-loop QCD corrections.

We note that the check on ratios of symbol letters employed in ref.~\cite{Abreu:2021oya} is not as strongly correlated with unstable evaluations in this case,
and therefore it cannot be used to effectively avoid a second double-precision evaluation most of the time.
We also note that without the scale evolution test, or equivalently if we take $\mu'
= \mu$ in \cref{eq:HardFunctionErrorEstimate}, some fraction of unstable points are falsely deemed stable, as revealed by comparing them to quadruple precision targets.

\section{Conclusions}
\label{sec:conclusions}

In this paper, we have demonstrated the feasibility of obtaining
compact analytic expressions for two-loop processes beyond five-point
massless kinematics. Specifically, we significantly simplified the
leading-color scattering amplitudes for the hadronic production of a
vector boson and two jets. The five-point one-mass kinematics of this
process introduces significantly more complex expressions compared to
the massless case, due to the additional degree of freedom and the
proliferation of Gram determinants. To manage this increased
complexity, we introduced advancements in analytic reconstruction
techniques while also leveraging established methods.

Our approach has the potential to simplify the functional
reconstruction of unknown integral coefficients of scattering
amplitudes, or, as done here, to optimize analytic expressions of
known ones. The method consists of two key computational steps.
First, we preface any reconstruction with a basis change in the space
of rational coefficient functions, disentangling overlapping physical
and spurious poles. This identifies simpler functions for
reconstruction. Second, for each coefficient function, we construct a
compact parameterization in partial-fraction form by numerically
analyzing its behavior in doubly-singular limits and subsequently
fitting a pole at a time \cite{Laurentis:2019bjh,
  DeLaurentis:2022otd}. These partial-fraction decompositions leverage
primary decompositions of ideals generated by denominator factors. We
introduce a tailored spinor-helicity ansatz optimized for processes
involving an external massive vector boson. This approach avoids
overparameterization, such as treating the process as a genuine
six-point interaction, while still allowing for chiral cancellations
between numerators and denominators.
We note that while for the present calculation we also performed an
intermediate reconstruction with generic finite-field samples, this
step could be avoided if sufficiently stable $p\kern0.2mm$-adic
evaluations of the amplitude coefficients in singular kinematics
limits were available. This would allow to directly target compact
representations.

To quantify the impact of the basis change algorithm, we compared it
to the partial-fraction decomposition used in the original computation
\cite{Abreu:2021asb}. We find that the two approaches yield similar
benefits in isolation. However, since we can still perform a partial-fraction 
decomposition after the basis change, this leads to
parametrizations of the coefficients that are about two orders of
magnitude smaller. To assess the complexity of the final results, we
compared the number of non-zero coefficients in the most complicated
rational functions, finding they are about three orders of magnitude
fewer in the new results compared to the original ones. This
observation aligns with a straightforward comparison of file sizes for
the coefficients, which are reduced from approximately 1 GB to about 1
MB. Additionally, the rational numbers appearing in the final results
for quark NMHV and gluon MHV coefficients are remarkably small,
consisting of no more than three digits. This serves as further
evidence that the chosen basis and representation are close to
optimal. For the gluon NMHV amplitudes, we observe larger numbers,
reaching up to 12 digits in some cases. While this suggests room for
improvement, we leave further refinements for future work. Overall,
the compactness of the results highlights the potential of our approach to
allow efficient computation of new multi-scale amplitudes in the future.

We implemented our results in a public C++ code
\cite{FivePointAmplitudes}, which we demonstrated yields numerically
stable and efficient evaluations. We envisage that the code will
enable phenomenological studies of the standard-candle vector-bosons
plus jets processes, and interesting theoretical investigations into
the perturbative properties of QCD.  Our compact rational coefficient
basis can also be used to study four-jet production in
electron-positron colliders.  This requires analytic continuation of
pentagon functions to the decay channel which we leave for future
work.

\section*{Acknowledgments}
We gratefully acknowledge the computing resources provided by the Paul
Scherrer Institut (PSI).
V.S.\ has received funding from the European Research Council (ERC)
under the European Union's Horizon 2020 research and innovation
programme grant agreement 101019620 (ERC Advanced Grant TOPUP).
G.D.L.'s work is supported in part by the U.K.\ Royal Society through
Grant URF\textbackslash R1\textbackslash 20109.
The Feynman diagrams have been produced with \texttt{FeynGame} \cite{Harlander:2020cyh}.

\pagebreak

\begin{appendix}
\section{Reference evaluations}
\label{sec:referenceEvaluations}

The perturbative expansions up to $\order{\alpha_s^2}$ of the hard functions \eqref{eq:hard-function} is defined as
\begin{equation}
  \mathcal{H}= 1 + \qty(\frac{\alpha_s}{2\pi}) \qty(H^{(1)[0]}+ \NF H^{(1)[1]}) + \left(\frac{\alpha_s}{2\pi}\right)^2 \qty(H^{(2)[0]}+ \NF H^{(2)[1]}+ \NF^2 H^{(2)[2]}) + \ldots{} \,,
\end{equation}
We present the reference evaluations of the hard functions at the point
\begin{equation} \label{eq:referencePointBis}
  \begin{aligned}
    p_1 &= \{-88.000000000000000,0,0,88.000000000000000\},\\
    p_2 &= \{-88.000000000000000,0,0,-88.000000000000000\},\\
    p_3 &= \{12.392045454545455,4.3855252841337518,8.2930798971534504,-8.0965909090909091\},\\
    p_4 &= \{54.130681818181818,37.047749948312009,-8.2930798971534504,38.585227272727273\},\\
    p_5 &= \{42.652688395266764,0,42.652688395266764,0\}, \\
    p_6 &= \{66.824584332005964,-41.433275232445761,-42.652688395266764,-30.488636363636364\},
  \end{aligned}
\end{equation}
with the renormalization scale set to $\mu=63$.  
The EW input parameters are
\begin{equation}
  \begin{aligned}
    M_W &= 80.385, \\
    G_W &= 2.0854, \\
  \end{aligned}
  \qquad\qquad
  \begin{aligned}
    M_Z &= 91.1876,\\
    G_Z &= 2.4952, \\
  \end{aligned}
  \qquad\qquad
  \cos^2\theta_w = \frac{M_W^2}{M_Z^2},
\end{equation}
and we set $N_c=3$.

For convenience for show the results obtained with the implementation in ref.~\cite{FivePointAmplitudes} 
for all partonic channels independent under permutations of final or initial particles.
We remind the reader that whenever we are using the $\to$ notation, the first two
particles are to be understood as crossed to be incoming.

\begin{table}[!h]
  \renewcommand{\arraystretch}{1.2}
  \centering
  \begin{adjustbox}{width=1\textwidth}
  \begin{tabular}{c*{5}{C{15ex}}}
    \toprule
    Channel&$H^{(1)[0]}$&$H^{(1)[1]}$&$H^{(2)[0]}$&$H^{(2)[1]}$&$H^{(2)[2]}$ \\
    \toprule
$u\,g\rightarrow g\,d\,{\bar{e}}\,{\nu_e}\,$ & $13.24581387$ & $-0.3456721333$ & $468.3898952$ & $-40.00503534$ & $0.4130259588$ \\
$u\,{\bar d}\rightarrow g\,g\,{\bar{e}}\,{\nu_e}\,$ & $4.486297129$ & $-0.6032840215$ & $201.0776410$ & $-17.21540765$ & $0.1839661237$ \\
$g\,g\rightarrow {\bar u}\,d\,{\bar{e}}\,{\nu_e}\,$ & $3.678108745$ & $0.3722315368$ & $334.4492762$ & $-22.87739872$ & $-0.2737219527$ \\
${\bar d}\,g\rightarrow {\bar u}\,g\,{\bar{e}}\,{\nu_e}\,$ & $9.061964404$ & $0.1460662704$ & $439.1778004$ & $-20.84738470$ & $-0.1157108089$ \\
\midrule
$u\,{\bar s}\rightarrow {\bar s}\,d\,{\bar{e}}\,{\nu_e}\,$ & $5.911796920$ & $-1.175054477$ & $179.2685001$ & $-32.88301674$ & $1.035564768$ \\
$u\,s\rightarrow s\,d\,{\bar{e}}\,{\nu_e}\,$ & $26.09665200$ & $-1.175054477$ & $794.9162055$ & $-83.47140348$ & $1.035564768$ \\
$u\,{\bar d}\rightarrow s\,{\bar s}\,{\bar{e}}\,{\nu_e}\,$ & $25.63492037$ & $-1.646086288$ & $1198.459061$ & $-53.76102385$ & $0.9355773391$ \\
${\bar s}\,s\rightarrow {\bar u}\,d\,{\bar{e}}\,{\nu_e}\,$ & $1.412046254$ & $0.2586879138$ & $144.6142698$ & $-0.09567425598$ & $-1.046433134$ \\
${\bar s}\,{\bar d}\rightarrow {\bar u}\,{\bar s}\,{\bar{e}}\,{\nu_e}\,$ & $12.91732892$ & $-0.1686080316$ & $263.9591583$ & $-20.06441123$ & $0.02132150125$ \\
$s\,{\bar d}\rightarrow {\bar u}\,s\,{\bar{e}}\,{\nu_e}\,$ & $-12.00989939$ & $-0.1686080316$ & $78.04535861$ & $0.8703127244$ & $0.02132150125$ \\
\midrule
$u\,{\bar u}\rightarrow {\bar u}\,d\,{\bar{e}}\,{\nu_e}\,$ & $5.795031380$ & $-1.167240037$ & $179.9367085$ & $-32.70138753$ & $1.024217089$ \\
$u\,u\rightarrow u\,d\,{\bar{e}}\,{\nu_e}\,$ & $20.69575727$ & $-1.719503108$ & $680.2428204$ & $-87.22443182$ & $2.420499565$ \\
$u\,{\bar d}\rightarrow u\,{\bar u}\,{\bar{e}}\,{\nu_e}\,$ & $13.21565293$ & $-1.858085286$ & $868.3068564$ & $-42.80969172$ & $1.957241009$ \\
${\bar u}\,{\bar d}\rightarrow {\bar u}\,{\bar u}\,{\bar{e}}\,{\nu_e}\,$ & $19.06059090$ & $-2.195796171$ & $759.2175342$ & $-105.1361731$ & $3.647765247$ \\
$u\,{\bar d}\rightarrow {\bar d}\,d\,{\bar{e}}\,{\nu_e}\,$ & $13.64985778$ & $-1.528986646$ & $322.1189979$ & $-50.79469666$ & $0.9604344545$ \\
$u\,d\rightarrow d\,d\,{\bar{e}}\,{\nu_e}\,$ & $34.30536185$ & $-1.313929309$ & $1161.767997$ & $-100.8936622$ & $1.299514826$ \\
${\bar d}\,d\rightarrow {\bar u}\,d\,{\bar{e}}\,{\nu_e}\,$ & $8.436163636$ & $-1.348380547$ & $423.5387330$ & $-47.78207291$ & $1.369263531$ \\
${\bar d}\,{\bar d}\rightarrow {\bar u}\,{\bar d}\,{\bar{e}}\,{\nu_e}\,$ & $40.96012110$ & $-1.333076014$ & $1459.547653$ & $-112.5647380$ & $1.355566521$ \\
    \bottomrule
  \end{tabular}
  \end{adjustbox}
  \caption{
   Reference evaluations for $pp \to W^+(\bar{e} \nu_e)\,jj$ partonic channels.
  }
  \label{tab:wjj-benchmark}
\end{table}

\begin{table}[!h]
  \renewcommand{\arraystretch}{1.2}
  \centering
  %\begin{adjustbox}{width=1\textwidth}
  \begin{tabular}{c*{5}{C{15ex}}}
    \toprule
    Channel&$H^{(1)[0]}$&$H^{(1)[1]}$&$H^{(2)[0]}$&$H^{(2)[1]}$&$H^{(2)[2]}$ \\
    \toprule
$u\,{\bar u}\rightarrow g\,g\,{\bar{e}}\,{e}\,$ & $3.628949549$ & $-0.6019425594$ & $188.7139074$ & $-15.51583012$ & $0.1810277175$ \\
$u\,g\rightarrow u\,g\,{\bar{e}}\,{e}\,$ & $4.806313359$ & $0.1431989986$ & $369.8020485$ & $-16.05516339$ & $-0.1147447276$ \\
${\bar u}\,g\rightarrow {\bar u}\,g\,{\bar{e}}\,{e}\,$ & $3.981495176$ & $0.1443900205$ & $336.4218513$ & $-15.50703369$ & $-0.1151460234$ \\
$g\,g\rightarrow {\bar u}\,u\,{\bar{e}}\,{e}\,$ & $1.732968275$ & $0.3789112560$ & $313.7567967$ & $-21.43701745$ & $-0.2730705488$ \\
$d\,{\bar d}\rightarrow g\,g\,{\bar{e}}\,{e}\,$ & $3.747634687$ & $-0.6014789153$ & $188.2808536$ & $-15.63648045$ & $0.1808307421$ \\
$d\,g\rightarrow d\,g\,{\bar{e}}\,{e}\,$ & $5.324698622$ & $0.1433146992$ & $381.6394957$ & $-16.61232352$ & $-0.1147837111$ \\
${\bar d}\,g\rightarrow {\bar d}\,g\,{\bar{e}}\,{e}\,$ & $3.059810412$ & $0.1441285472$ & $316.3896738$ & $-14.51701299$ & $-0.1150579241$ \\
$g\,g\rightarrow {\bar d}\,d\,{\bar{e}}\,{e}\,$ & $1.999140581$ & $0.3811468354$ & $317.0698902$ & $-21.66390613$ & $-0.2734732041$ \\
\midrule
$u\,{\bar u}\rightarrow {\bar d}\,d\,{\bar{e}}\,{e}\,$ & $16.20297806$ & $-1.616436452$ & $390.2429499$ & $-53.88163277$ & $0.9117971458$ \\
$u\,d\rightarrow u\,d\,{\bar{e}}\,{e}\,$ & $23.94758679$ & $-1.721054994$ & $718.8993994$ & $-90.37240317$ & $2.335444353$ \\
$u\,{\bar d}\rightarrow u\,{\bar d}\,{\bar{e}}\,{e}\,$ & $-5.157376308$ & $-1.725766671$ & $217.3132100$ & $-16.87324830$ & $2.384483569$ \\
${\bar u}\,d\rightarrow {\bar u}\,d\,{\bar{e}}\,{e}\,$ & $-0.8838781447$ & $-1.816733693$ & $362.2366980$ & $-36.18370364$ & $2.603647894$ \\
${\bar u}\,{\bar d}\rightarrow {\bar u}\,{\bar d}\,{\bar{e}}\,{e}\,$ & $27.51804319$ & $-1.849726905$ & $999.5714713$ & $-112.3729992$ & $2.695916188$ \\
$d\,{\bar d}\rightarrow {\bar u}\,u\,{\bar{e}}\,{e}\,$ & $15.93071076$ & $-1.626095947$ & $388.9085368$ & $-53.45188192$ & $0.9026503299$ \\
$u\,{\bar u}\rightarrow {\bar c}\,c\,{\bar{e}}\,{e}\,$ & $16.99214471$ & $-1.664206929$ & $405.9893321$ & $-55.70586798$ & $0.9648399409$ \\
$u\,c\rightarrow u\,c\,{\bar{e}}\,{e}\,$ & $35.73520739$ & $-1.594311580$ & $1462.297864$ & $-118.5520843$ & $2.054494583$ \\
$u\,{\bar c}\rightarrow u\,{\bar c}\,{\bar{e}}\,{e}\,$ & $0.6731015458$ & $-1.707367423$ & $394.7480141$ & $-39.46678849$ & $2.308203971$ \\
${\bar u}\,c\rightarrow {\bar u}\,c\,{\bar{e}}\,{e}\,$ & $-5.290212522$ & $-1.760035521$ & $206.4503190$ & $-16.29958348$ & $2.484287313$ \\
${\bar u}\,{\bar c}\rightarrow {\bar u}\,{\bar c}\,{\bar{e}}\,{e}\,$ & $23.96985862$ & $-1.813977521$ & $868.5013057$ & $-98.16206871$ & $2.604607335$ \\
$s\,d\rightarrow s\,d\,{\bar{e}}\,{e}\,$ & $37.29600787$ & $-1.619215278$ & $1537.011250$ & $-123.9750819$ & $2.124002439$ \\
$s\,{\bar d}\rightarrow s\,{\bar d}\,{\bar{e}}\,{e}\,$ & $0.8904772978$ & $-1.716146067$ & $413.4271095$ & $-40.68638289$ & $2.329047506$ \\
${\bar s}\,d\rightarrow {\bar s}\,d\,{\bar{e}}\,{e}\,$ & $-6.819281264$ & $-1.749963802$ & $171.3540755$ & $-12.89870801$ & $2.436861339$ \\
${\bar s}\,{\bar d}\rightarrow {\bar s}\,{\bar d}\,{\bar{e}}\,{e}\,$ & $23.77022652$ & $-1.784909189$ & $888.1774209$ & $-96.27614625$ & $2.537979361$ \\
$d\,{\bar d}\rightarrow {\bar s}\,s\,{\bar{e}}\,{e}\,$ & $16.58448708$ & $-1.649072331$ & $401.9466966$ & $-54.97400388$ & $0.9377068489$ \\
\midrule
$u\,u\rightarrow u\,u\,{\bar{e}}\,{e}\,$ & $34.68778319$ & $-1.555627016$ & $1398.527044$ & $-114.8424132$ & $1.928400870$ \\
$u\,{\bar u}\rightarrow {\bar u}\,u\,{\bar{e}}\,{e}\,$ & $15.01368635$ & $-1.574355326$ & $370.2066142$ & $-51.87996832$ & $0.9992641659$ \\
${\bar u}\,{\bar u}\rightarrow {\bar u}\,{\bar u}\,{\bar{e}}\,{e}\,$ & $24.26958584$ & $-1.765978117$ & $879.6346659$ & $-96.87178341$ & $2.473041125$ \\
$d\,d\rightarrow d\,d\,{\bar{e}}\,{e}\,$ & $36.14622584$ & $-1.576878190$ & $1468.746310$ & $-119.8053289$ & $1.987999614$ \\
$d\,{\bar d}\rightarrow {\bar d}\,d\,{\bar{e}}\,{e}\,$ & $14.21098469$ & $-1.542139168$ & $356.7549911$ & $-50.18946411$ & $0.9617384975$ \\
${\bar d}\,{\bar d}\rightarrow {\bar d}\,{\bar d}\,{\bar{e}}\,{e}\,$ & $23.92072448$ & $-1.734411127$ & $889.5475431$ & $-94.53750896$ & $2.395462859$ \\
    \bottomrule
  \end{tabular}
  %\end{adjustbox}
  \caption{
   Reference evaluations of $pp \to Z(\bar{e} e)\,jj$ partonic channels.
  \label{tab:zjj-charged-benchmark}
  }
\end{table}

\begin{table}[!h]
  \renewcommand{\arraystretch}{1.2}
  \centering
  %\begin{adjustbox}{width=1\textwidth}
  \begin{tabular}{c*{5}{C{15ex}}}
    \toprule
    Channel&$H^{(1)[0]}$&$H^{(1)[1]}$&$H^{(2)[0]}$&$H^{(2)[1]}$&$H^{(2)[2]}$ \\
    \toprule
    $u\,{\bar u}\rightarrow g\,g\,{\bar \nu_e}\,{\nu_e}\,$ & $4.616164189$ & $-0.6025888043$ & $202.9021572$ & $-17.39489635$ & $0.1829226195$ \\
    $u\,g\rightarrow u\,g\,{\bar \nu_e}\,{\nu_e}\,$ & $6.110887853$ & $0.1430352973$ & $401.8763990$ & $-17.26017459$ & $-0.1146895711$ \\
    ${\bar u}\,g\rightarrow {\bar u}\,g\,{\bar \nu_e}\,{\nu_e}\,$ & $7.965404433$ & $0.1449400225$ & $425.3173590$ & $-19.51445045$ & $-0.1153313378$ \\
    $g\,g\rightarrow {\bar u}\,u\,{\bar \nu_e}\,{\nu_e}\,$ & $3.530288694$ & $0.3757333466$ & $333.1883556$ & $-22.78362992$ & $-0.2738373843$ \\
    $d\,{\bar d}\rightarrow g\,g\,{\bar \nu_e}\,{\nu_e}\,$ & $4.509314636$ & $-0.6031608019$ & $201.4010164$ & $-17.24722005$ & $0.1837811741$ \\
    $d\,g\rightarrow d\,g\,{\bar \nu_e}\,{\nu_e}\,$ & $5.967458877$ & $0.1428879852$ & $400.0634667$ & $-17.08582809$ & $-0.1146399367$ \\
    ${\bar d}\,g\rightarrow {\bar d}\,g\,{\bar \nu_e}\,{\nu_e}\,$ & $8.798476196$ & $0.1457956486$ & $435.8473278$ & $-20.52709902$ & $-0.1156196273$ \\
    $g\,g\rightarrow {\bar d}\,d\,{\bar \nu_e}\,{\nu_e}\,$ & $3.651638394$ & $0.3728586110$ & $334.2234813$ & $-22.86060740$ & $-0.2737426232$ \\
    \midrule
    $u\,{\bar u}\rightarrow {\bar d}\,d\,{\bar \nu_e}\,{\nu_e}\,$ & $14.79368691$ & $-1.619828845$ & $343.3964703$ & $-52.65540340$ & $0.8610897824$ \\
    $u\,d\rightarrow u\,d\,{\bar \nu_e}\,{\nu_e}\,$ & $25.54933256$ & $-1.699156155$ & $752.4931485$ & $-92.94204309$ & $2.266295390$ \\
    $u\,{\bar d}\rightarrow u\,{\bar d}\,{\bar \nu_e}\,{\nu_e}\,$ & $-4.739598568$ & $-1.730387658$ & $269.5220476$ & $-16.56981735$ & $2.421714538$ \\
    ${\bar u}\,d\rightarrow {\bar u}\,d\,{\bar \nu_e}\,{\nu_e}\,$ & $-2.333211885$ & $-1.817041304$ & $368.1562135$ & $-30.93411025$ & $2.581692196$ \\
    ${\bar u}\,{\bar d}\rightarrow {\bar u}\,{\bar d}\,{\bar \nu_e}\,{\nu_e}\,$ & $32.66990971$ & $-1.969841014$ & $1378.431541$ & $-130.2459919$ & $3.052355821$ \\
    $d\,{\bar d}\rightarrow {\bar u}\,u\,{\bar \nu_e}\,{\nu_e}\,$ & $14.76826187$ & $-1.624387958$ & $343.7535083$ & $-53.13569938$ & $0.8598542234$ \\
    $u\,{\bar u}\rightarrow {\bar c}\,c\,{\bar \nu_e}\,{\nu_e}\,$ & $18.00241146$ & $-1.668983991$ & $401.2897881$ & $-60.53194945$ & $1.035041878$ \\
    $u\,c\rightarrow u\,c\,{\bar \nu_e}\,{\nu_e}\,$ & $35.58944330$ & $-1.533823196$ & $1463.241933$ & $-109.2065471$ & $1.901861105$ \\
    $u\,{\bar c}\rightarrow u\,{\bar c}\,{\bar \nu_e}\,{\nu_e}\,$ & $1.357340251$ & $-1.700946139$ & $415.9641031$ & $-41.52234864$ & $2.288603183$ \\
    ${\bar u}\,c\rightarrow {\bar u}\,c\,{\bar \nu_e}\,{\nu_e}\,$ & $-2.849461263$ & $-1.799217885$ & $307.6012267$ & $-21.50850700$ & $2.622770714$ \\
    ${\bar u}\,{\bar c}\rightarrow {\bar u}\,{\bar c}\,{\bar \nu_e}\,{\nu_e}\,$ & $23.73921542$ & $-1.857508340$ & $820.2425951$ & $-95.44805217$ & $2.711621850$ \\
    $s\,d\rightarrow s\,d\,{\bar \nu_e}\,{\nu_e}\,$ & $43.19604759$ & $-1.428425233$ & $1869.812061$ & $-125.0127358$ & $1.647352563$ \\
    $s\,{\bar d}\rightarrow s\,{\bar d}\,{\bar \nu_e}\,{\nu_e}\,$ & $1.099843048$ & $-1.701729843$ & $420.5953789$ & $-41.43265686$ & $2.283405328$ \\
    ${\bar s}\,d\rightarrow {\bar s}\,d\,{\bar \nu_e}\,{\nu_e}\,$ & $-7.226102744$ & $-1.873373481$ & $255.3315894$ & $-6.781438301$ & $2.832723583$ \\
    ${\bar s}\,{\bar d}\rightarrow {\bar s}\,{\bar d}\,{\bar \nu_e}\,{\nu_e}\,$ & $24.55489192$ & $-1.909836910$ & $860.0064337$ & $-101.4781115$ & $2.849210985$ \\
    $d\,{\bar d}\rightarrow {\bar s}\,s\,{\bar \nu_e}\,{\nu_e}\,$ & $16.95797060$ & $-1.654449385$ & $383.2989502$ & $-58.43345206$ & $0.9739424630$ \\
    \midrule
    $u\,u\rightarrow u\,u\,{\bar \nu_e}\,{\nu_e}\,$ & $35.28333302$ & $-1.503354412$ & $1422.436856$ & $-109.0504016$ & $1.792474679$ \\
    $u\,{\bar u}\rightarrow {\bar u}\,u\,{\bar \nu_e}\,{\nu_e}\,$ & $15.81065643$ & $-1.582150470$ & $367.5110059$ & $-56.00827133$ & $1.060262824$ \\
    ${\bar u}\,{\bar u}\rightarrow {\bar u}\,{\bar u}\,{\bar \nu_e}\,{\nu_e}\,$ & $24.93724947$ & $-1.812945568$ & $869.5208377$ & $-96.84587137$ & $2.593120460$ \\
    $d\,d\rightarrow d\,d\,{\bar \nu_e}\,{\nu_e}\,$ & $41.09708924$ & $-1.417585686$ & $1730.181473$ & $-121.3708838$ & $1.576114437$ \\
    $d\,{\bar d}\rightarrow {\bar d}\,d\,{\bar \nu_e}\,{\nu_e}\,$ & $14.46602121$ & $-1.547677754$ & $340.2794629$ & $-52.87853322$ & $0.9977864769$ \\
    ${\bar d}\,{\bar d}\rightarrow {\bar d}\,{\bar d}\,{\bar \nu_e}\,{\nu_e}\,$ & $25.67668049$ & $-1.867923098$ & $906.4547341$ & $-102.4933716$ & $2.740064053$ \\
    \bottomrule
  \end{tabular}
  %\end{adjustbox}
  \caption{
    Reference evaluations of $pp \to Z(\bar{\nu}_e \nu_e)\,jj$ partonic channels.
    \label{tab:zjj-neutral-benchmark}
  }
\end{table}

\clearpage
\section{Six-point primary decompositions}
\label{sec:PrimDec}

In this appendix, we report on the primary decompositions of ideals
generated by pairs of the first 40 denominators in
eq.~\eqref{eq:AllDenominatorFactors}, that is, those that appear at
one loop.  We provide a set of $I_{1\leq i\leq 65}$ from which any
generates the complete set of ideals $\langle \mathcal{D}_i,
\mathcal{D}_j \rangle$, by application of either a permutation of the
external legs, or parity (angle/bracket swap).  Although beyond the
scope of this work, our set is closed under the full set of
permutations of 6 legs. All ideals are therefore ideals in $R_6$, as
defined in ref.~\cite{DeLaurentis:2022otd}.  There are 56 generating
prime ideals, which we denote as $P_{i}$. In two cases a primary
component is not prime, which we denote as $Q_{i}$. As there are fewer
generating primes than denominator pair ideals, it is clear that some
primes appear repeatedly, though potentially under some
permutation/parity operation.  We denote permutations of momentum
labels are denoted by an index set, so that $P(i_1,..., i_6)$ means
the swapping momenta, $\{ p_1,p_2,p_3,p_4,p_5,p_6 \} \rightarrow \{
p_{i_1},p_{i_2},p_{i_3},p_{i_4},p_{i_5},p_{i_6} \}$ in the
expressions. Parity conjugation of an ideal is denoted by a bar.

Computer readable expressions for these decompositions are provided in
the ancillary file \texttt{test\_primary\_decompositions.py}. This
script can be executed with \texttt{pytest}, and checks the primality
of all the $P_i$, except for eight cases that have $\Delta$ as
generator, for which the test gives inconclusive results. The test
completes in under one hour on a 13$^{\text{th}}$ Gen Intel i9 CPU
running on a single core. Furthermore, through a longer-running
variation of this test, we were able to verify primality for six more
cases, except the two generated by pairs of $\Delta$'s. Although
unlikely, these might involve non-trivial decompositions including
additional polynomials of higher degree, which our derivation was not
sensitive to. We provide them as a conjecture.

We find the following decompositions:
\begin{align}
I_1 & = \big \langle ⟨12⟩, ⟨13⟩ \big \rangle_{R_6} = P_1 \cap P_2\\
I_2 & = \big \langle ⟨12⟩, ⟨34⟩ \big \rangle_{R_6} = P_3 \\
I_3 & = \big \langle ⟨12⟩, [12] \big \rangle_{R_6} = P_4 \\
I_4 & = \big \langle ⟨12⟩, [13] \big \rangle_{R_6} = P_5 \\
I_5 & = \big \langle ⟨12⟩, [34] \big \rangle_{R_6} = P_6 \\
I_6 & = \big \langle ⟨12⟩, ⟨1|2+3|1] \big \rangle_{R_6} = P_1 \cap P_2 \cap P_5 \\
I_7 & = \big \langle ⟨12⟩, ⟨1|3+4|1] \big \rangle_{R_6} = P_2 \cap P_7 \\
I_8 & = \big \langle ⟨12⟩, ⟨3|1+2|3] \big \rangle_{R_6} = P_1 \cap P_8 \\
I_9 & = \big \langle ⟨12⟩, ⟨3|1+4|3] \big \rangle_{R_6} = P_9 \\
I_{10} & = \big \langle ⟨12⟩, ⟨3|4+5|3] \big \rangle_{R_6} = P_{10} \\
I_{11} & = \big \langle ⟨12⟩, ⟨1|3+4|2] \big \rangle_{R_6} = P_2 \cap P_7(213456) \\
I_{12} & = \big \langle ⟨12⟩, ⟨1|2+4|3] \big \rangle_{R_6} = P_1(124356) \cap P_2 \cap P_6 \\
I_{13} & = \big \langle ⟨12⟩, ⟨3|2+4|1] \big \rangle_{R_6} = P_{11} \\
I_{14} & = \big \langle ⟨12⟩, ⟨3|1+2|4] \big \rangle_{R_6} = P_1 \cap P_8(123456) \\
I_{15} & = \big \langle ⟨12⟩, ⟨3|1+5|4] \big \rangle_{R_6} = P_{12} \\
I_{16} & = \big \langle ⟨12⟩, s_{123} \big \rangle_{R_6} = P_1 \cap P_8 \\
I_{17} & = \big \langle ⟨12⟩, s_{134} \big \rangle_{R_6} = P_{13} \\
I_{18} & = \big \langle ⟨12⟩, Δ_{12|34|56} \big \rangle_{R_6} = \big \langle ⟨12⟩, (s_{34}-s_{56})^2 \big \rangle_{R_6} = Q_{14} \\
I_{19} & = \big \langle ⟨12⟩, Δ_{13|24|56} \big \rangle_{R_6} = P_{15}
\end{align}

\begin{align}
I_{20} & = \big \langle ⟨1|2+3|1], ⟨1|2+4|1] \big \rangle_{R_6} = P_2 \cap \bar P_2 \cap P_{16} \\
I_{21} & = \big \langle ⟨1|2+3|1], ⟨1|4+5|1] \big \rangle_{R_6} = P_2 \cap \bar P_2 \cap P_7(162345) \cap P_7(162345) \\
I_{22} & = \big \langle ⟨1|2+3|1], ⟨2|1+3|2] \big \rangle_{R_6} = P_1 \cap \bar P_1 \cap P_{17} \\
I_{23} & = \big \langle ⟨1|2+3|1], ⟨2|1+4|2] \big \rangle_{R_6} = P_{18} \\
I_{24} & = \big \langle ⟨1|2+3|1], ⟨2|3+4|2] \big \rangle_{R_6} = P_{19} \\
I_{25} & = \big \langle ⟨1|2+3|1], ⟨2|4+5|2] \big \rangle_{R_6} = P_{20} \\
I_{26} & = \big \langle ⟨1|2+3|1], ⟨4|2+3|4] \big \rangle_{R_6} = P_{21} \\
I_{27} & = \big \langle ⟨1|2+3|1], ⟨4|2+5|4] \big \rangle_{R_6} = P_{22} \\
I_{28} & = \big \langle ⟨1|2+3|1], ⟨4|5+6|4] \big \rangle_{R_6} = P_{23} \\
I_{29} & = \big \langle ⟨1|2+3|1], ⟨1|3+4|2] \big \rangle_{R_6} = P_2 \cap P_{24} \\
I_{30} & = \big \langle ⟨1|2+3|1], ⟨1|2+3|4] \big \rangle_{R_6} = P_1 \cap P_2 \cap P_7(142356) \\
I_{31} & = \big \langle ⟨1|2+3|1], ⟨1|2+5|4] \big \rangle_{R_6} = P_2 \cap P_{25} \\
I_{32} & = \big \langle ⟨1|2+3|1], ⟨2|1+4|3] \big \rangle_{R_6} = P_{26} \\
I_{33} & = \big \langle ⟨1|2+3|1], ⟨2|1+3|4] \big \rangle_{R_6} = P_1 \cap P_{27} \\
I_{34} & = \big \langle ⟨1|2+3|1], ⟨2|1+5|4] \big \rangle_{R_6} = P_{28} \\
I_{35} & = \big \langle ⟨1|2+3|1], ⟨4|1+2|5] \big \rangle_{R_6} = P_{29} \\
I_{36} & = \big \langle ⟨1|2+3|1], ⟨4|2+3|5] \big \rangle_{R_6} = P_{30} \\
I_{37} & = \big \langle ⟨1|2+3|1], s_{123} \big \rangle_{R_6} = P_1 \cap \bar P_1(123456) \cap P_8(321654) \cap \bar P_8(321654) \\
I_{38} & = \big \langle ⟨1|2+3|1], s_{124} \big \rangle_{R_6} = P_{31} \\
I_{39} & = \big \langle ⟨1|2+3|1], s_{145} \big \rangle_{R_6} = P_{32} \\
I_{40} & = \big \langle ⟨1|2+3|1], Δ_{12|34|56} \big \rangle_{R_6} = P_{33} \\
I_{41} & = \big \langle ⟨1|2+3|1], Δ_{14|25|36} \big \rangle_{R_6} = P_{34} \\
I_{42} & = \big \langle ⟨1|2+3|1], Δ_{23|14|56} \big \rangle_{R_6} = P_{35} \\
I_{43} & = \big \langle ⟨1|3+4|2], ⟨1|3+5|2] \big \rangle_{R_6} = P_2 \cap \bar P_2(234561) \cap P_{36}  \\
  \begin{split}
I_{44} & = \big \langle ⟨1|3+4|2], ⟨1|2+4|3] \big \rangle_{R_6} = P_1(165432) \cap \bar P_1(234561) \cap P_2 \cap  \\
& \phantom{=} \kern54mm P_6(142356) \cap \bar P_8(561234)
  \end{split}
\\
I_{45} & = \big \langle ⟨1|3+4|2], ⟨1|2+5|3] \big \rangle_{R_6} = P_2 \cap P_{37} \\
I_{46} & = \big \langle ⟨1|3+4|2], ⟨2|3+4|1] \big \rangle_{R_6} = P_{38} \\
I_{47} & = \big \langle ⟨1|3+4|2], ⟨2|3+5|1] \big \rangle_{R_6} = P_{39} \\
I_{48} & = \big \langle ⟨1|3+4|2], ⟨2|1+4|3] \big \rangle_{R_6} = P_{40} \\
I_{49} & = \big \langle ⟨1|3+4|2], ⟨2|1+5|3] \big \rangle_{R_6} = P_{41} \\
I_{50} & = \big \langle ⟨1|3+4|2], ⟨3|1+2|4] \big \rangle_{R_6} = P_6(132456) \cap P_{42}\\
I_{51} & = \big \langle ⟨1|3+4|2], ⟨3|1+5|4] \big \rangle_{R_6} = P_{43} \\
I_{52} & = \big \langle ⟨1|3+4|2], ⟨3|1+4|5] \big \rangle_{R_6} = P_1(134256) \cap \bar P_1(256134) \cap P_{44} \\
I_{53} & = \big \langle ⟨1|3+4|2], ⟨3|2+4|5] \big \rangle_{R_6} = P_{45} \\
I_{54} & = \big \langle ⟨1|3+4|2], s_{123} \big \rangle_{R_6} = P_{46} \\
I_{55} & = \big \langle ⟨1|3+4|2], s_{134} \big \rangle_{R_6} = P_1(134256) \cap \bar P_1(256134) \cap \bar P_8(341562) \cap P_8(562341)
\end{align}

\begin{align}
I_{56} & = \big \langle ⟨1|3+4|2], s_{135} \big \rangle_{R_6} = P_{47} \\
I_{57} & = \big \langle ⟨1|3+4|2], Δ_{12|34|56} \big \rangle_{R_6} = \big \langle ⟨1|3+4|2], (s_{134}-s_{234})^2 \big \rangle_{R_6} = Q_{48} \label{eq:spabdeltanonradical} \\
I_{58} & = \big \langle ⟨1|3+4|2], Δ_{12|35|46} \big \rangle_{R_6} = P_{49} \\
I_{59} & = \big \langle ⟨1|3+4|2], Δ_{13|24|56} \big \rangle_{R_6} = P_{50} \\
I_{60} & = \big \langle ⟨1|3+4|2], Δ_{13|25|46} \big \rangle_{R_6} = P_{51} \\
I_{61} & = \big \langle s_{123}, s_{124} \big \rangle_{R_6} = P_{52} \\
I_{62} & = \big \langle s_{123}, Δ_{12|34|56} \big \rangle_{R_6} = P_{53} \\
I_{63} & = \big \langle s_{123}, Δ_{14|25|36} \big \rangle_{R_6} = P_{54} \\
I_{64} & = \big \langle Δ_{12|34|56}, Δ_{12|35|46} \big \rangle_{R_6} = P_{55} \\
I_{65} & = \big \langle Δ_{12|34|56}, Δ_{13|25|46} \big \rangle_{R_6} = P_{56}
\end{align}
where
\begin{align}
P_1 & = \big \langle ⟨12⟩, ⟨13⟩, ⟨23⟩ \big \rangle_{R_6} \\
P_2 & = \big \langle \lambda_{1}^{\alpha} \big \rangle_{R_6} \\  % |1⟩
P_7 & = \big \langle ⟨12⟩, ⟨1|3+4|1], ⟨2|3+4|1] \big \rangle_{R_6} \\
P_8 & = \big \langle ⟨12⟩, \lambda_{1}^{\alpha}[13]+\lambda_{2}^{\alpha}[23]   \big \rangle_{R_6} \\  % |1+2|3]
\sqrt{Q_{14}} & = P_{14} = \big \langle ⟨12⟩, (s_{34}-s_{56}) \big \rangle_{R_6} \\
P_{16} & = \big \langle ⟨1|2+3|1], ⟨1|2+4|1], ⟨1|2+3|2+4|1⟩, \lambda_{i}^{\alpha}\leftrightarrow \tilde\lambda_{i}^{\dot\alpha} \big \rangle_{R_6} \\  % |i\rangle \leftrightarrow [i|
P_{17} & = \big \langle ⟨1|2+3|1], ⟨2|1+3|2], \tilde\lambda_{1}^{\dot\alpha}⟨12⟩⟨13⟩+\tilde\lambda_{2}^{\dot\alpha}⟨21⟩⟨23⟩+\tilde\lambda_{3}^{\dot\alpha}⟨31⟩⟨23⟩, \lambda_{i}^{\alpha}\leftrightarrow \tilde\lambda_{i}^{\dot\alpha} \big \rangle_{R_6} \\
P_{24} & = \big \langle ⟨1|2+3|1], ⟨1|3+4|2], [1|2+3|3+4|2] \big \rangle_{R_6} \\
P_{25} & = \big \langle ⟨1|2+3|1], ⟨1|3+4|2], [1|2+3|2+5|4] \big \rangle_{R_6} \\
P_{27} & = \big \langle ⟨1|2+3|1], ⟨2|1+3|4], \lambda_{1}^{\alpha}[13][14]-\lambda_{2}^{\alpha}[21][34]-\lambda_{3}^{\alpha}[31][34] \big \rangle_{R_6} \\
P_{36} & = \big \langle ⟨1|3+4|2], ⟨1|3+5|2], ⟨1|3+4|3+5|1⟩, [2|3+4|3+5|2] \big \rangle_{R_6} \\
P_{37} & = \big \langle ⟨1|3+4|2], ⟨1|2+5|3], [2|5+6|4+6|3] \big \rangle_{R_6} \\
P_{42} & = \big \langle ⟨1|3+4|2], ⟨3|1+2|4], s_{14}-s_{23} \big \rangle_{R_6}\\
  \begin{split}
P_{44} & = \big \langle ⟨1|3+4|2], ⟨3|1+4|5], ⟨6|1+5|4], \lambda_{1}^{\alpha}[15][24]+\lambda_{3}^{\alpha}[23][45]+\lambda_{4}^{\alpha}[24][45], \\
 & \phantom{=\big\langle} \; \tilde\lambda_{2}^{\dot\alpha}⟨13⟩⟨26⟩+\tilde\lambda_{3}^{\dot\alpha}⟨13⟩⟨36⟩+\tilde\lambda_{4}^{\dot\alpha}⟨14⟩⟨36⟩ \big \rangle_{R_6}
\end{split}
\\
\sqrt{Q_{48}} & = P_{48} = \big \langle ⟨1|3+4|2], (s_{134}-s_{234}) \big \rangle_{R_6} \label{eq:spabdeltaradical} 
\end{align}

Finally, while we do not carry out an exhaustive study of the primary decompositions
for denominator pair ideals involving the new denominator at two loops,
we observe that, at least for some cases, these can
be expressed entirely in terms of the prime ideals already encountered
in the one-loop analysis, e.g.,
\begin{equation}
\begin{gathered}
  \big \langle ⟨12⟩, ⟨2|3|5+6|1|2]-⟨3|4|5+6|1|3] \big \rangle_{R_6} \; = \;
  \big \langle ⟨12⟩, ⟨4|2+3|1] \big \rangle_{R_6} \; \cap \; \big \langle ⟨12⟩, [34]
  \big \rangle_{R_6} \\ \kern80mm \cap \; \big \langle ⟨12⟩, ⟨13⟩, ⟨23⟩
  \big \rangle_{R_6} \; \cap \; \big \langle \lambda_{1}^{\alpha} \big \rangle_{R_6} \, .
\end{gathered}
\end{equation}
Given the correspondence between letters and denominators, this suggests that
new letters might be identified by searching the space of intersections of prime
ideals for new ideals which are generated by two polynomials.

\section{Renormalization}
\label{sec:subtraction}

For completeness, we summarize here the conventions of
ref.~\cite{Abreu:2021asb} for the definition of finite remainders.

The UV renormalization is performed in the $\overline{\text{MS}}$ scheme by relating the bare coupling $\alpha_s^0$
to the renormalized $\alpha_s$ through
\begin{equation}\label{eq:renormCoupling}
    \alpha_s^0\mu_0^{2\epsilon}S_{\epsilon}
  =\alpha_s(\mu)\mu^{2\epsilon}\left(
  1-\frac{\beta_0}{\epsilon}\frac{\alpha_s(\mu)}{2\pi}
  +\left(\frac{\beta_0^2}{\epsilon^2}-\frac{\beta_1}{2\epsilon}\right)
  \left(\frac{\alpha_s(\mu)}{2\pi}\right)^2+\mathcal{O}
  \left(\alpha_s^3(\mu)\right)\right),
\end{equation}
where  $\mu_0$ is the dimensional
regularization scale, $S_\epsilon=(4\pi)^{\epsilon}e^{-\epsilon\gamma_E}\,$, and 
$\gamma_E$ the Euler-Mascheroni constant.  
We will suppress the dependence on $\mu$ from all quantities. 
The coefficients of the
QCD $\beta$ function $\beta_i$ are 
\begin{equation}\label{eq:betai}
  \begin{aligned}
    \beta_0&=\frac{11}{6}\NC-\frac{2}{6}\NF \,,\quad\\
    \beta_1&=\frac{17}{6}\NC^2-\frac{13}{12}\NC\NF +\mathcal{O}(\NC^{-1})\,,
  \end{aligned}
\end{equation}
omitting subleading color terms.

The color-space operator ${\bf I}$ in \cref{eqn:remainder} has a perturbative
expansion given by
\begin{equation}
  {\bf I}_\kappa = 1 + \left(  \frac{\alpha_s}{2\pi}\right) \left(\frac{N_c}{2}  \right) {\bf I}^{(1)}_\kappa + \left( \frac{\alpha_s}{2\pi} \right)^2 \left( \frac{N_c}{2} \right)^2 {\bf I}^{(2)}_\kappa + \cdots.
\end{equation}
In the LCA, the ${\bf I}_{\kappa}$ are diagonal in color space. That is, the partial amplitudes
$A$ in \cref{eqn:colorAmp} do not mix in the definition of finite remainders.
For $A_\kappa(1,2,3,4)$ the operator $\mathbf{I}^{(1)}_{\kappa}$ is given by
\begin{align}
  {\bf I}^{(1)}_{\gluon}(\epsilon)=&
  \frac{e^{\gamma_E\epsilon}}{\Gamma(1-\epsilon)}
  \left(
  \left(\frac{1}{\epsilon^2}+\frac{1}{\epsilon}\frac{\beta_0}{\NC}\right)
  \left( -s_{23}\right)^{-\epsilon}+
  \left(\frac{1}{\epsilon^2}+\frac{1}{\epsilon}\frac{\beta_0}{2\NC}+\frac{3}{4\epsilon}\right)
  \left(\left( -s_{12}\right)^{-\epsilon}+\left( -s_{34}\right)^{-\epsilon}\right)
  \right)\,,\nonumber\\
  {\bf I}^{(1)}_{\quark}(\epsilon)=&
  \frac{e^{\gamma_E\epsilon}}{\Gamma(1-\epsilon)}
  \left(\frac{1}{\epsilon^2}+\frac{3}{2\epsilon}\right)
  \left(\left( -s_{12}\right)^{-\epsilon}+\left( -s_{34}\right)^{-\epsilon}\right)\,,
\end{align}
where $s_{ij}=(p_i+p_j)^2$ and we take the Feynman prescription $s_{ij} \rightarrow s_{ij} + i 0$ wherever relevant.
And the operator~${\bf I}^{(2)}_{\kappa}$ is given by
\begin{align}\begin{split} \label{eqn:Iop}
    {\bf I}^{(2)}_{\kappa}(\epsilon)=&
  \frac{1}{2}{\bf I}^{(1)}_{\kappa}(\epsilon)
  {\bf I}^{(1)}_{\kappa}(\epsilon)
  +\frac{2\beta_0}{\NC\epsilon}{\bf I}^{(1)}_{\kappa}(\epsilon) - 
  \frac{e^{-\gamma_E\epsilon}\Gamma(1-2\epsilon)}
  {\Gamma(1-\epsilon)}
  \left(\frac{2\beta_0}{\NC\epsilon}+K\right)
  {\bf I}^{(1)}_{\kappa}(2\epsilon) \\
  &- \frac{e^{\gamma_E\epsilon}}{\epsilon\Gamma(1-\epsilon)}
  {\bf H}_{\kappa}\,,
\end{split}\end{align}
where 
\begin{equation}
K=\frac{67}{9}-\frac{\pi ^2}{3}-\frac{10}{9}\frac{\NF}{\NC}\,.
\end{equation}
The operator ${\bf H}_{\kappa}$
only depends on the number of external gluons and quarks:
\begin{equation}
  {\bf H}_{\gluon}=2H_\gluon+2H_\Quark\,,\qquad
  {\bf H}_{\Quark}=4H_\Quark\,,
\end{equation}
with
\begin{align}\begin{split}\label{eq:hexp}
  H_\gluon &= \left(\frac{\zeta_3}{2}+\frac{5}{12}+
  \frac{11\pi^2}{144}\right)
  -\left(\frac{\pi^2}{72}+\frac{89}{108}\right)\frac{N_f}{\NC}
  +\frac{5}{27}\left(\frac{N_f}{\NC}\right)^2\,,\\
  H_\Quark &=
  \left(\frac{7\zeta_3}{4}+\frac{409}{864}
  -\frac{11\pi^2}{96}\right)
  +\left(\frac{\pi^2}{48}-\frac{25}{216}\right)\frac{N_f}{\NC}\,.
\end{split}\end{align}

\end{appendix}

\bibliographystyle{JHEP}
\bibliography{main.bib}

\end{document}